\documentclass[%
reprint,
superscriptaddress,
 amsmath,amssymb,
 aps,
pre,
]{revtex4-1}

\usepackage{graphicx}
\usepackage{dcolumn}
\usepackage{bm}

\begin{document}

\preprint{APS/123-QED}

\title{Chimera states in systems of nonlocal nonidentical
phase-coupled oscillators}

\author{Jianbo Xie}
\email{swordwave@berkeley.edu}
\affiliation{Department of Physics, University of California at Berkeley, Berkeley CA 94720, USA}
%
\author{Hsien-Ching Kao}
\email{hkao@wolfram.com}
\affiliation{Wolfram Research Inc., Champaign, IL 61820, USA}

\author{Edgar Knobloch}%
\email{knobloch@berkeley.edu}
\affiliation{Department of Physics, University of California at Berkeley, Berkeley CA 94720, USA}


\date{\today}

\begin{abstract}
Chimera states consisting of domains of coherently and incoherently oscillating nonlocally-coupled phase oscillators in systems with spatial inhomogeneity are studied. The inhomogeneity is introduced through the dependence of the oscillator frequency on its location. Two types of spatial inhomogeneity, localized and spatially periodic, are considered and their effects on the existence and properties of multi-cluster and traveling chimera states are explored. The inhomogeneity is found to break up splay states, to pin the chimera states to specific locations and to trap traveling chimeras. Many of these states can be studied by constructing an evolution equation for a complex order parameter. Solutions of this equation are in good agreement with the results of numerical simulations.

\begin{description}
\item[PACS numbers]
May be entered using the \verb+\pacs{#1}+ command.
\end{description}
\end{abstract}

\pacs{Valid PACS appear here}
\maketitle


\section{Introduction}\label{intro}
Networks of coupled oscillators have been extensively studied for many years, owing to their wide applicability in physics, chemistry, and biology. As examples we mention laser arrays, Josephson junctions, populations of fireflies, etc. \cite{Winfree1980,Kuramoto1984book,Nixon2012,Strogatz1993,Mirollo1990}. The phase-only models have proved to provide useful models for systems with weak coupling. The best known model of this type is the Kuramoto model in which the oscillators are described by phase variables $\theta_i$ and coupled to others through a sinusoidal function \cite{ Kuramoto1984PTPS,Crawford1994,Strogatz2000}. These models exhibit a transition to collective synchronization as the coupling strength increases, a process that has been described as a phase transition. A general form of these systems is as follows: 
\begin{equation}
\frac{d\theta_i}{dt} = \omega_i - \frac{k}{N}\sum\limits_{j=1}^{N}G_{ij}\sin(\theta_i-\theta_j+\alpha). \label{phase_eq_discrete}
\end{equation}
Here $\omega_i$ is the natural frequency of oscillator $i$, $G_{ij}$ represents the coupling between oscillators $i$ and $j$, $\alpha$ is a phase lag and $k$ is the overall coupling strength. A general treatment of this system is not easy, and two types of simplifications are commonly used. One of them is to assume global coupling among the oscillators, and assign the natural frequencies $\omega_i$ randomly and independently from some prespecified distribution \cite{Strogatz2000,RMP2005,Martens2009}. The second tractable case arises when all the oscillators are assumed to be identical, and the coupling $G_{ij}$ is taken to be local, eg., nearest-neighbor coupling \cite{Wiley2006}. The intermediate case of nonlocal coupling is harder but the phenomena described by the resulting model are much richer.

In this paper we suppose that the oscillators are arranged on a ring. In the continuum limit $N\to\infty$ the system is described by the nonlocal equation
\begin{equation}
\frac{\partial \theta}{\partial t}=\omega(x)-\int G(x-y)\sin[\theta(x,t)-\theta(y,t)+\alpha]\,dy\label{phase_eq}
\end{equation}
for the phase distribution $\theta(x,t)$. When the oscillators are identical ($\omega$ is a constant) and $G(x) = \frac{\kappa}{2}\exp(-\kappa|x|)$ this system admits a new type of state in which a fraction of the oscillators oscillate coherently (i.e., in phase) while the phases of the remaining oscillators remain incoherent \cite{Kuramoto2002}. In this paper we think of this state, nowadays called a {\it chimera} state \cite{Abrams2004}, as a localized structure embedded in a ``turbulent'' background. Subsequent studies of this unexpected state with different coupling functions $G(x)$ have identified a variety of different one-cluster and multi-cluster chimera states \cite{Abrams2006,Abrams2008,Sethia2008,Laing2009Chaos,Omelchnko2010,Wolfrum2011PRE,Laing2012Chaos,Zhu2012,Ujjwal2013,O2013,BickarXiv}, consisting of clusters or groups of adjacent oscillators oscillating in phase with a common frequency. The clusters are almost stationary in space, although their position (and width) fluctuates under the influence of the incoherent oscillators on either side. Recently, a new type of chimera state has been discovered, a traveling chimera state \cite{XKK2014}. In this state the leading edge plays the role of a synchronization front, which kicks oscillators into synchrony with the oscillators behind it, while the trailing front kicks oscillators out of synchrony; these two fronts travel with the same speed, forming a bound state.


It is natural to ask whether these states persist in the presence of spatial inhomogeneity, in particular, in the presence of spatial inhomogeneity in the natural frequency distribution ($\omega=\omega(x)$). In \cite{Laing2009Chaos} Laing demonstrated the robustness of the chimera state with respect to inhomogeneity in a two population model, which can be considered to be the simplest model exhibiting chimera states \cite{Abrams2008}. However, in this model there is no spatial structure to either population. Consequently, we focus in this paper on a ring of adjacent oscillators with a prescribed but nonuniform frequency distribution $\omega(x)$, where the continuous variable $x$ ($-\pi < x\le\pi$) represents position along the ring. To be specific, two classes of inhomogeneity are considered, a bump inhomogeneity $\omega(x) = \omega_0 \exp (-\kappa |x|)$, $\kappa>0$, and a periodic inhomogeneity $\omega(x) = \omega_0 \cos(lx)$, where $l$ is a positive integer.  In each case we follow \cite{XKK2014} and study the coupling functions
\begin{eqnarray}
&&G^{(1)}_n(x)\equiv\cos(nx), \nonumber  \\
&&G^{(2)}_n(x)\equiv \cos(nx)+\cos[(n+1)x],\nonumber
\end{eqnarray}
where $n$ is an arbitrary positive integer. These choices are motivated by biological systems in which coupling between nearby oscillators is often attractive while that between distant oscillators may be repelling \cite{Ermentrout2010}. There are several advantages to the use of these two types of coupling. The first is that these couplings allow us to obtain chimera states with random initial conditions. The second is that we can identify a large variety of new states for suitable parameter values, including (a) splay states, (b) stationary multi-cluster states with evenly distributed coherent clusters, (c) stationary multi-cluster states with unevenly distributed clusters, (d) a fully coherent state traveling with a constant speed, and (e) a single cluster traveling chimera state \cite{XKK2014}.

In this paper, we analyze the effect of the two types of inhomogeneity in $\omega$ on each of these states and describe the results in terms of the parameters $\omega_0$ and $\kappa$ or $l$ describing the strength and length scale of the inhomogeneity, while varying the parameter $\beta\equiv \frac{\pi}{2}-\alpha$ representing the phase lag $\alpha$. In Section \ref{eff_eq}, we briefly review the notion of a local order parameter for studying chimera states and introduce the self-consistency equation for this quantity. In Sections \ref{rot}, \ref{trav_coh} and \ref{trav} we investigate, respectively, the effect of inhomogeneity on rotating states (including splay states and stationary chimera states), traveling coherent states and the traveling chimera state. We conclude in Section \ref{conclusion} with a brief summary of the results and directions for future research. 

\section{Effective equation}\label{eff_eq}

Equation (\ref{phase_eq}) is widely used in studies of chimera states. An equivalent description can be obtained by constructing an equation for the local order parameter $z(x,t)$ defined as the local spatial average of $\exp[i\theta(x,t)]$,
\begin{equation}
z(x,t)\equiv\lim_{\delta\rightarrow 0^+}\frac{1}{\delta}\int_{-\delta/2}^{\delta/2}e^{i\theta(x+y,t)}\,dy.\label{localmean}
\end{equation}
The evolution equation for $z$ then takes the form \cite{Wolfrum2011,Pikovsky2008,Laing2009PhysD}
\begin{equation}
z_t=i\omega(x) z+\frac{1}{2}\left(e^{-i\alpha}Z-e^{i\alpha}z^2Z^*\right),\label{effective_eq}
\end{equation}
where $Z(x,t)\equiv K[z](x,t)$ and $K$ is a compact linear operator defined via the relation
\begin{equation}
K[u](x,t)\equiv\int_{-\pi}^{\pi}G(x-y)u(y,t)\,dy.
\end{equation}

A derivation of Eq.~(\ref{effective_eq}) based on the Ott--Antonsen Ansatz \cite{OA2008} is given in the Appendix. Equation~(\ref{effective_eq}) can also be obtained directly from Eq.~(\ref{phase_eq}) using the change of variable
\begin{equation}
z(x,t)\equiv \exp\left[i\theta(x,t)\right].\label{changeofvar}
\end{equation}

An important class of solutions of Eq.~(\ref{phase_eq}) consists of stationary rotating solutions, i.e., states of the form 
\begin{equation}
z(x,t)=\tilde{z}(x)e^{-i\Omega t},
\end{equation}
whose common frequency $\Omega$ satisfies the nonlinear eigenvalue relation
\begin{equation}
i\left[\Omega+\omega(x)\right]\tilde{z}+\frac{1}{2}\left[e^{-i\alpha}\tilde{Z}(x)-\tilde{z}^2e^{i\alpha}\tilde{Z}^*(x)\right] = 0.\label{tw1}
\end{equation}
Here $\tilde{z}(x)$ describes the spatial profile of the rotating solution and $\tilde{Z}\equiv K[\tilde{z}]$.

Solving Eq.~(\ref{tw1}) as a quadratic equation in $\tilde{z}$ we obtain
\begin{equation}
\tilde{z}(x)=e^{i\beta}\frac{\Omega+\omega(x)-\mu(x)}{\tilde{Z}^*(x)}=\frac{e^{i\beta}\tilde{Z}(x)}{\Omega+\omega(x)+\mu(x)}.\label{zx}
\end{equation} 
The function $\mu$ is chosen to be $[(\Omega+\omega)^2-|\tilde{Z}|^2]^{1/2}$ when $|\Omega+\omega|>|\tilde{Z}|$ and $i[|\tilde{Z}|^2-(\Omega+\omega)^2]^{1/2}$ when $|\Omega+\omega|<|\tilde{Z}|$. This choice is dictated by stability considerations, and in particular the requirement that the essential spectrum of the linearization about the rotating solution is either stable or neutrally stable \cite{Wolfrum2011}. The coherent (incoherent) region corresponds to the subdomain of $(-\pi,\pi]$ where $|\Omega+\omega(x)|$ falls below (above) $|\tilde{Z}(x)|$. Substitution of expression (\ref{zx}) into the definition of $\tilde{Z}(x)$ now leads to the self-consistency relation
\begin{equation}
\tilde{Z}(x)=\left<G(x-y)e^{i\beta}\frac{\Omega+\omega(y)-\mu(y)}{\tilde{Z}^*(y)}\right>.\label{self-consistency}
\end{equation}
Here the bracket $\left<\cdot\right>$ is defined as the integral over the interval $[-\pi,\pi]$, i.e.,
\begin{equation}
\left<u\right>\equiv\int_{-\pi}^{\pi}u(y)\,dy.
\end{equation}
In the following we write
\begin{equation}
\tilde{Z}(x)=R(x)e^{i\Theta(x)}.\label{local-op}
\end{equation}
and refer to $R(x)$ and $\Theta(x)$ as the amplitude and phase of the complex order parameter $\tilde{Z}(x)$.




\section{Stationary rotating solutions}\label{rot}

As shown in \cite{XKK2014}, Eq.~(\ref{phase_eq}) with constant natural frequency $\omega$ exhibits both stationary rotating solutions (splay states and stationary chimera states) and traveling solutions (traveling coherent states and traveling chimera states) for suitable coupling functions $G(x)$. In this section, we investigate the effect of spatial inhomogeneity in $\omega$ (i.e., $\omega=\omega(x)$) on the splay states and on stationary chimera states, focusing on the case $G^{(1)}_n(x)=\cos nx$ studied in \cite{XKK2014}. We find that when the inhomogeneity is sufficiently weak, the above solutions persist. However, as the magnitude of the inhomogeneity increases, new types of solutions are born. The origin and spatial structure of these new states can be understood with the help of the self-consistency relation (\ref{self-consistency}). Since Eq.~(\ref{self-consistency}) is invariant under the transformation $z\rightarrow ze^{i\phi}$ with $\phi$ an arbitrary real constant, the local order parameter $\tilde{Z}(x)$ for the coupling function $G(x)=\cos nx$ can be written as
\begin{equation}
\tilde{Z}(x) = a\cos nx+b\sin nx,\label{ansatz0}
\end{equation}
where $a$ is positive and $b\equiv b_r+ib_i$ with $b_r$ and $b_i$ both real. Substituting the Ansatz (\ref{ansatz0}) into Eq.~(\ref{self-consistency}) we obtain the following pair of integral-algebraic equations
\begin{eqnarray}
ae^{-i\beta}&=&\left<\frac{\cos(ny)(\tilde{\Omega}(y)-\mu(y))}{a\cos ny+b^*\sin ny}\right>,\label{sc-full1}\\
be^{-i\beta}&=&\left<\frac{\sin(ny)(\tilde{\Omega}(y)-\mu(y))}{a\cos ny+b^*\sin ny}\right>,
\label{sc-full2}
\end{eqnarray}
where $\mu(y)=(\tilde{\Omega}(y)^2-|a\cos(ny)+b\sin(ny)|^2)^{1/2}$ and $\tilde{\Omega}(y)\equiv\Omega+\omega(y)$. These equations may also be written in the more convenient form
\begin{eqnarray}
&\left<\tilde{\Omega}(y)-\mu(y)\right>=e^{-i\beta}(a^2+|b|^2),\label{sc-full12}\\
&\left<\frac{\left(a^2-b^2\right)\sin(2ny)-2ab\cos(2ny)}{\tilde{\Omega}(y)+\mu(y)}\right>=0.
\label{sc-full22}
\end{eqnarray}

In the following we consider two choices for the inhomogeneity $\omega(x)$, a bump $\omega(x)=\omega_0\exp{(-\kappa |x|)}$, $\kappa>0$, and a periodic inhomogeneity $\omega(x)=\omega_0\cos(lx)$ where $l$ is a positive integer. The resulting equations possess an important and useful symmetry, $x\rightarrow -x$, $b\rightarrow -b$. Moreover, for $\omega(x)$ satisfying $\omega(x+2\pi/l)=\omega(x)$ with $l$ an integer, a solution $\tilde{Z}(x)$ of the self-consistency relation implies that $\tilde{Z}(x+2m\pi/l)$ is also a solution. Here $m$ is an arbitrary integer.

In the following we use Eqs.~(\ref{sc-full12}) and (\ref{sc-full22}) repeatedly to study the changes in both the splay states and the stationary chimera states as the magnitude $\omega_0>0$ of the inhomogeneity increases, and compare the resulting predictions with numerical simulation of $N=512$ oscillators evenly distributed in $[-\pi,\pi]$. 

\subsection{Bump inhomogeneity: $\omega=\omega_0\exp{(-\kappa |x|)}$}

We now consider the case where $\omega(x)$ has a bump defect at $x=0$. For the spatial profile of the defect we pick $\omega(x) = \omega_0 \exp{(-\kappa |x|)}$, where $\omega_0>0$ and $\kappa>0$ are parameters that can be varied.

\subsubsection{Effect on splay states}

When $G(x) = \cos(nx)$ and $\omega$ is a constant, Eq.~(\ref{phase_eq}) exhibits so-called splay state solutions with $\theta(x,t)=qx-\Omega t$, where $\Omega$ is the overall rotation frequency and $q$ is an integer called the twist number. The phase in this type of solution drifts with speed $c=\Omega/q$ to the right but the order parameter is stationary. Consequently we think of the splay states as a stationary rotating states. Linear stability analysis for $G(x)=\cos nx$ shows that the splay state is stable when $|q|=n$ \cite{XKK2014}, a result that is easily confirmed in simulations starting from randomly distributed initial phases.  In the following we consider the case $n=1$ and simulate $N=512$ oscillators evenly distributed in $[-\pi,\pi]$ for different values of $\omega_0$ and $\kappa$. 
\begin{figure}
\includegraphics[height=3cm]{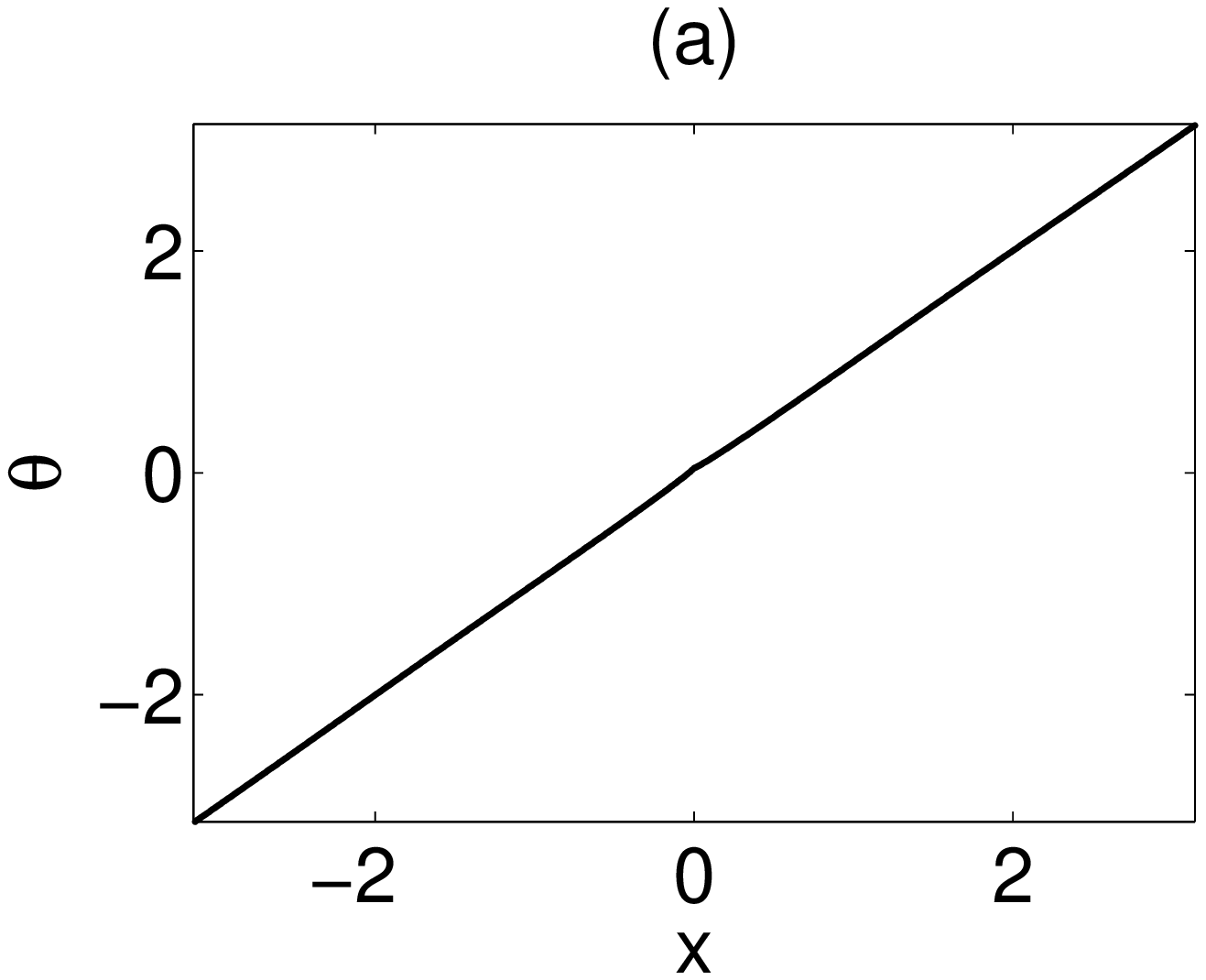}
\includegraphics[height=3cm]{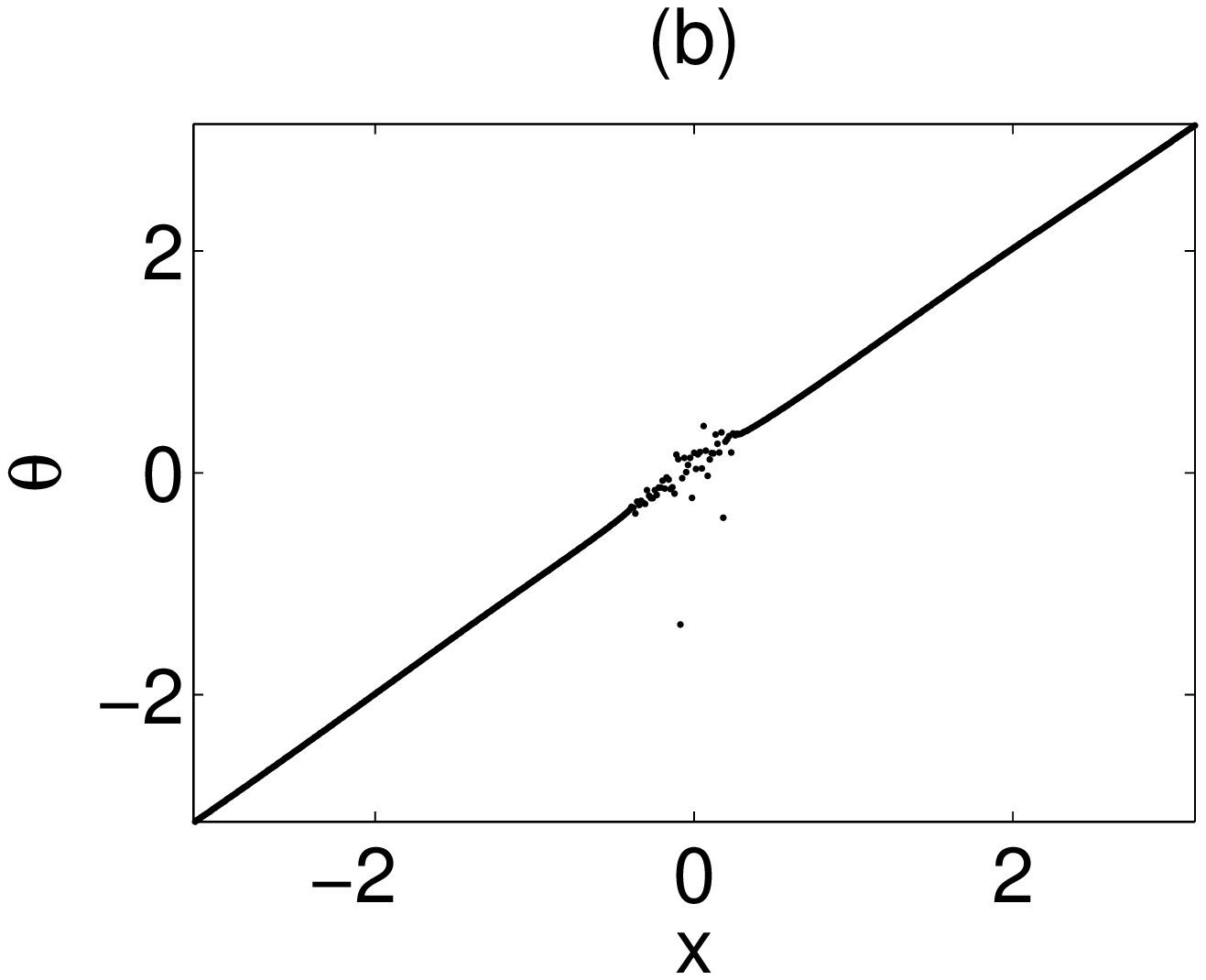}
\includegraphics[height=3cm]{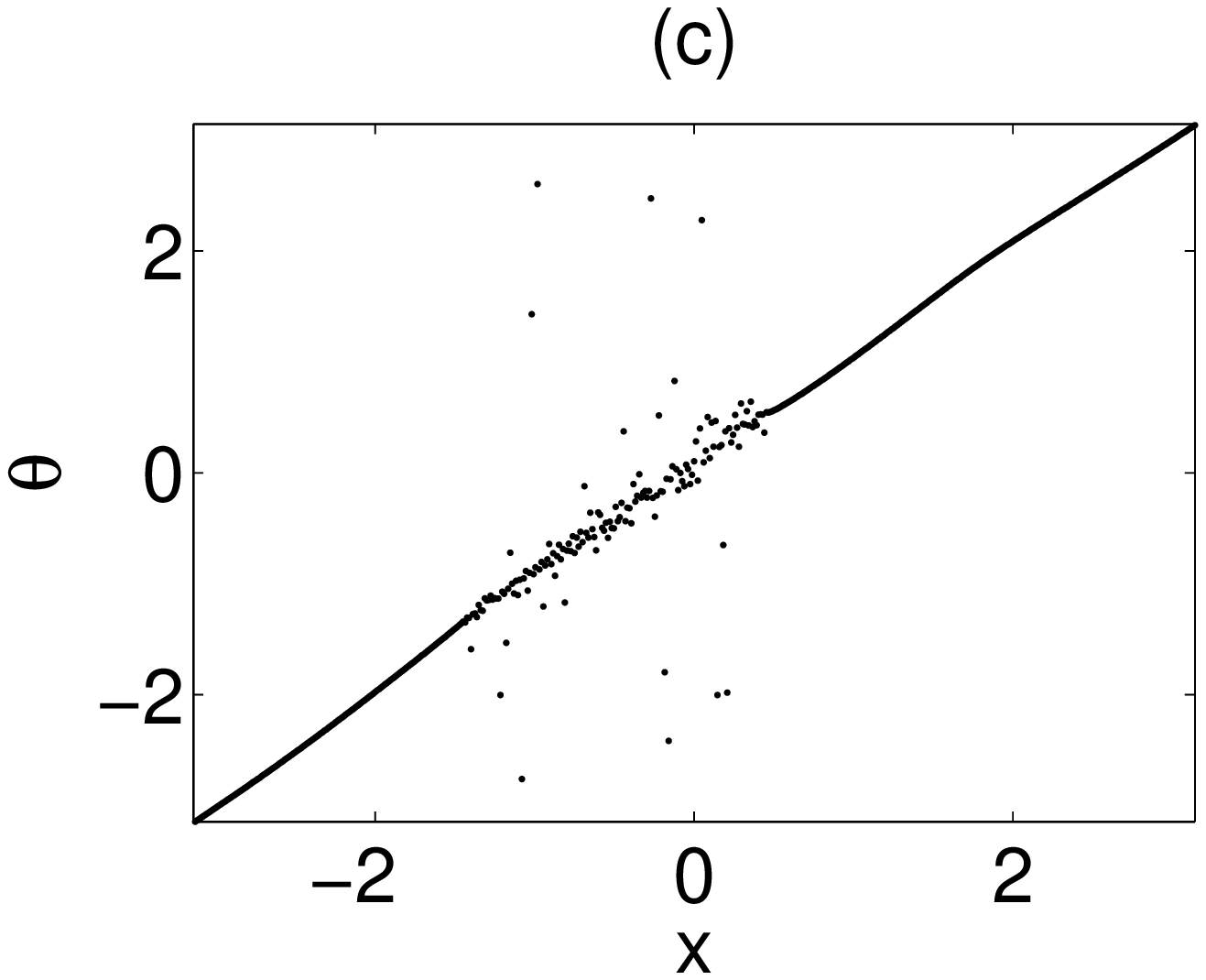}
\includegraphics[height=3cm]{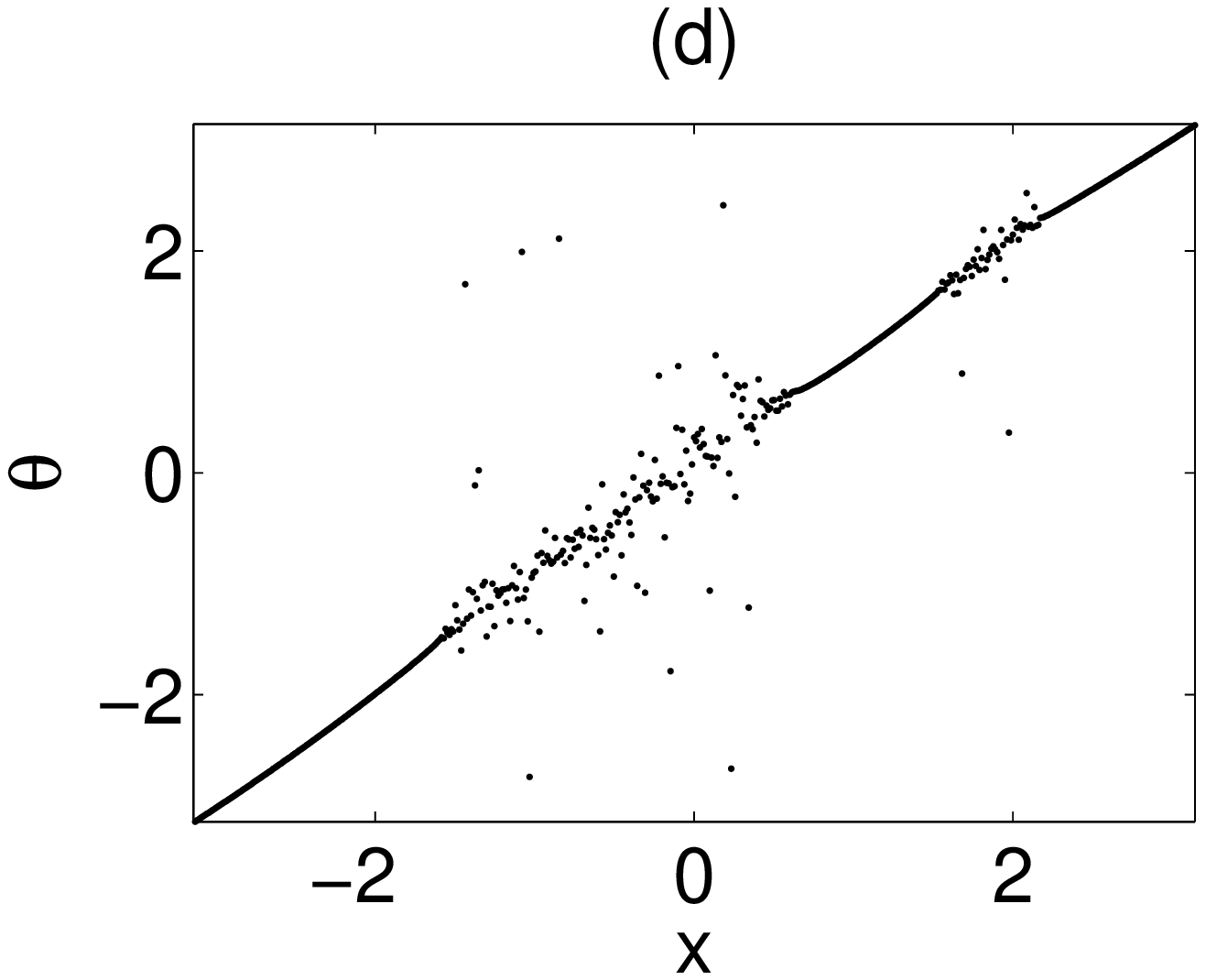}
\caption{The phase distribution $\theta(x)$ for splay states observed with $G(x) = \cos(x)$, $\omega(x) = \omega_0 \exp(-2|x|)$ and $\beta = 0.05$. (a) $\omega_0 = 0.006$. (b) $\omega_0 = 0.02$. (c) $\omega_0 = 0.05$. (d) $\omega_0 = 0.1$. The states travel to the right ($\Omega>0$). }
\label{fig:cosx_bump_splay_four}
\end{figure} 

\begin{figure}
\includegraphics[height=3cm]{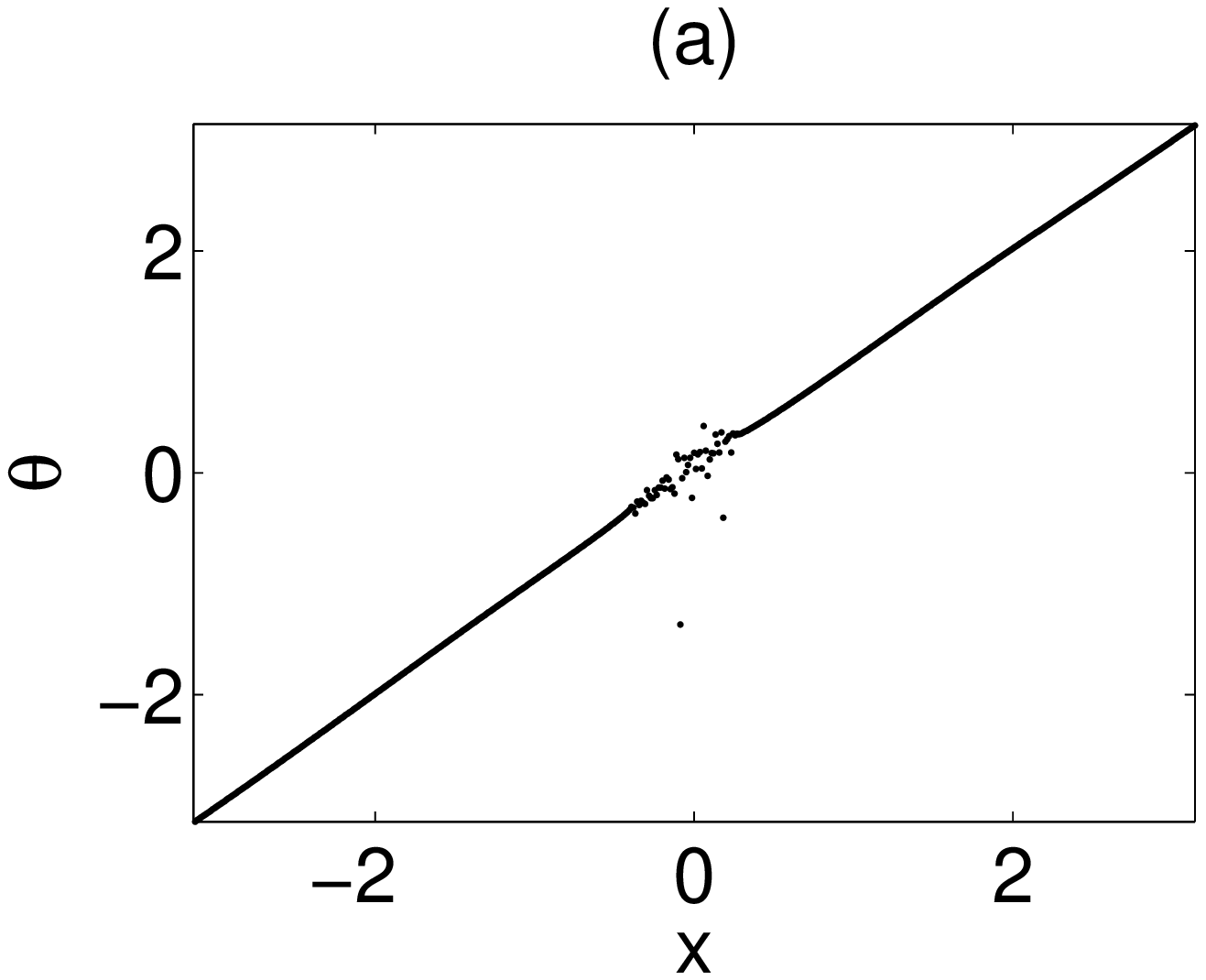}
\includegraphics[height=3cm]{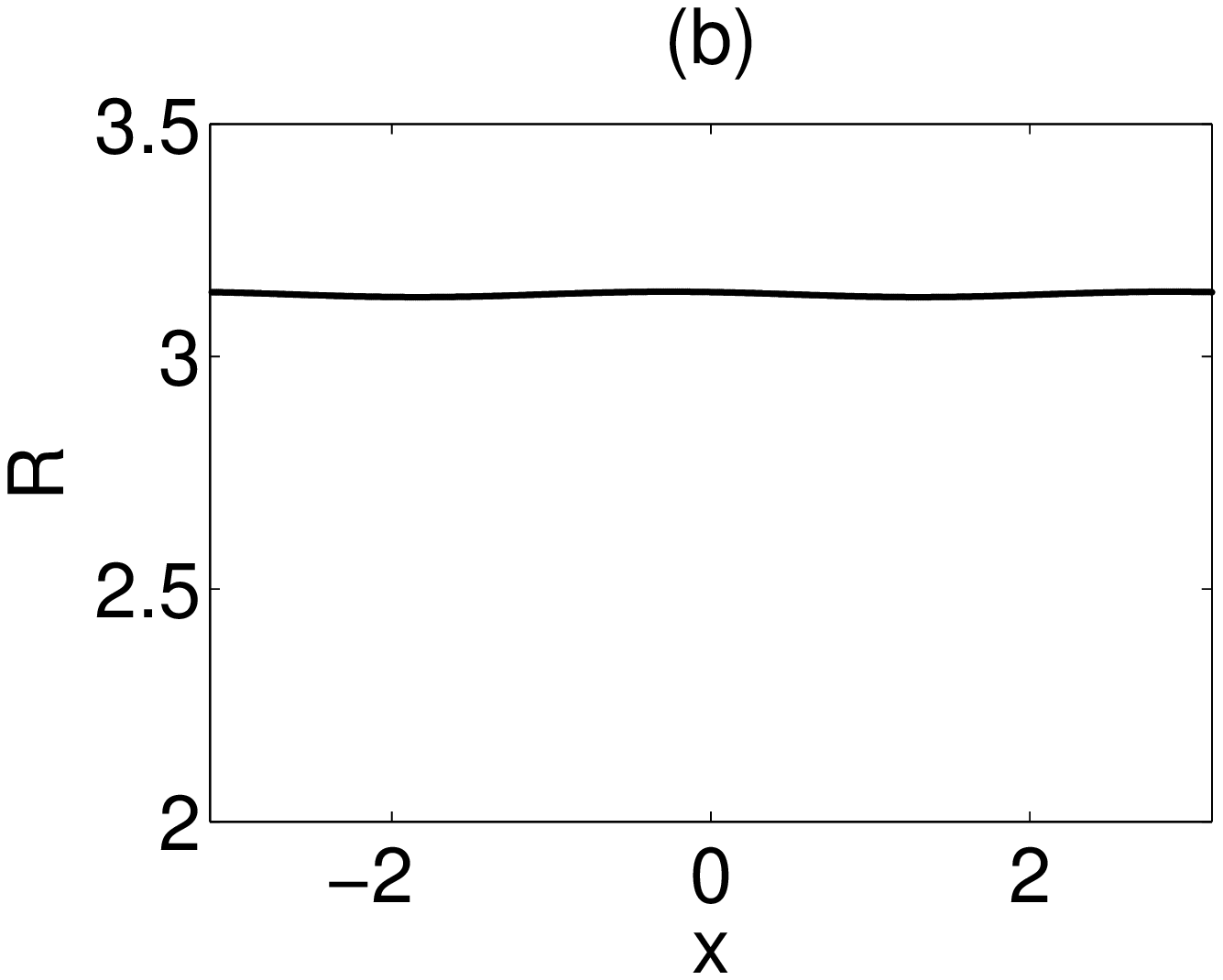}
\includegraphics[height=3cm]{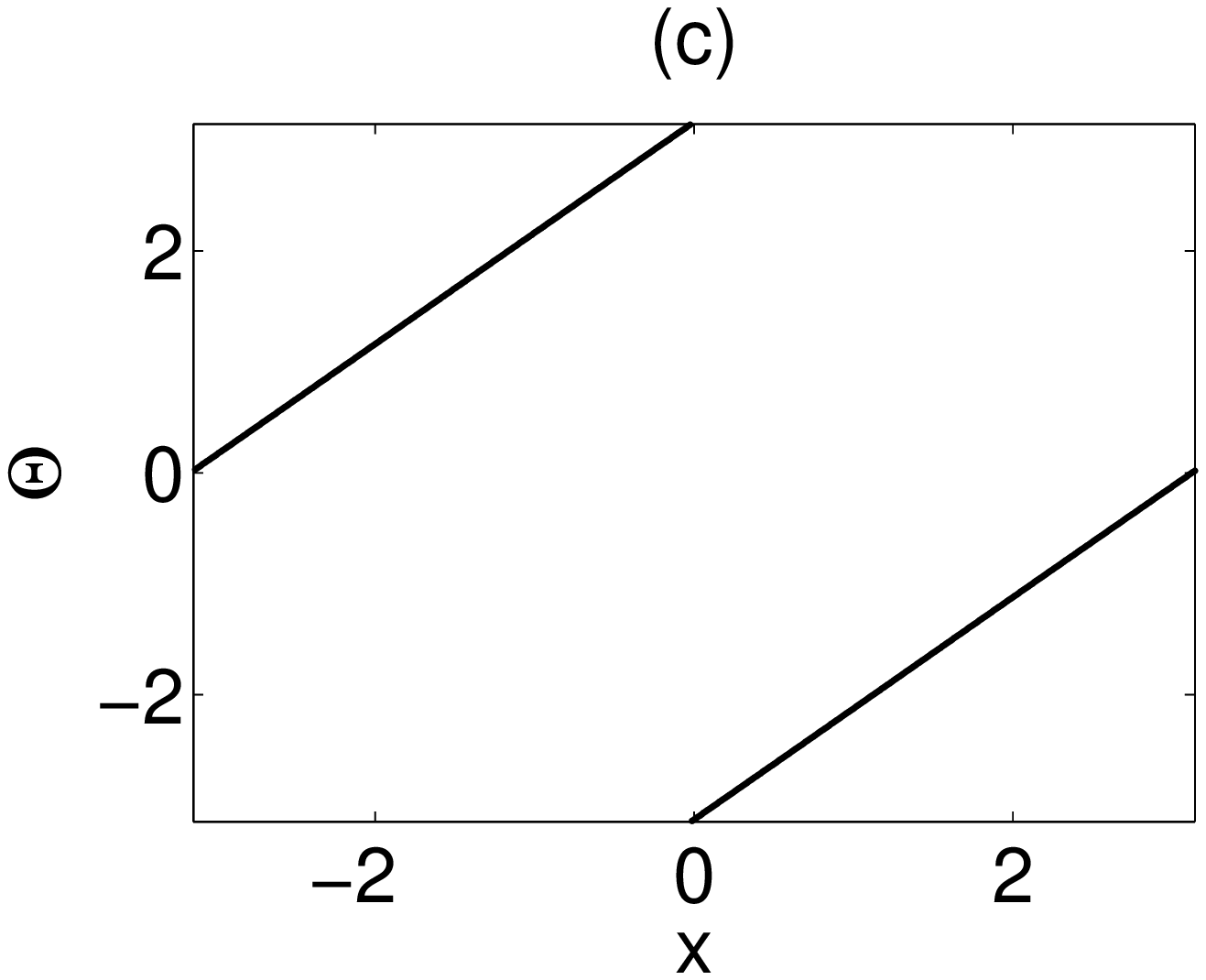}
\includegraphics[height=3cm]{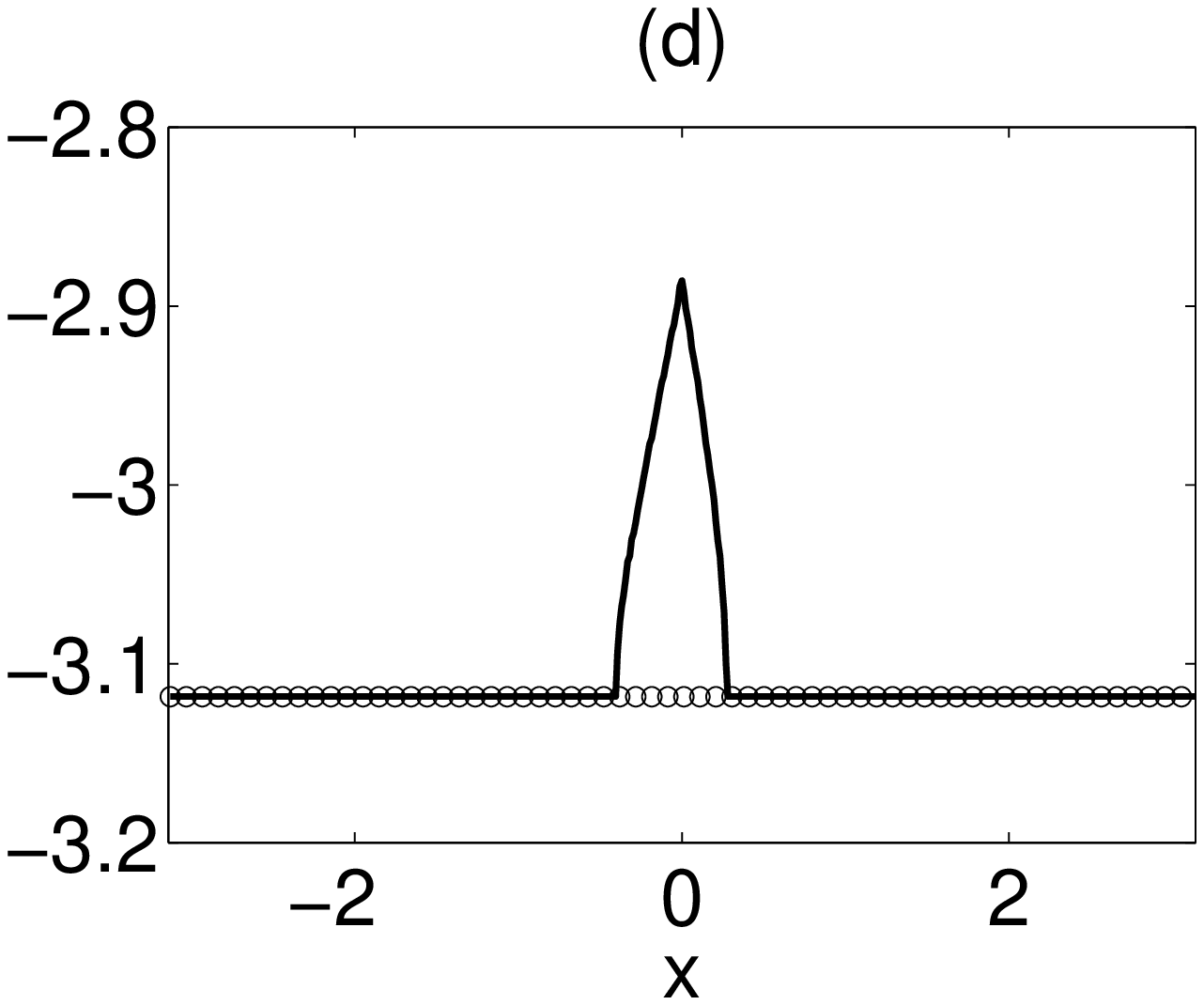}
\caption{(a) A snapshot of the phase distribution $\theta(x,t)$ for $G(x) = \cos(x)$ and $\omega(x) = 0.02 \exp(-2|x|)$. (b) The corresponding $R(x)$. (c) The corresponding $\Theta(x)$. (d) The oscillator frequency ${\bar \theta}_t$ averaged over the time interval $0<t<1000$ (solid line). In the coherent region ${\bar \theta}_t$ coincides with the global oscillation frequency $-\Omega$ (open circles). All simulations are done with $\beta = 0.05$ and $N=512$.}
\label{fig:cosx_bump_splay}
\end{figure}

Figure~\ref{fig:cosx_bump_splay_four} shows how the splay solutions change as $\omega_0$ increases. When $\omega_0$ becomes nonzero but remains small, the splay states persist but their phase $\theta(x,t)$ no longer varies uniformly in space (Fig.~\ref{fig:cosx_bump_splay_four}(a)). Instead
\begin{equation}
\theta(x,t)\equiv \phi(x)-\Omega t,\label{splay_ansatz}
\end{equation}
where $\phi(x)$ is a continuous function of $x$ with $\phi(\pi)-\phi(-\pi)=2\pi q$. We refer to this type of state as a near-splay state. As $\omega_0$ becomes larger an incoherent region appears in the vicinity of $x=0$, with width that increases with increasing $\omega_0$ (Figs.~\ref{fig:cosx_bump_splay_four}(b,c)). We refer to this type of state as a chimera splay state. As $\omega_0$ increases further and exceeds a second threshold, a new region of incoherence is born (Fig.~\ref{fig:cosx_bump_splay_four}(d)). Figure \ref{fig:cosx_bump_splay} provides additional information about the partially coherent near-splay state in Fig.~\ref{fig:cosx_bump_splay_four}(b). The figure shows the real order parameters $R(x)$ and $\Theta(x)$ together with ${\bar \theta}_t$, the local rotation frequency averaged over a long time interval (Fig.~\ref{fig:cosx_bump_splay}(d)), and reveals that in the coherent region the oscillation frequencies are identical (with ${\bar \theta}_t=-\Omega$), with an abrupt but continuous change in the frequency distribution within the incoherent region. The figures also reveals that as $\omega_0$ increases the incoherent region develops a stronger and stronger asymmetry with respect to $x=0$, the bump maximum. This is a consequence of the asymmetry introduced by the direction of travel, i.e., the sign of the frequency $\Omega$ in Eq.~(\ref{splay_ansatz}) as discussed further below.

\begin{figure}
\includegraphics[height=4cm]{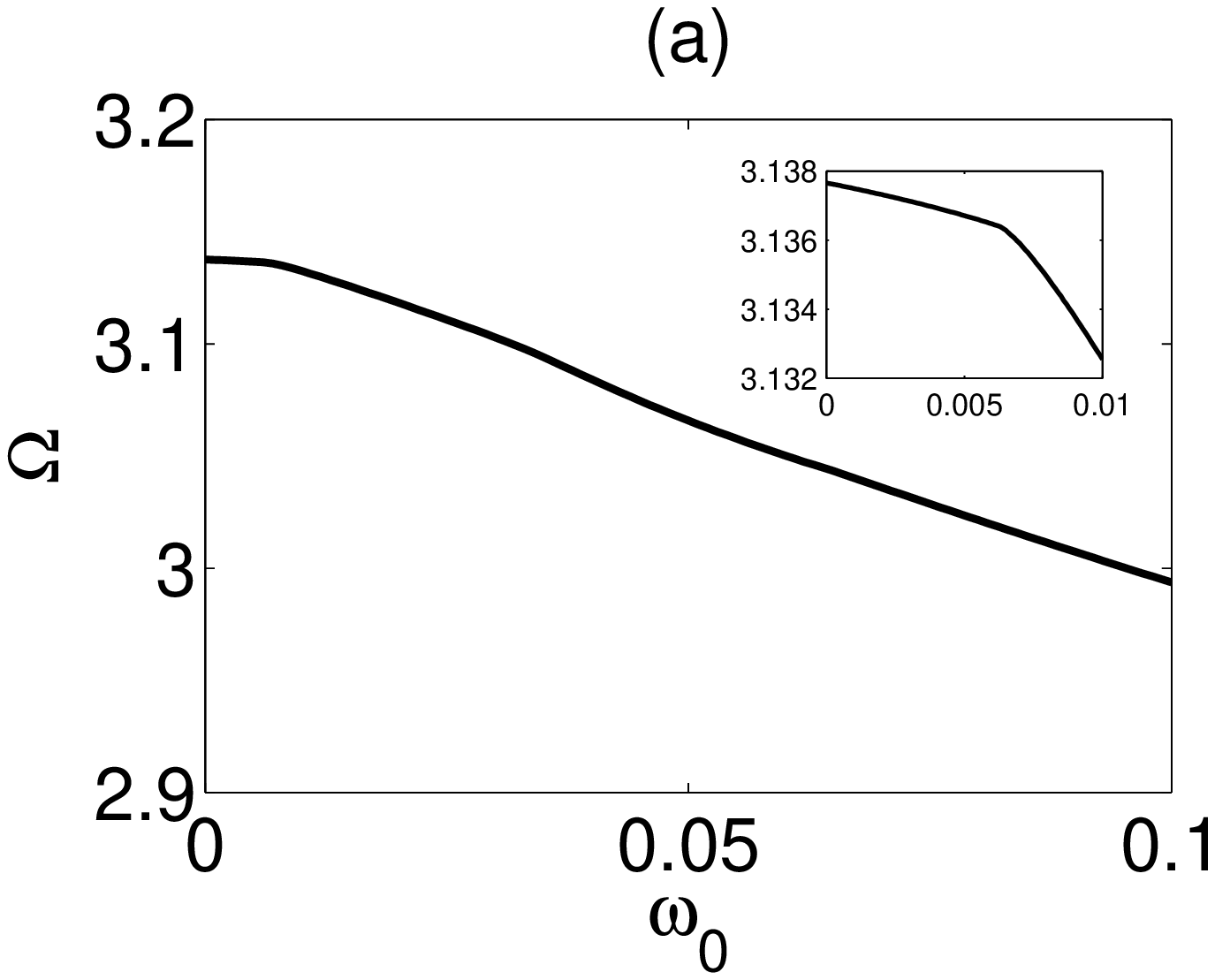}
\includegraphics[height=4cm]{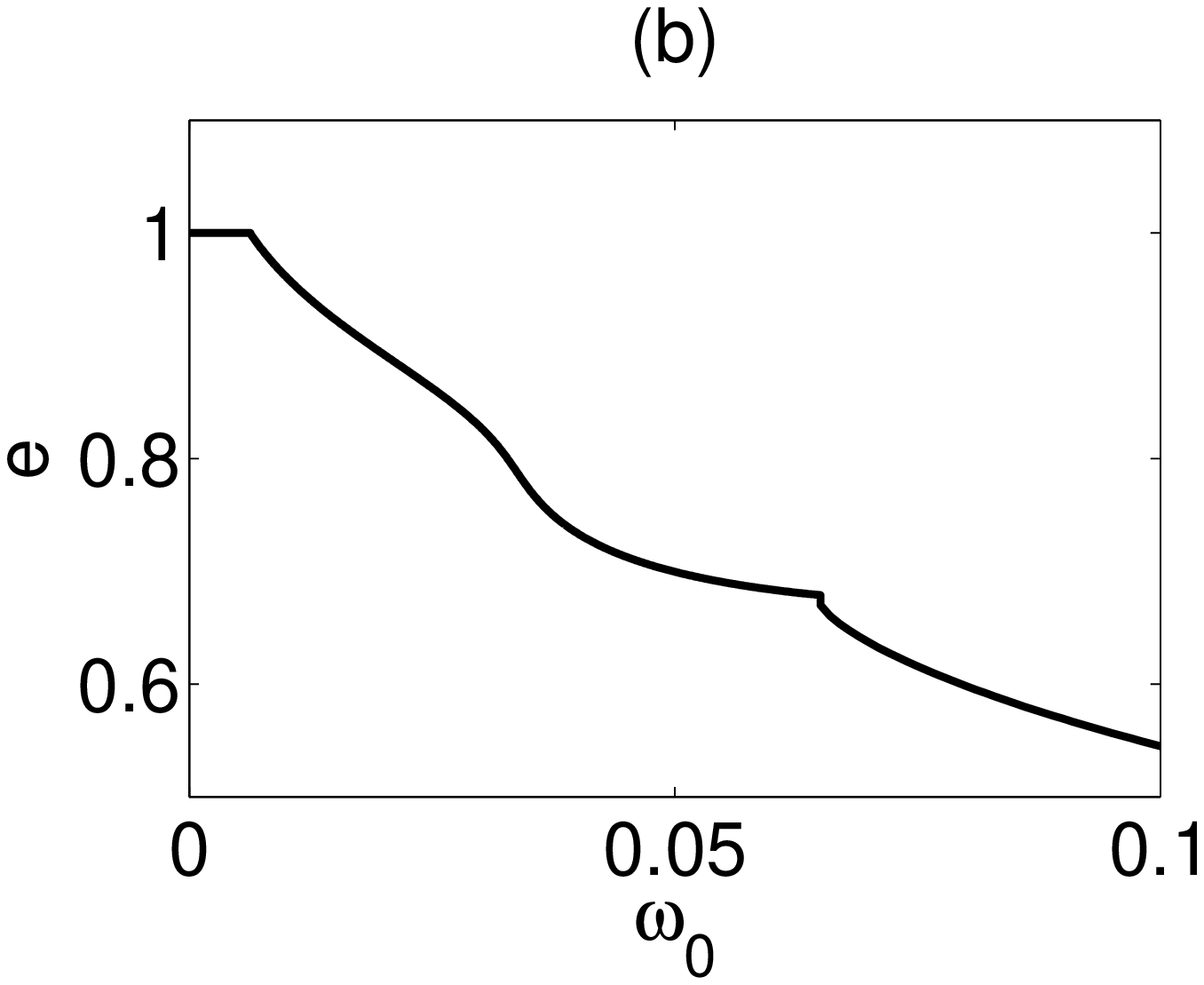}
\includegraphics[height=4cm]{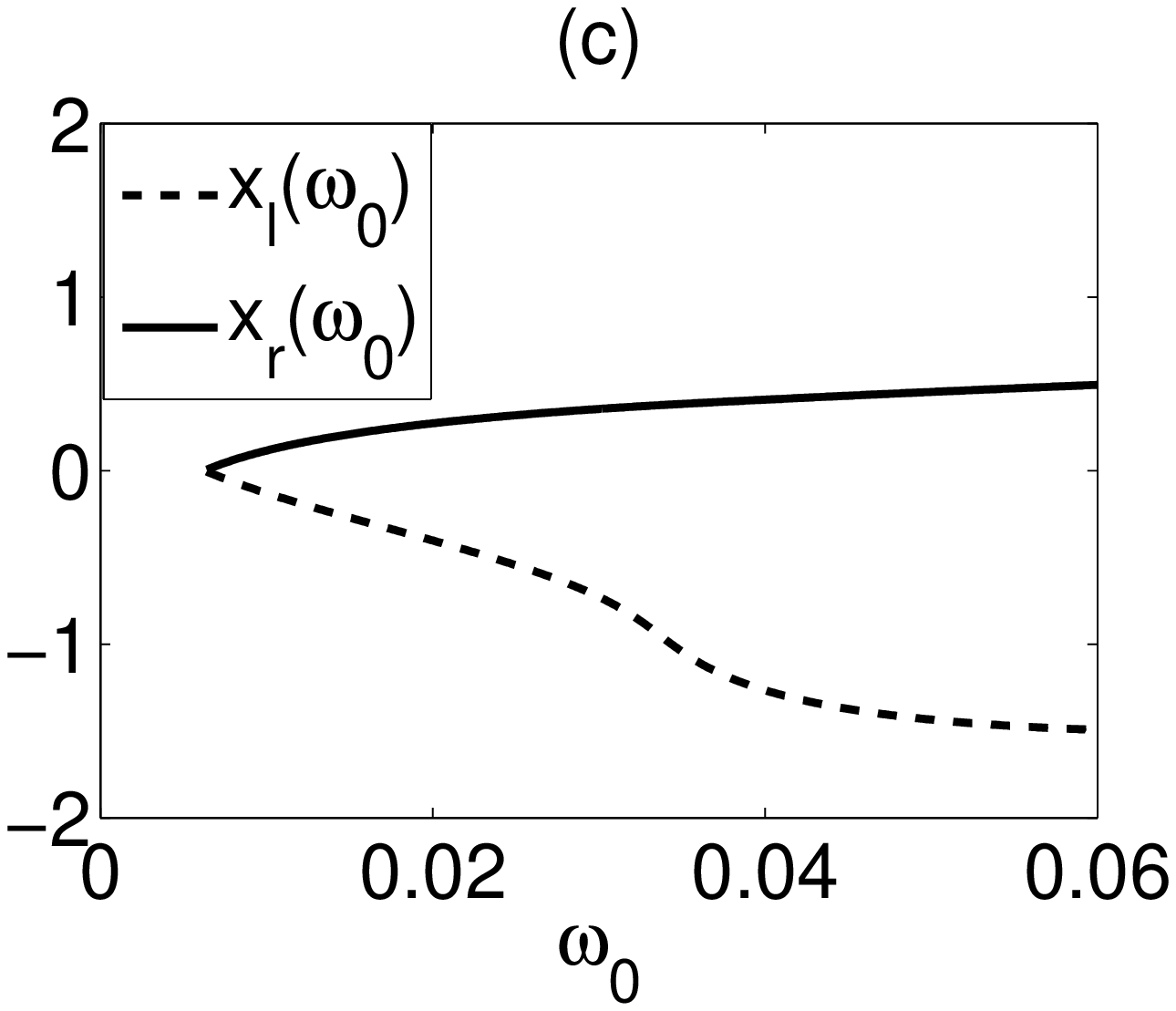}
\caption{(a) The overall frequency $\Omega$, (b) the fraction $e$ of the domain occupied by the coherent oscillators, and (c) the width of the incoherent region $x_l\le x\le x_r$, all as functions of the parameter $\omega_0$. The calculation is for $\kappa = 2$, $\beta = 0.05$ and $N=512$.}
\label{fig:cosx_bump_omega0_continue_splay}
\end{figure}

\begin{figure}
\includegraphics[height=4cm]{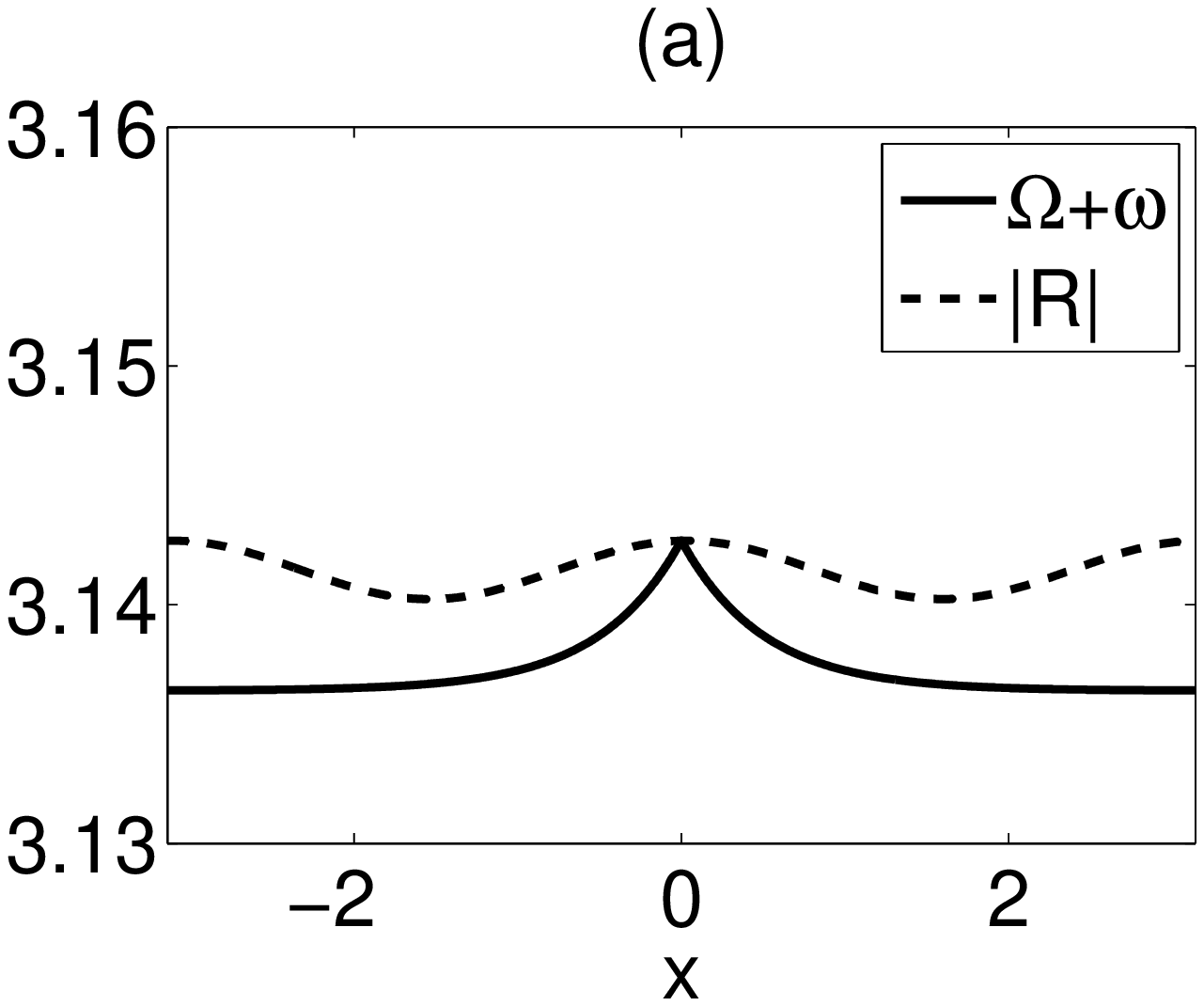}
\includegraphics[height=4cm]{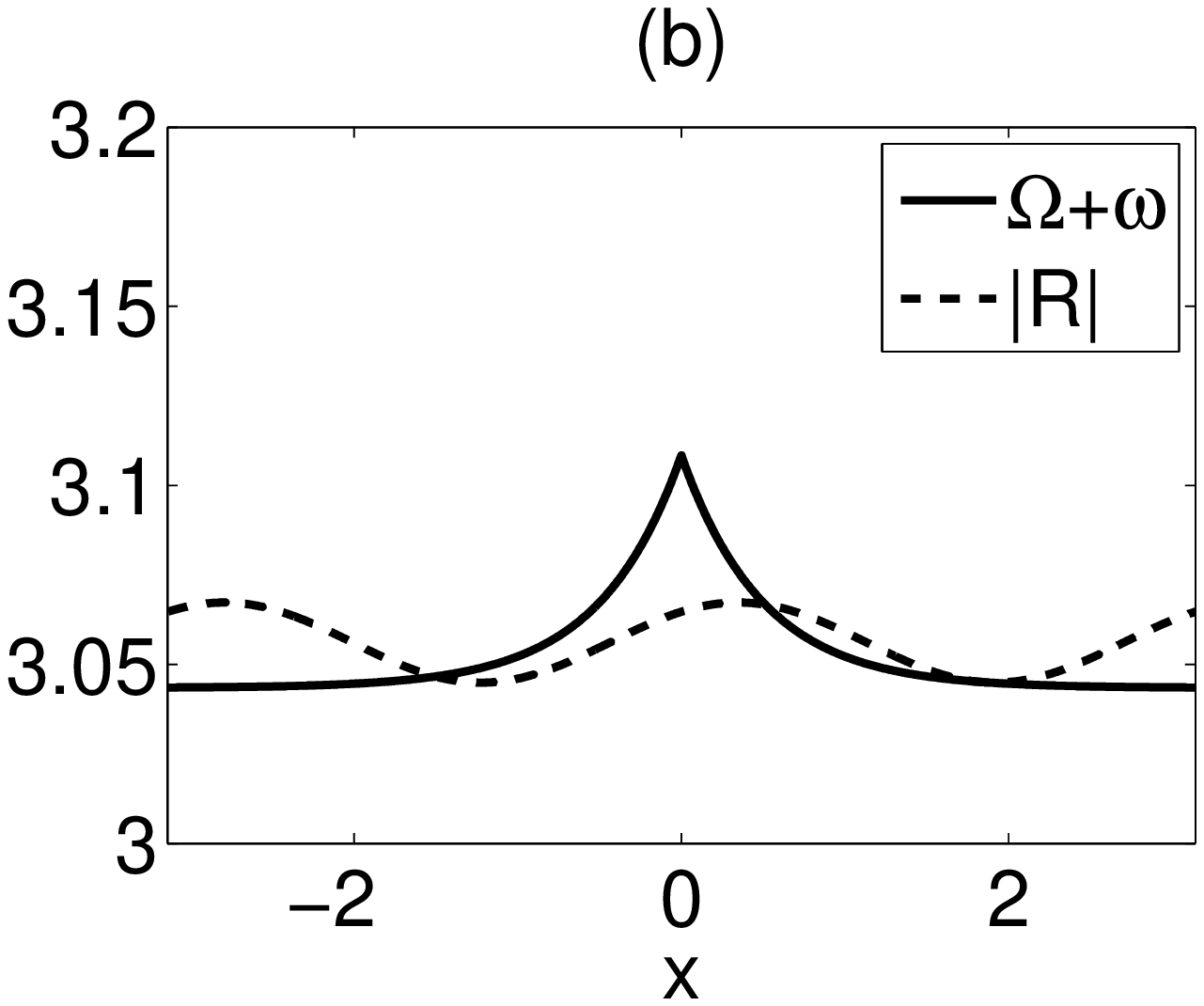}
\caption{Comparison of $\Omega + \omega(x)$ and $R(x)$ at the critical values $\omega_0$ for the appearance of new regions of incoherence around (a) $x=0$ for $\omega_0 = 0.0063$ and (b) $x=1.83$ for $\omega_0 = 0.065$. The calculation is for $\kappa = 2$, $\beta = 0.05$ and $N = 512$.}
\label{fig:transitions}
\end{figure}

These states and the transitions between them can be explained within the framework of the self-consistent analysis. To compute the solution branches and the transition thresholds, we numerically continue solutions of Eqs.~(\ref{sc-full12}) and (\ref{sc-full22}) with respect to the parameter $\omega_0$. When $\omega_0 = 0$, the splay state (with positive slope) corresponds to $a = \pi$, $b_r = 0$, $b_i = \pi$, with $\Omega = \pi$. The order parameter is therefore $R \exp{(i\Theta)} = \pi \exp{(\pm i x)}$.  We use this splay state as the starting point for continuation. As $\omega_0$ increases, a region of incoherence develops in the phase pattern in the vicinity of $x=0$. From the point of view of the self-consistent analysis, the incoherent region corresponds to the region where the natural frequency exceeds the amplitude $R(x)$ of the complex order parameter. The boundaries between the coherent and incoherent oscillators are thus determined by the relation $\Omega + \omega(x) = |a\cos(x) + (b_r + ib_i)\sin(x)|$.  For $\omega_0 = 0.02$, the left and right boundaries are thus $x_l \approx -0.4026$ and $x_r \approx 0.2724$, respectively. 
These predictions are in good agreement with the values measured in direct numerical simulation (Fig.~\ref{fig:cosx_bump_splay}(a)). Figure~\ref{fig:cosx_bump_omega0_continue_splay}(a) shows the overall frequency $\Omega$ obtained by numerical continuation of the solution of Eqs.~(\ref{sc-full12}) and (\ref{sc-full22}) in the parameter $\omega_0$, while Figs.~\ref{fig:cosx_bump_omega0_continue_splay}(b,c) show the corresponding results for the fraction $e$ of the domain occupied by the coherent oscillators and the extent of the (first) region of incoherence, i.e., the interval $x_l\le x\le x_r$, also as functions of $\omega_0$. From the figure we can see a clear transition at $\omega_0 \approx 0.0063$ from a single domain-filling coherent state to a ``splay state with one incoherent cluster'' in $x_l\le x\le x_r$, followed by a subsequent transition at $\omega_0 \approx 0.065$ from this state to a ``splay state with two incoherent clusters''. These transitions occur when the profiles of $\Omega + \omega(x)$ and $R(x)$ touch as $\omega_0$ increases and these points of tangency therefore correspond to the locations where coherence is first lost. Figure~\ref{fig:transitions} shows that tangencies between $\Omega + \omega(x)$ and $R(x)$ occur when $\omega_0 \approx 0.0063$ and $0.065$, implying that intervals of incoherent oscillators appear first at $x=0$ (i.e., the bump maximum) and subsequently at $x\approx 1.83$, as $\omega_0$ increases. These predictions are in excellent agreement with the direct numerical simulations shown in Fig.~\ref{fig:cosx_bump_splay_four}. Moreover, the critical values of $\omega_0$ predicted by the self-consistency analysis are fully consistent with the simulation results when $\beta$ is increased quasi-statically (not shown). In each case we repeated the simulations for decreasing $\beta$ but found no evidence of hysteresis in these transitions.

\subsubsection{Effect on stationary chimera states}

\begin{figure}
\includegraphics[height=4cm]{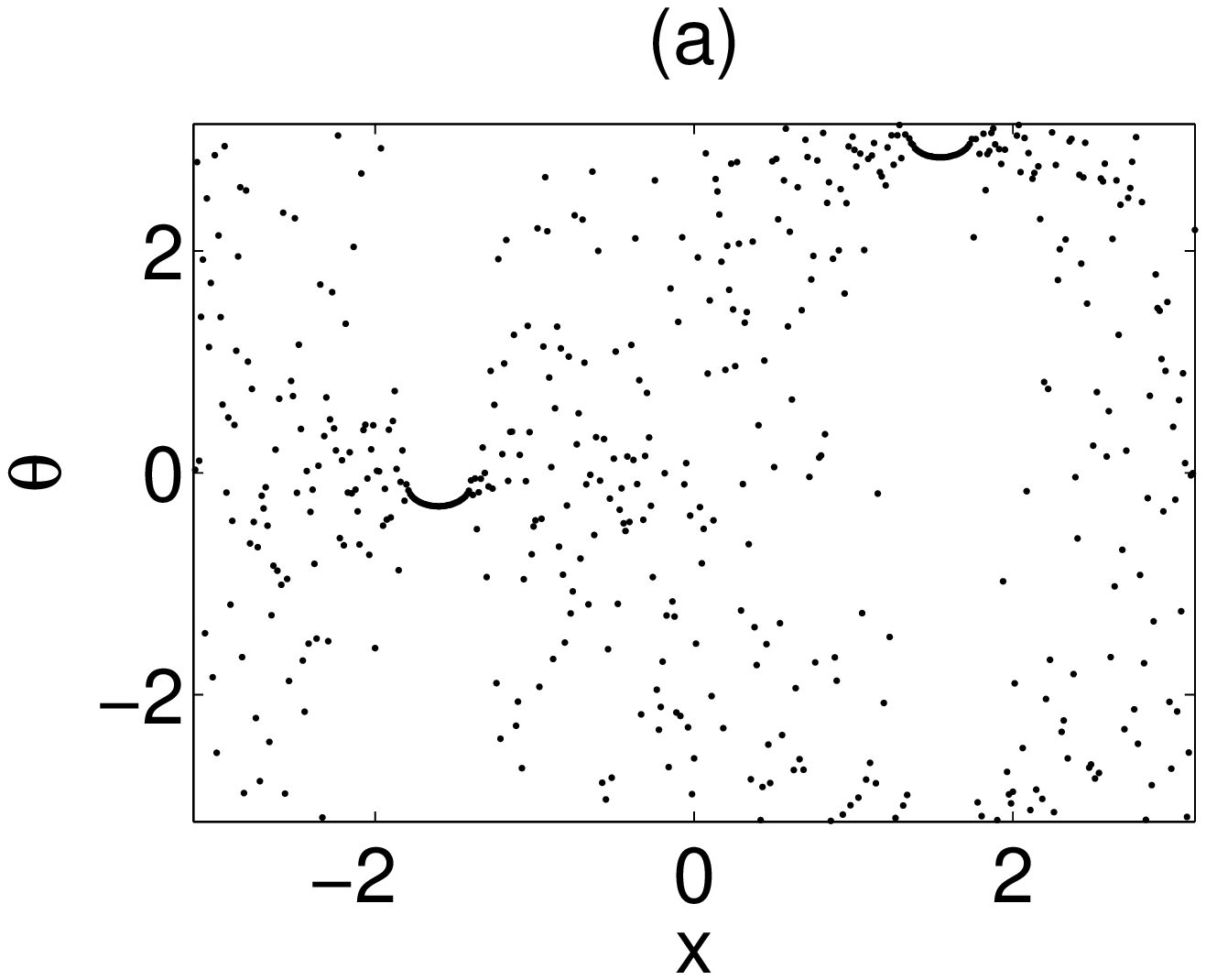}
\includegraphics[height=4cm]{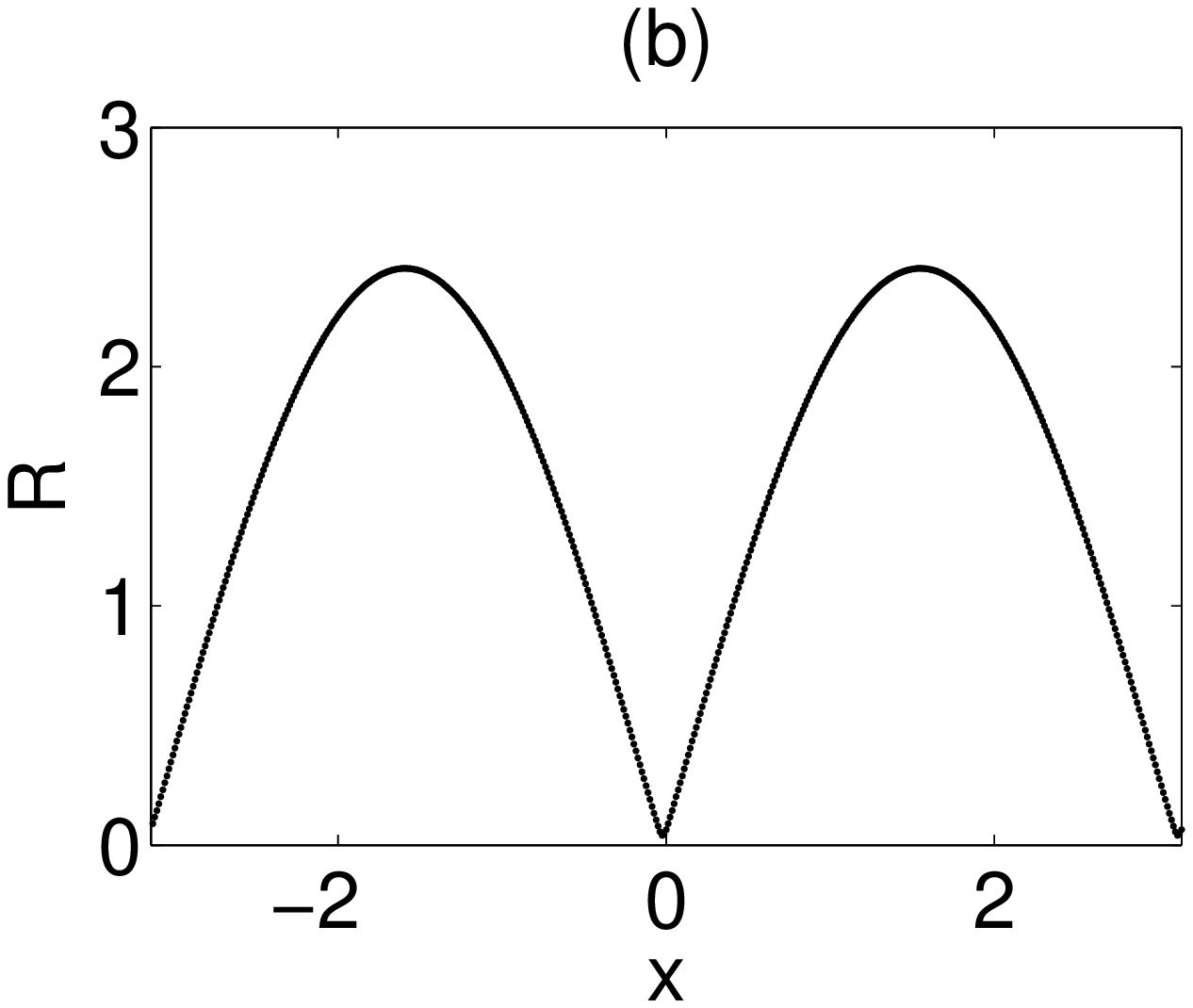}
\includegraphics[height=4cm]{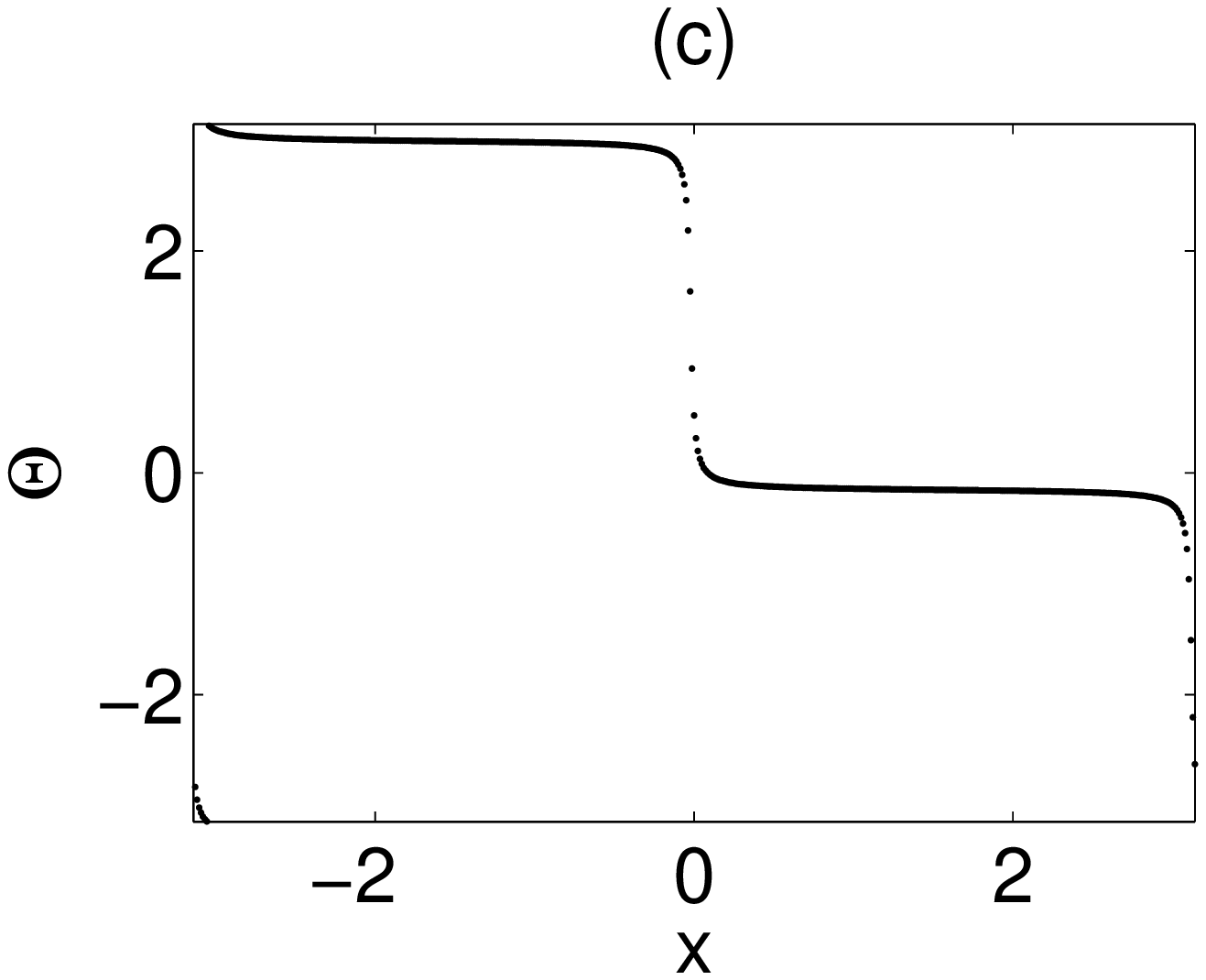}
\caption{(a) A snapshot of the phase distribution $\theta(x,t)$ in a 2-cluster chimera state for $G(x) = \cos(x)$ and $\omega(x) = 0.1 \exp(-2|x|)$. (b) The corresponding order parameter $R(x)$. (c) The corresponding order parameter $\Theta(x)$. Note that the oscillators in the two clusters oscillate with the same frequency but $\pi$ out of phase. The calculation is done with $\kappa = 2$, $\beta = 0.05$ and $N=512$.}
\label{fig:cosx_bump_2cluster}
\end{figure}

\begin{figure}
\includegraphics[height=4cm]{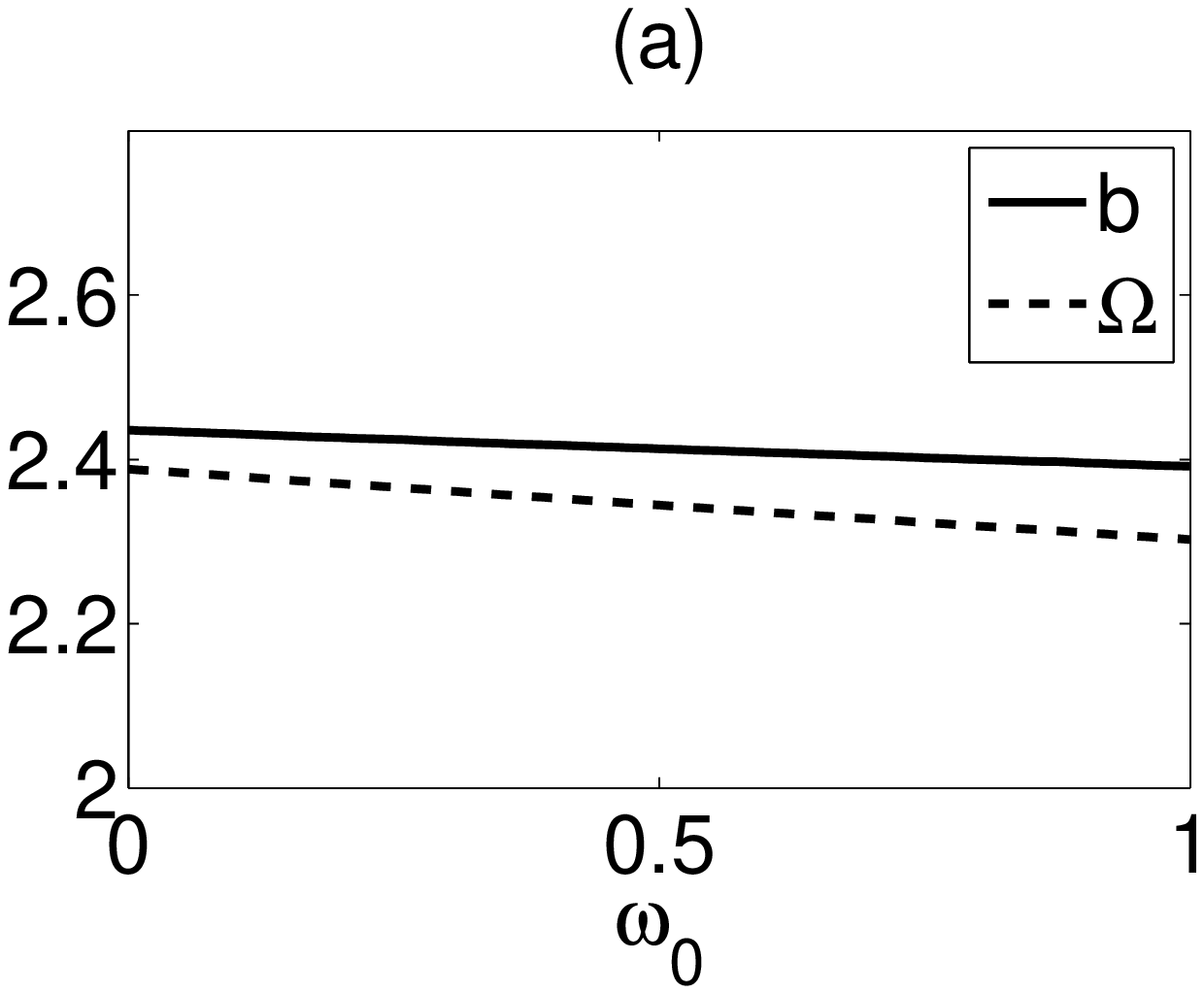}
\includegraphics[height=4cm]{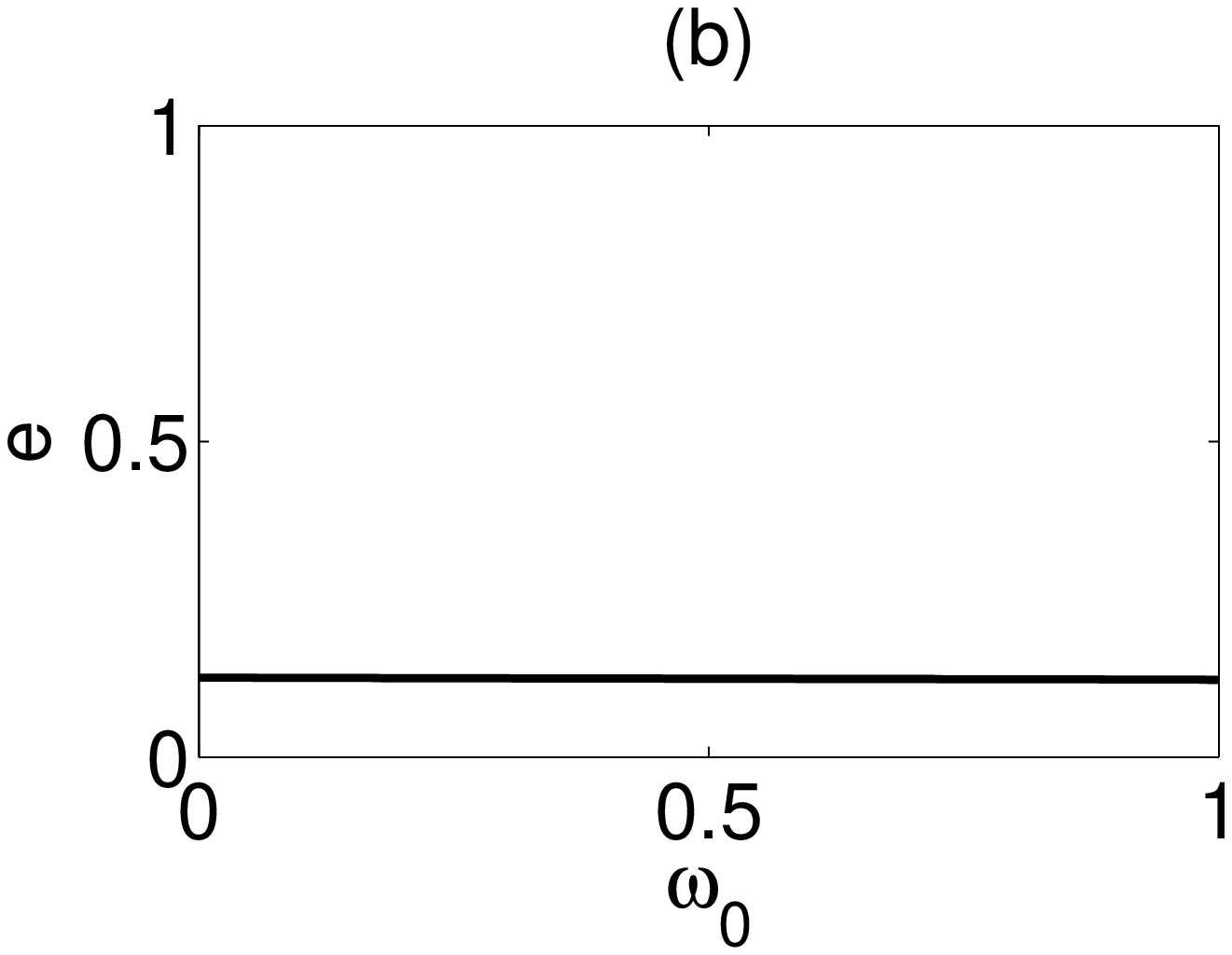}
\caption{The dependence of (a) $\Omega$ and the order parameter amplitude $b$ on $\omega_0$. (b) The fraction $e$ of coherent oscillators as a function of $\omega_0$. The calculation is done with $\kappa = 2$, $\beta = 0.05$ and $N = 512$.}
\label{fig:cosx_bump_omega0_continue_2cluster}
\end{figure}

Chimera states with $2n$ evenly distributed coherent clusters are readily observed when $G(x) = \cos(nx)$ and $\omega$ is a constant. These states persist when a bump is introduced into the frequency distribution $\omega(x)$. Figure~\ref{fig:cosx_bump_2cluster} shows the phase distribution and the corresponding local order parameters $R(x)$ and $\Theta(x)$ for $G(x) = \cos(x)$ when $\omega(x) = 0.1 \exp(-2|x|)$. The figure shows that the two clusters persist, but are now always located near $x = -\frac{\pi}{2}$ and $\frac{\pi}{2}$. This is a consequence of the fact that the presence of the bump breaks the translation invariance of the system. Figures~\ref{fig:cosx_bump_2cluster}(b,c) show that the local order parameter $Z(x)$ has the symmetry $\tilde{Z}(-x) = - \tilde{Z}(x)$. This symmetry implies $\tilde{Z}$ should take the form $R\exp{(i\Theta)} = b\sin(x)$, where $b=b_r$ is real. The corresponding self-consistency equation takes the form
\begin{equation}
b^2e^{-i\beta}=\left<\Omega+\omega(y)-\sqrt{(\Omega+\omega(y))^2-b^2\sin^2y}\right>.\label{bump_scequation_2cluster}
\end{equation}
The result of numerical continuation of the solutions of Eq.~(\ref{bump_scequation_2cluster}) are shown in Fig.~\ref{fig:cosx_bump_omega0_continue_2cluster}. The 2-cluster chimera state persists to large values of $\omega_0$, with the size of the coherent clusters largely insensitive to the value of $\omega_0$. This prediction has been corroborated using direct simulation of Eq.~(\ref{phase_eq_discrete}) with $N=512$ oscillators.

 
\begin{figure}
\includegraphics[height=3.5cm]{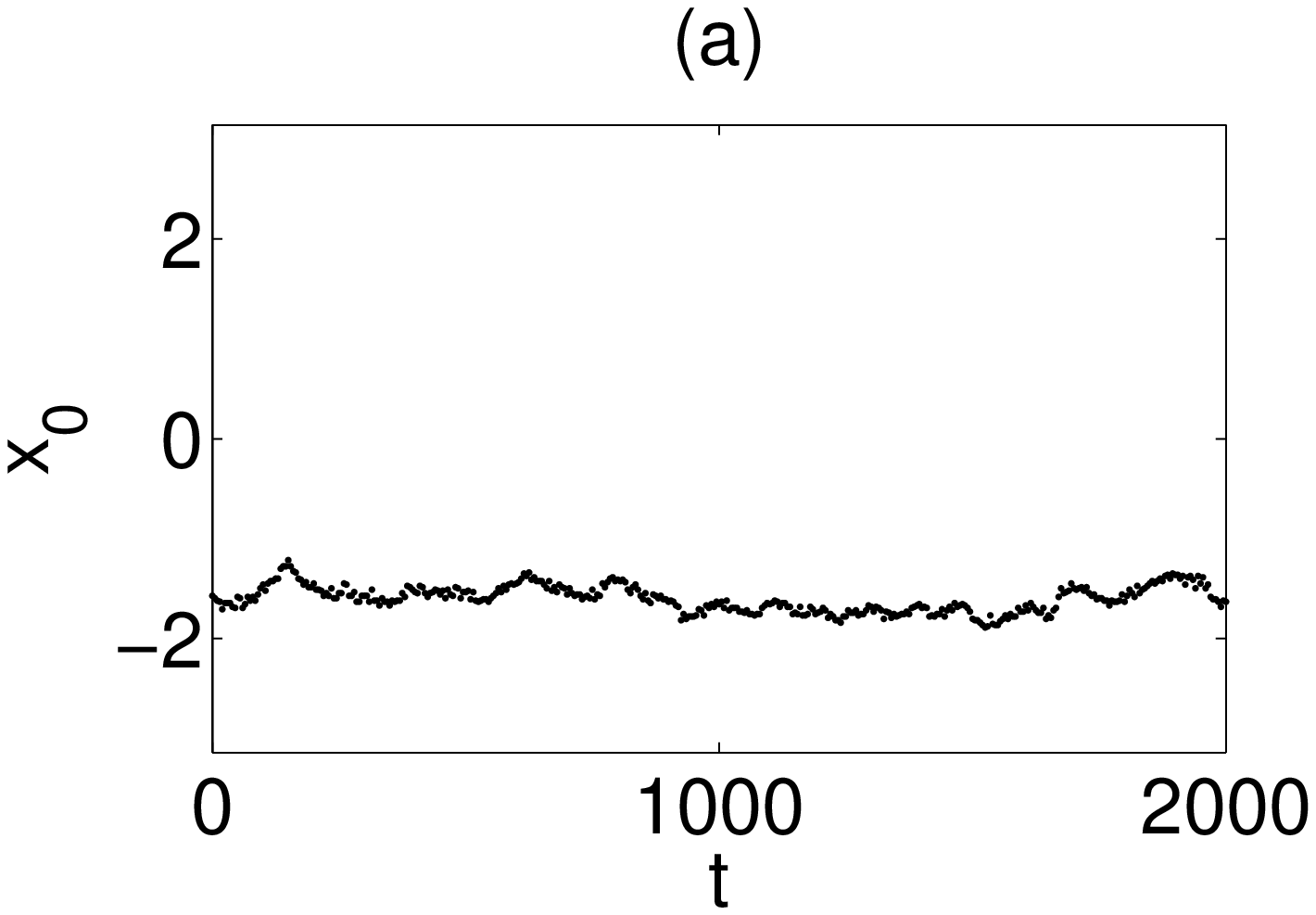}
\includegraphics[height=3.5cm]{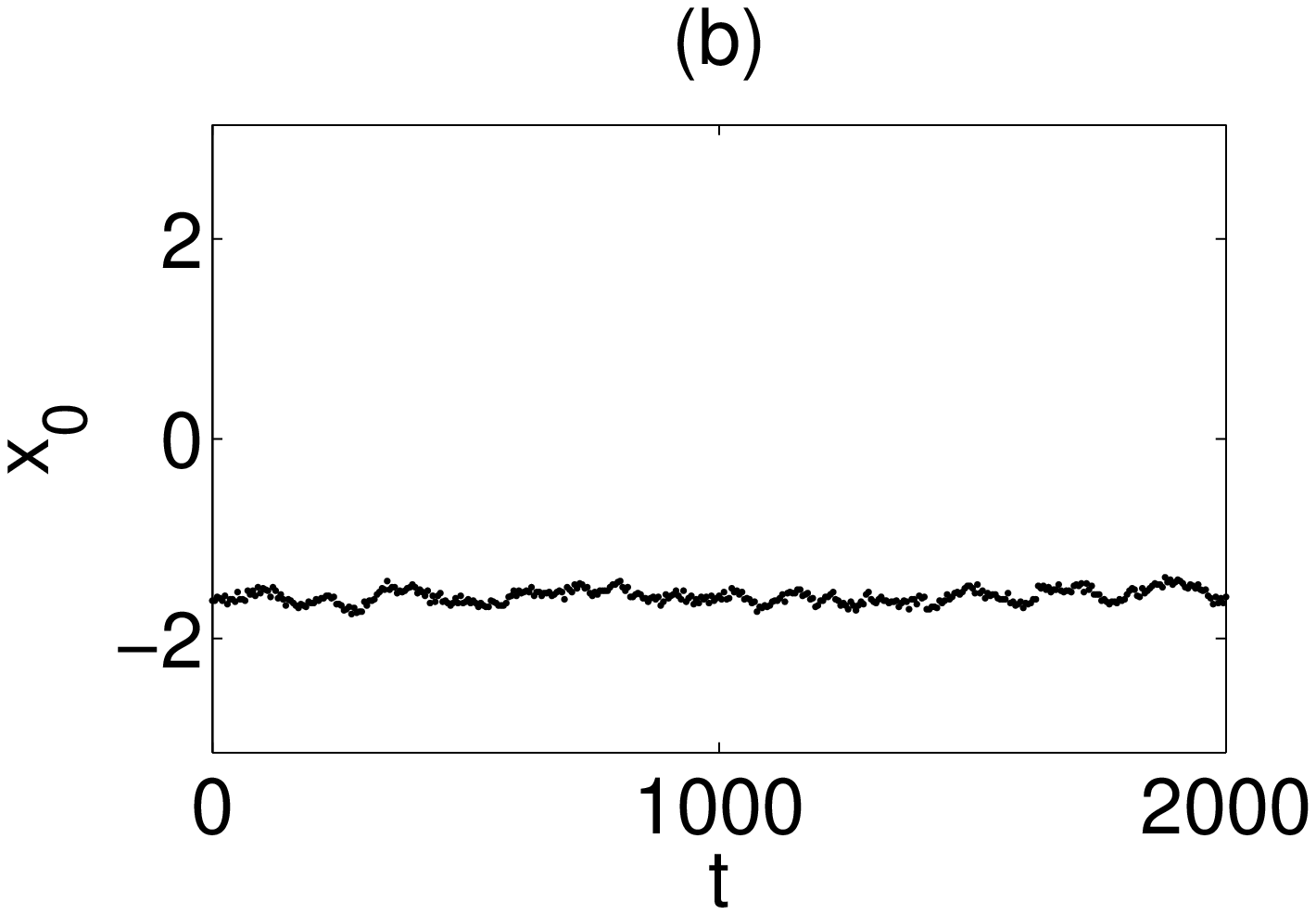}
\includegraphics[height=3.5cm]{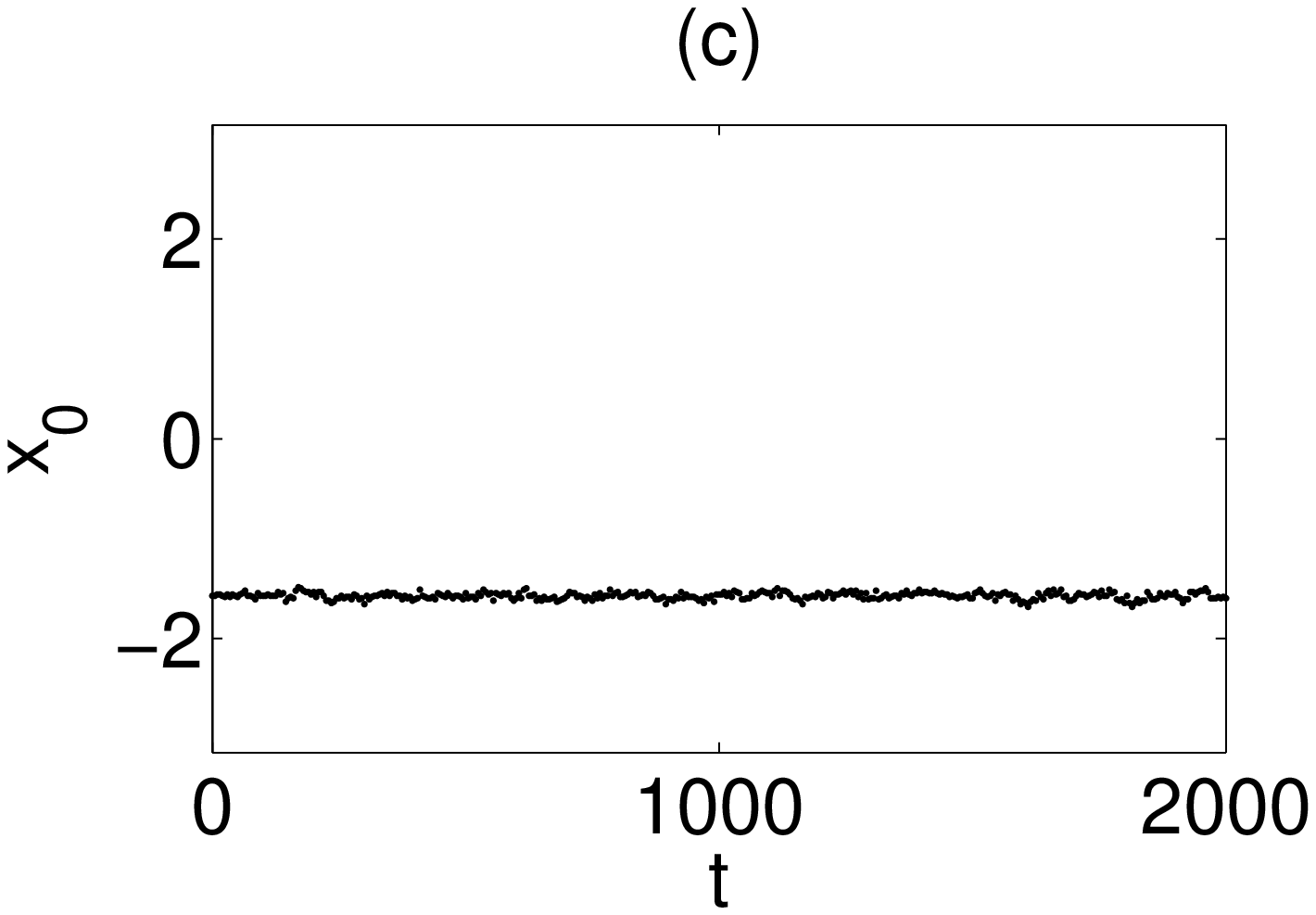}
\caption{The position $x_0(t)$ of the coherent cluster as a function of time when $\omega_0 = 0.1$, $\beta = 0.05$, $N = 512$ and (a) $\kappa = 10$, (b) $\kappa = 6$ and (c) $\kappa = 2$.}
\label{fig:bump_pos_vs_t}
\end{figure}

\begin{figure}
\includegraphics[height=4cm]{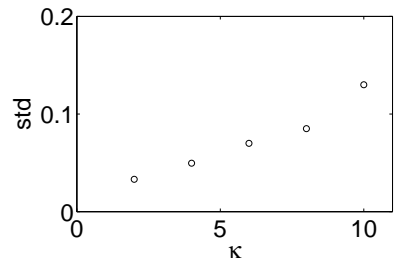}
\caption{The standard deviation of the position $x_0(t)$ as a function of the parameter $\kappa$ for $\omega_0 = 0.1$, $\beta = 0.05$ and $N = 512$.}
\label{fig:std_of_pos}
\end{figure}

When $\omega$ is constant, finite size effects cause the phase pattern to fluctuate in location. In \cite{XKK2014}, we demonstrate that this fluctuation is well modeled by Brownian motion in which the variance is proportional to $t$, even though the original system is strictly deterministic. As mentioned above, when $\omega(x)$ is spatially dependent, the translation symmetry is broken and the coherent cluster has a preferred location. Figure \ref{fig:cosx_bump_2cluster} suggests that local maxima of the order parameter $R$ can be used to specify the location of coherent clusters. Consequently we plot in Fig.~\ref{fig:bump_pos_vs_t} the position $x_0(t)$ of the right coherent cluster as a function of time for three different values of the parameter $\kappa$. We see that the inhomogeneity pins the coherent cluster to a particular location, and that the cluster position executes apparently random oscillations about this preferred location, whose amplitude increases with increasing $\kappa$, i.e., with decreasing width of the bump. Figure~\ref{fig:std_of_pos} shows the standard deviation of the position $x_0(t)$ of the coherent cluster as a function of the parameter $\kappa$.
\begin{figure}
\includegraphics[height=4cm]{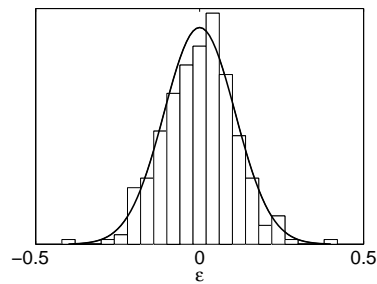}
\caption{The histogram of the residues $\varepsilon \equiv x_0(t+\Delta t) - Ax_0(t) -B$ is well approximated by a normal distribution when $\Delta t=100$.
}
\label{fig:hist}
\end{figure} 

The behavior shown in Figs.~\ref{fig:bump_pos_vs_t} and \ref{fig:std_of_pos} can be modeled using an Ornstein-Uhlenbeck process, i.e., a linear stochastic ordinary differential equation of the form
\begin{equation}
d x_0 = \lambda ( \mu - x_0)dt + \sigma dW_t,\label{eq:OE}
\end{equation}
where $\lambda$ represents the strength of the attraction to the preferred location $\mu$, and $\sigma$ indicates the strength of the noise. Models of this type are expected to apply on an appropriate timescale only: the time increment $\Delta t$ between successive steps of the stochastic process must be large enough that the position of the cluster can be thought of as the result of a large number of pseudo-random events and hence normally distributed, but not so large that nonlinear effects become significant. Figure \ref{fig:hist} reveals that for an appropriate interval of $\Delta t$ the fluctuations $x_0(t+\Delta t) - Ax_0(t) -B$ are indeed normally distributed, thereby providing support for the applicability of Eq.~(\ref{eq:OE}) to the present system.

Equation (\ref{eq:OE}) has the solution
\begin{equation}
x_0(t+\Delta t) =A x_0(t)+B+CN_{0,1}, \label{eq:linear}
\end{equation}
where $A=e^{-\lambda\Delta t}$, $B=\mu(1-A)$ and $C = \sigma \sqrt{\frac{1-A^2}{2\lambda}}$. To fit the parameters to the data in Fig.~\ref{fig:bump_pos_vs_t} we notice that the relationship between consecutive observations $x_0$ is linear with an {\it i.i.d.} error term $CN_{0,1}$, where $N_{0,1}$ denotes the normal distribution with zero mean and unit variance (Fig.~\ref{fig:hist}) and $C$ is a constant. A least-squares fit to the data $(x_0(t), x_0(t+\Delta t))$ gives the parameters $\lambda$, $\mu$ and $\sigma$.  We find that for $\omega_0 =0.1$, $\kappa=2$ and $\beta = 0.05$, the choice $\Delta t = 100$ works well and yields the empirical model parameters $\lambda \approx 0.016$, $\mu \approx -1.57$ and $\sigma \approx 0.02$, a result that is in good agreement with the simulation results for $\kappa=2$ summarized in Fig.~\ref{fig:std_of_pos}.



\subsection{Periodic $\omega(x)$}

In this section we consider the case $\omega(x)=\omega_0\cos(lx)$. When $\omega_0$ is small, the states present for $\omega_0=0$ persist, but with increasing $\omega_0$ one finds a variety of intricate dynamical behavior. In the following we set $G(x)=\cos x$ and $\beta \equiv \frac{\pi}{2} - \alpha = 0.05$, with $\omega_0$ and $l$ as parameters to be varied.

\subsubsection{Splay states, near-splay states and chimera splay states}

\begin{figure}
\includegraphics[width=3.5cm]{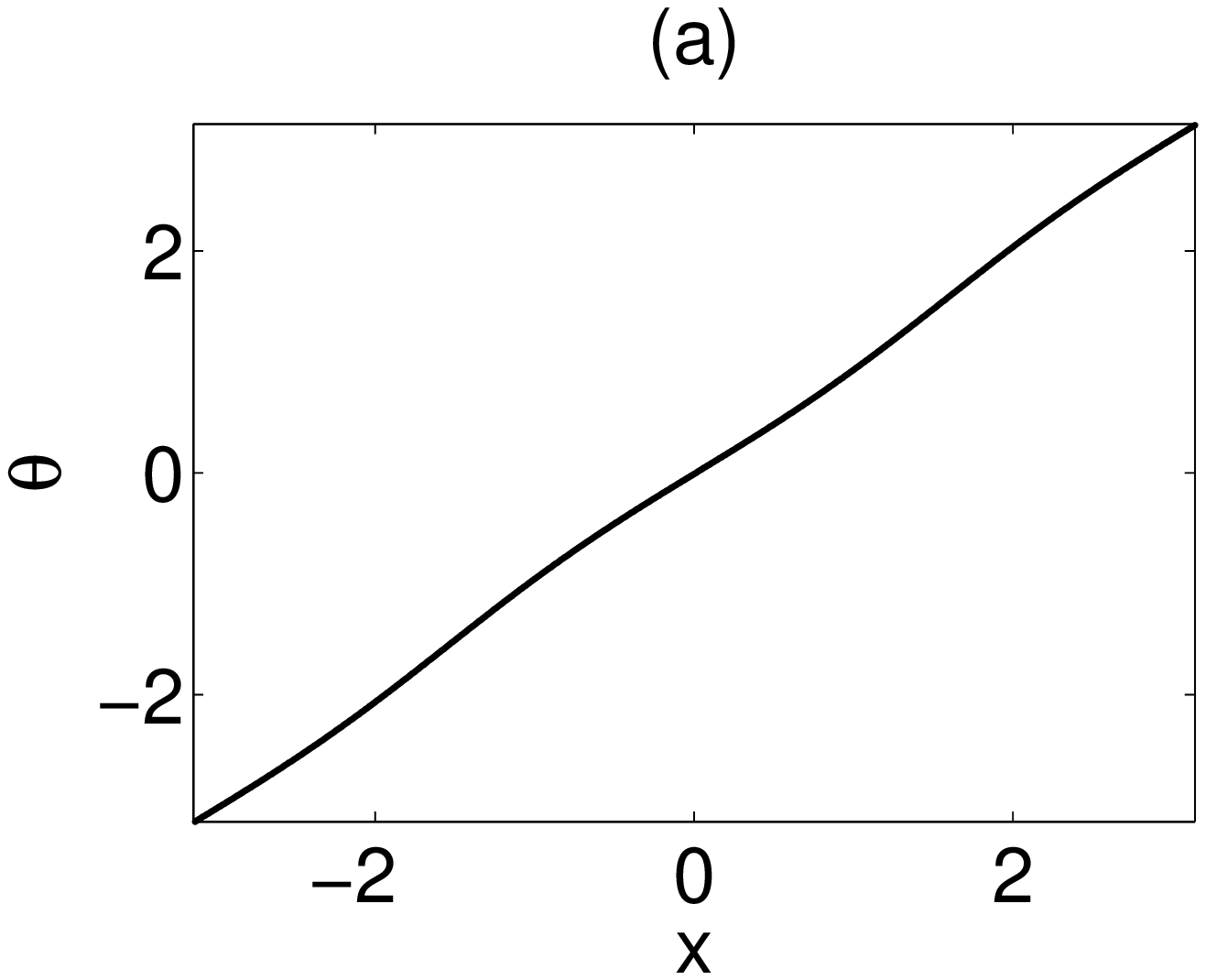}
\includegraphics[width=3.5cm]{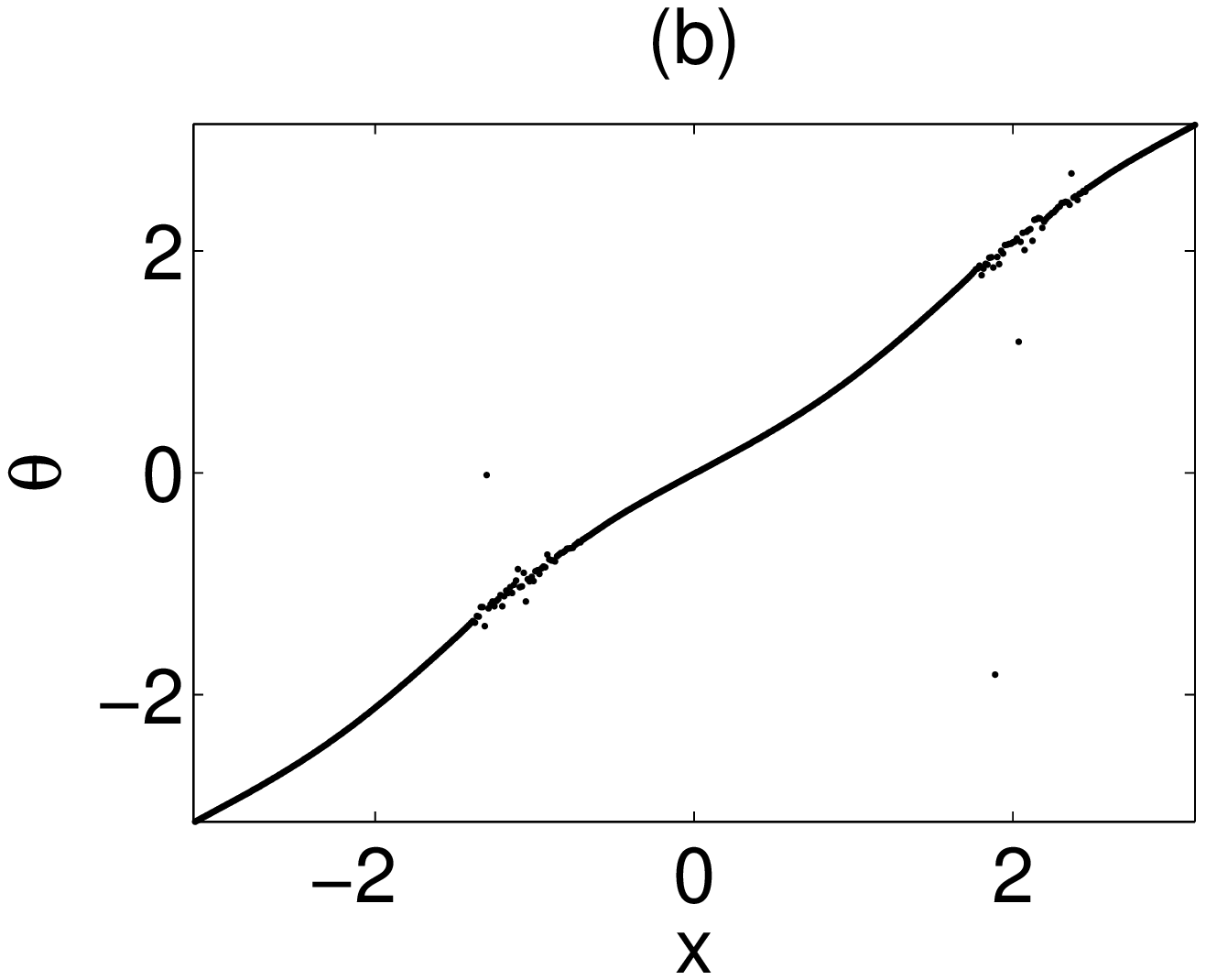}
\includegraphics[width=3.5cm]{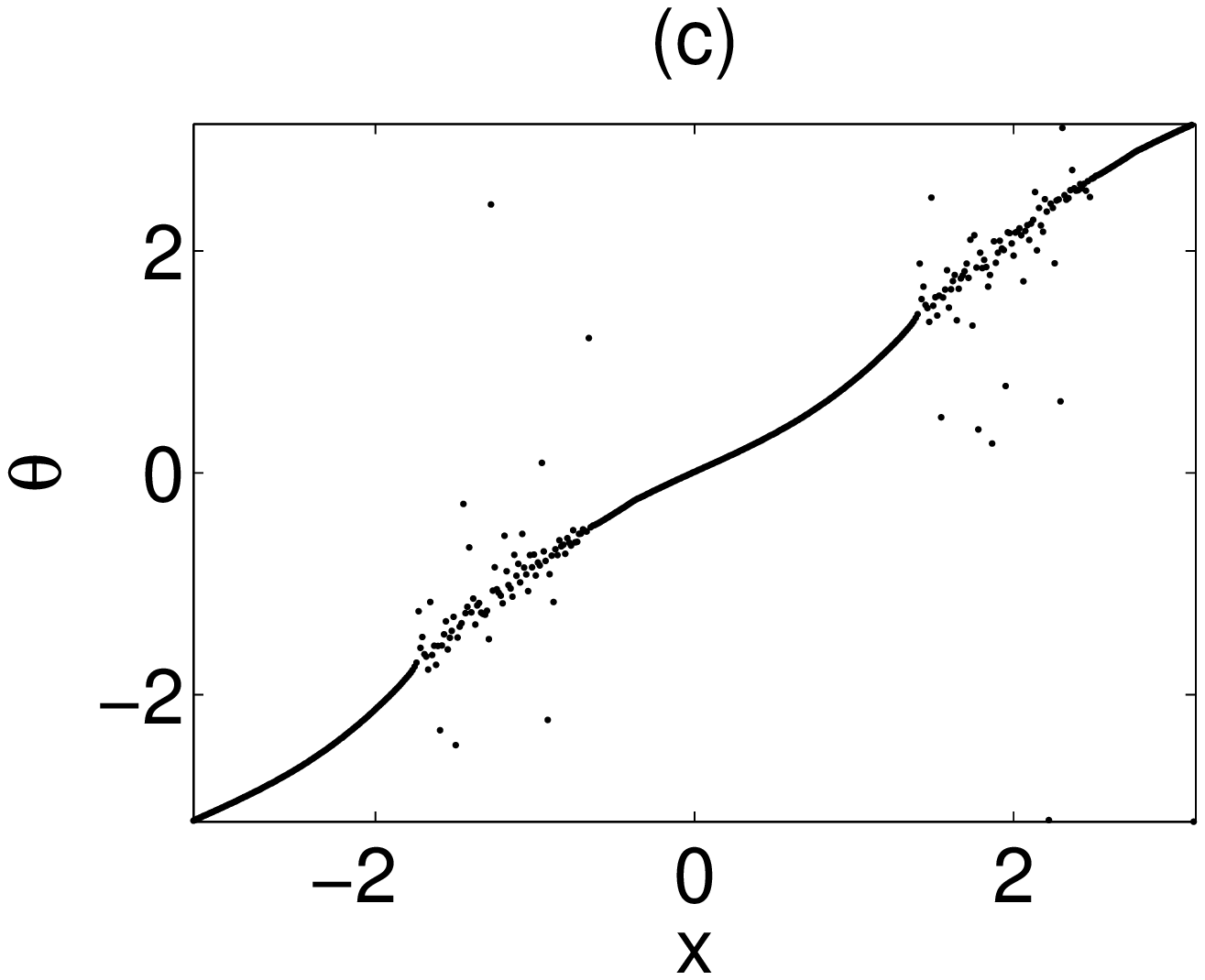}
\includegraphics[width=3.5cm]{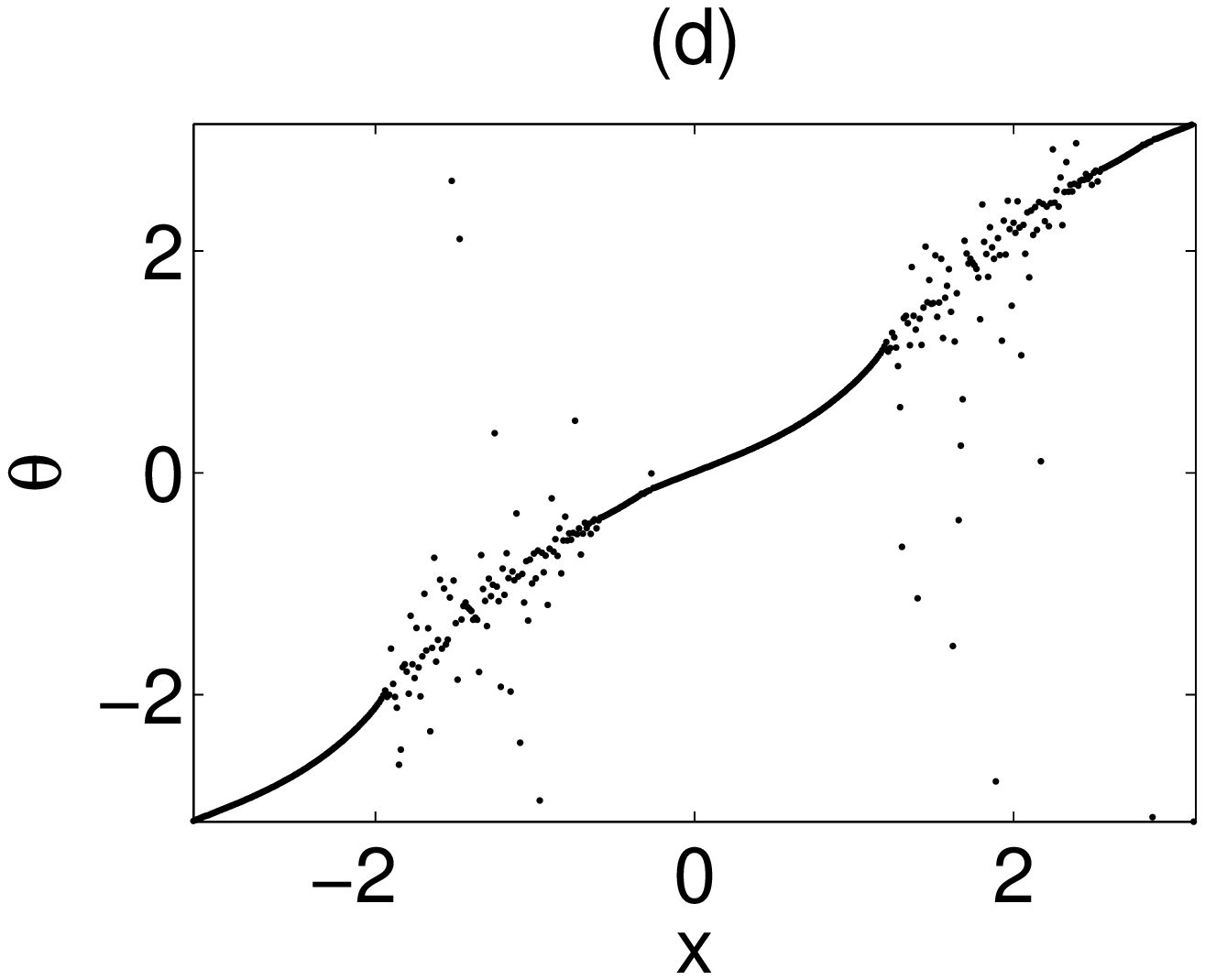}
\caption{A near-splay state for (a) $\omega(x) = 0.1 \cos(2x)$.  Chimera splay states for (b) $\omega(x) = 0.2 \cos(2x)$, (c) $\omega(x) = 0.3 \cos(2x)$ and (d) $\omega(x) = 0.4 \cos(2x)$. In all cases $\beta = 0.05$ and $N=512$.}
\label{fig:cosx_inho_nearsplay_wam1234}
\end{figure}

\begin{figure}
\includegraphics[height=3.5cm]{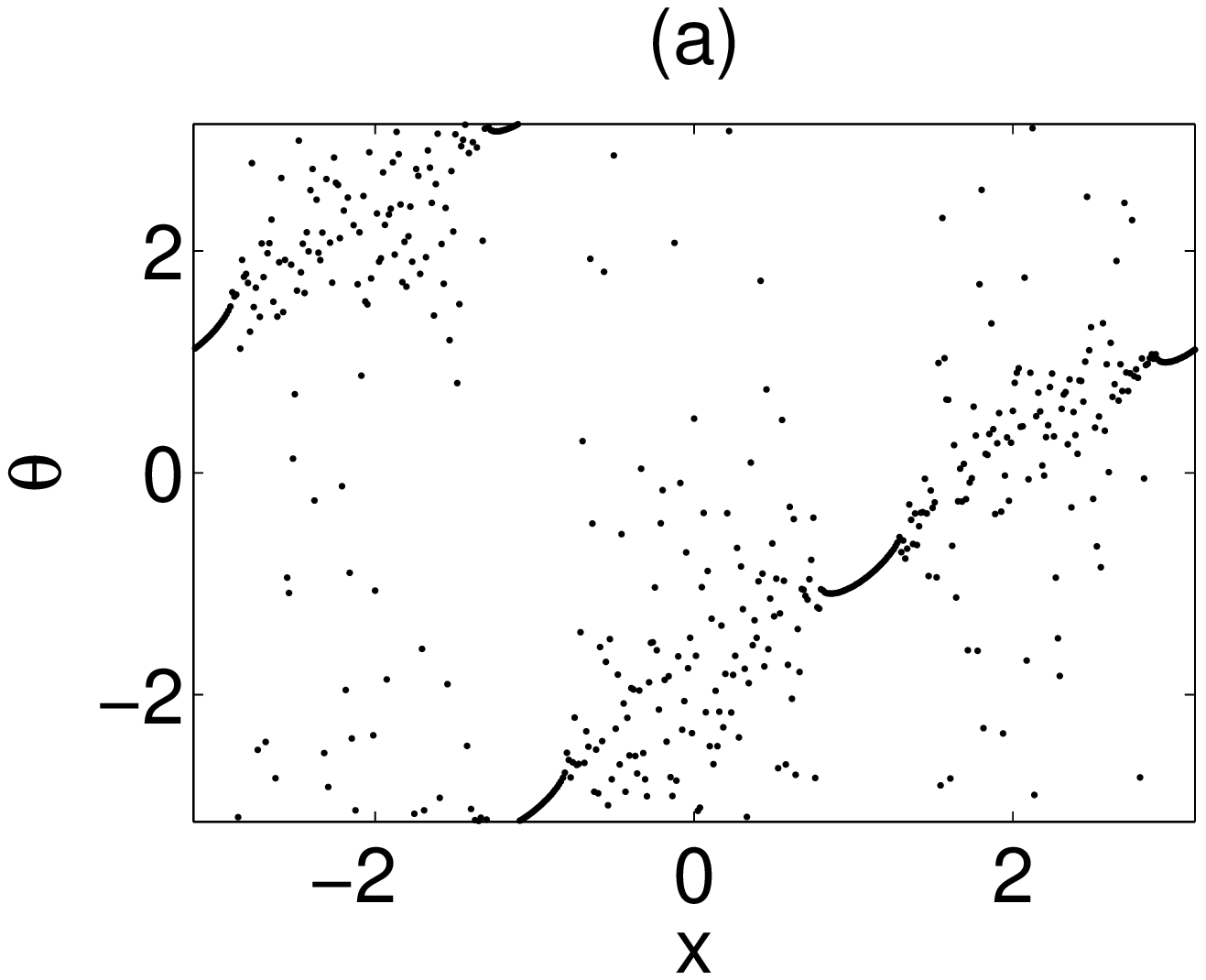}
\includegraphics[height=3.5cm]{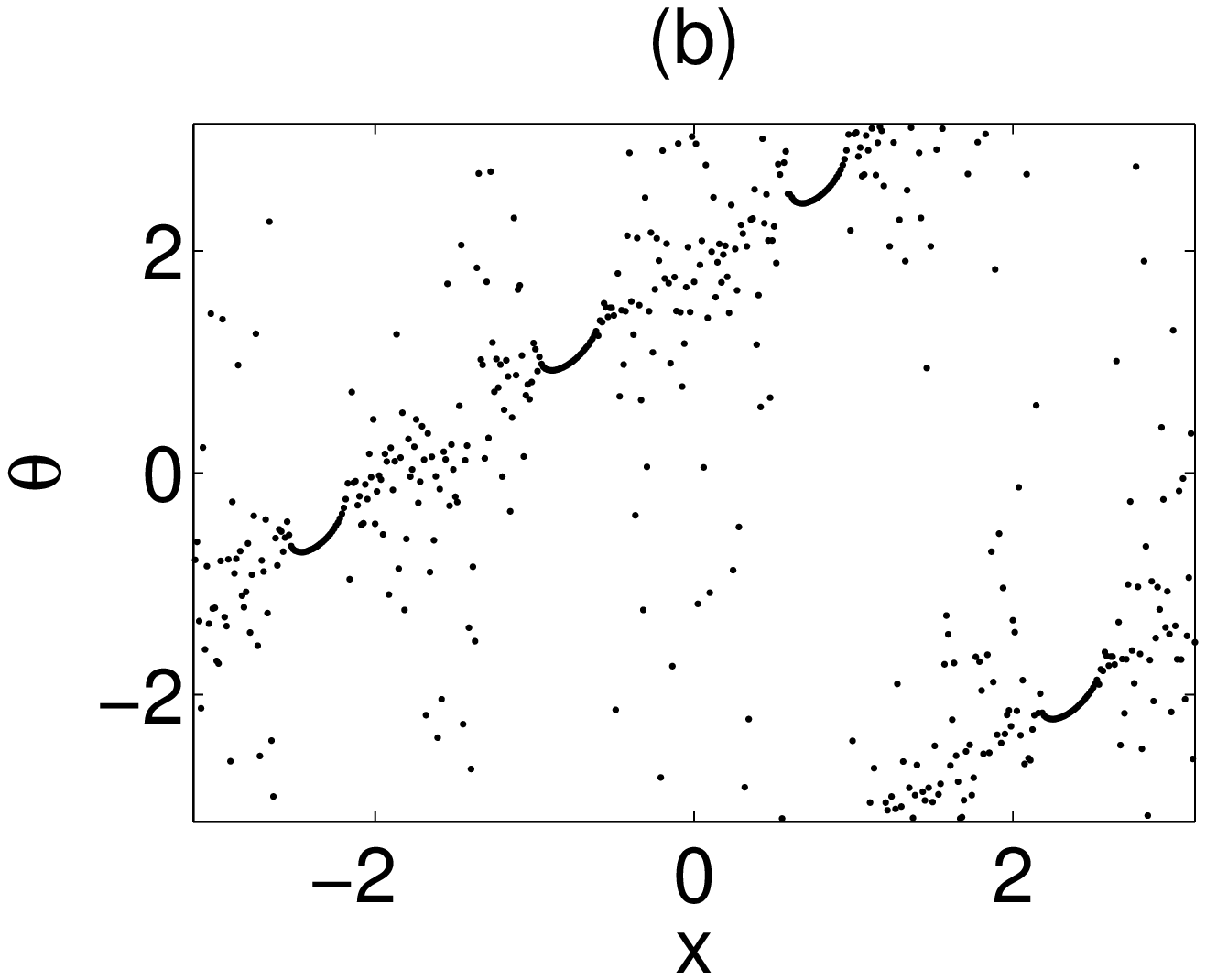}
\includegraphics[height=3.5cm]{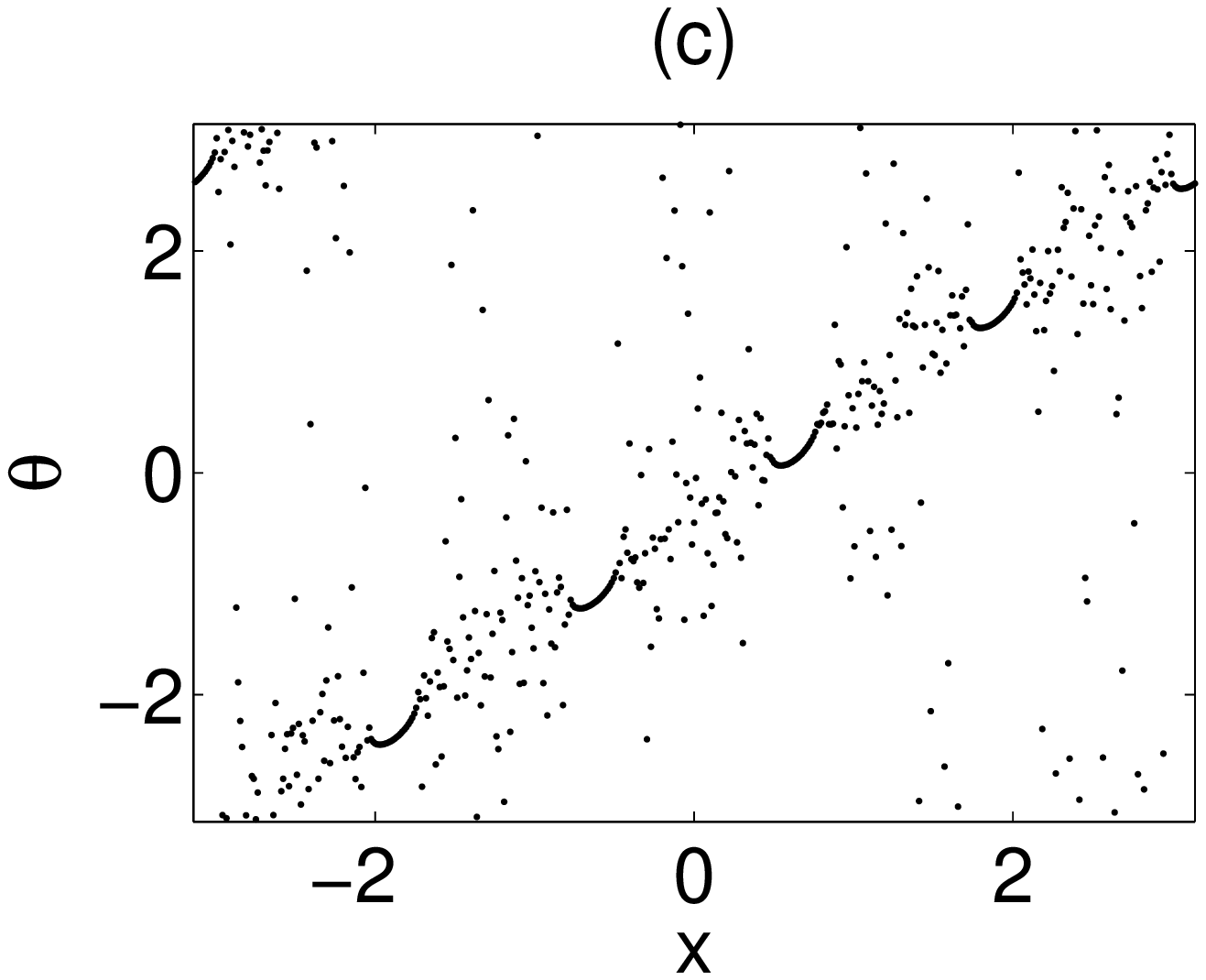}
\caption{Chimera splay states for $G(x) = \cos(x)$ and (a) $\omega(x) = 0.2 \cos(3x)$ (3-cluster state), (b) $\omega(x) = 0.2 \cos(4x)$ (4-cluster state), and (c) $\omega(x) = 0.2 \cos(5x)$ (5-cluster state). In all cases $\beta = 0.05$ and $N=512$.}
\label{fig:cosx_inho_l_cluster_345}
\end{figure}

When $\omega(x)$ is a constant, splay states are observed in which the phase $\theta(x,t)$ varies linearly with $x$. When $\omega_0$ is nonzero but small, the splay states persist as the near-splay states described by Eq.~(\ref{splay_ansatz}); Fig.~\ref{fig:cosx_inho_nearsplay_wam1234}(a) shows an example of such a state when $\omega(x)=0.1\cos(2x)$. As $\omega_0$ becomes larger, incoherent regions appear and these increase in width as $\omega_0$ increases further (e.g., Fig.~\ref{fig:cosx_inho_nearsplay_wam1234}(b)--(d)). We refer to this type of state as a chimera splay state, as in the bump inhomogeneity case. Figures~\ref{fig:cosx_inho_nearsplay_wam1234} show the phase distribution obtained for different values of $\omega_0$. For these solutions, the slope of the coherent region is no longer constant but the oscillators rotate with a constant overall frequency $\Omega$. This type of solution is also observed for other values of $l$. Figure~\ref{fig:cosx_inho_l_cluster_345} shows examples of chimera splay states for $l = 3$, $4$ and $5$, with $l$ coherent clusters in each case.

\begin{figure}
\includegraphics[height=4cm]{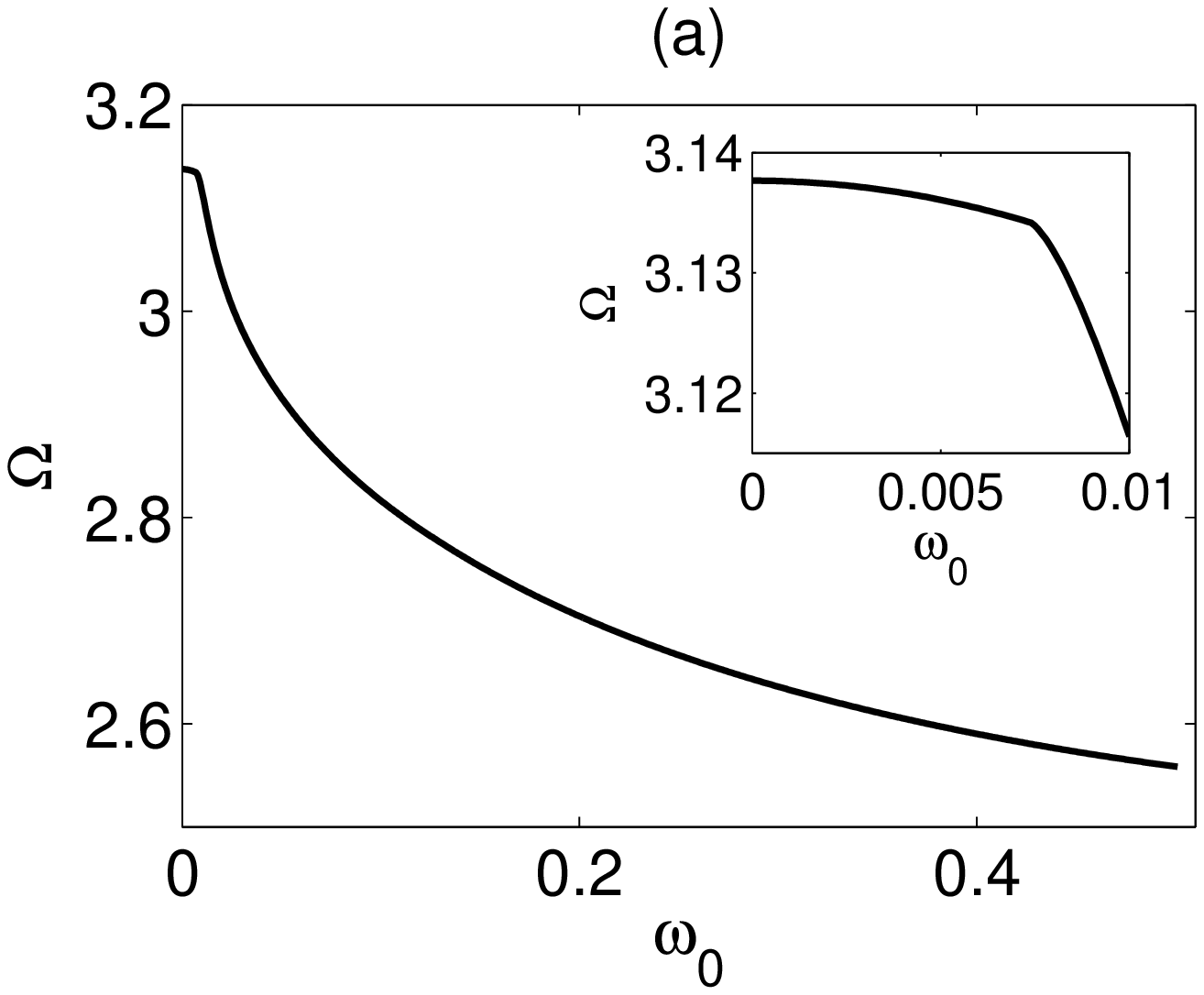}
\includegraphics[height=4cm]{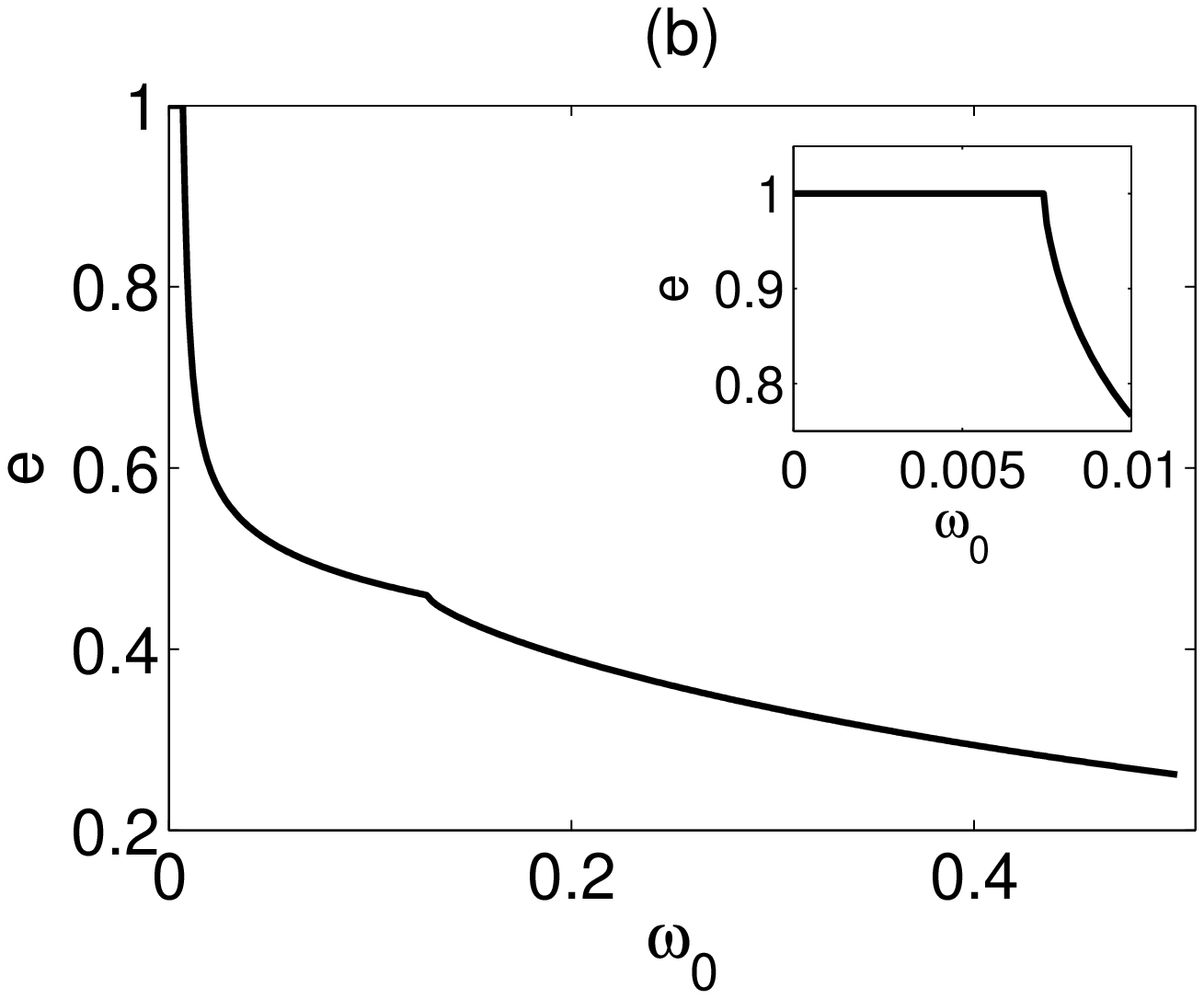}
\caption{The dependence of (a) $\Omega$ and (b) the fraction $e$ of the domain occupied by the coherent oscillators on $\omega_0$, with $\omega(x)=\omega_0\cos(x)$ and $\beta=0.05$.}
\label{fig:splay1}
\end{figure}

\begin{figure}
\includegraphics[height=4cm]{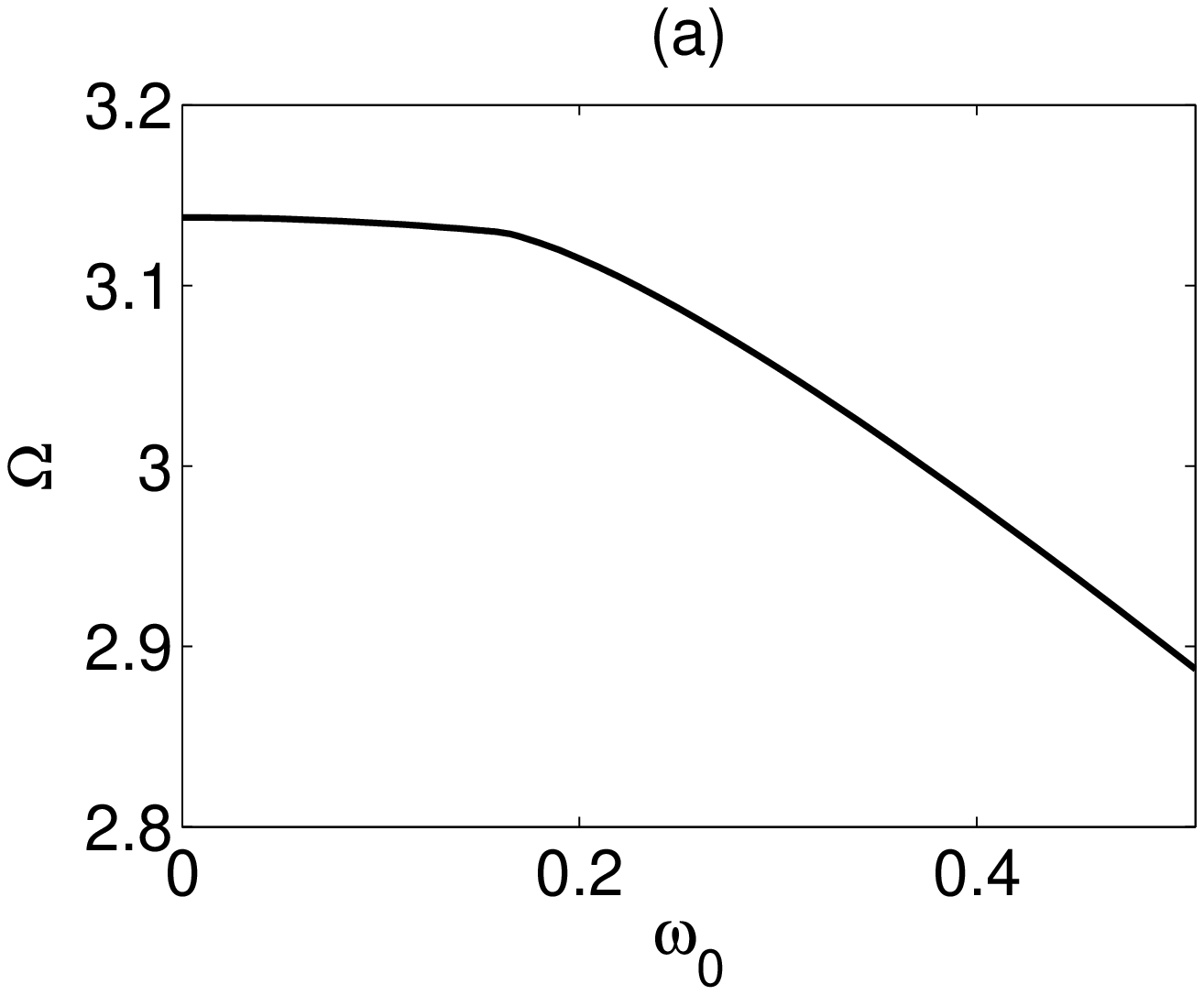}
\includegraphics[height=4cm]{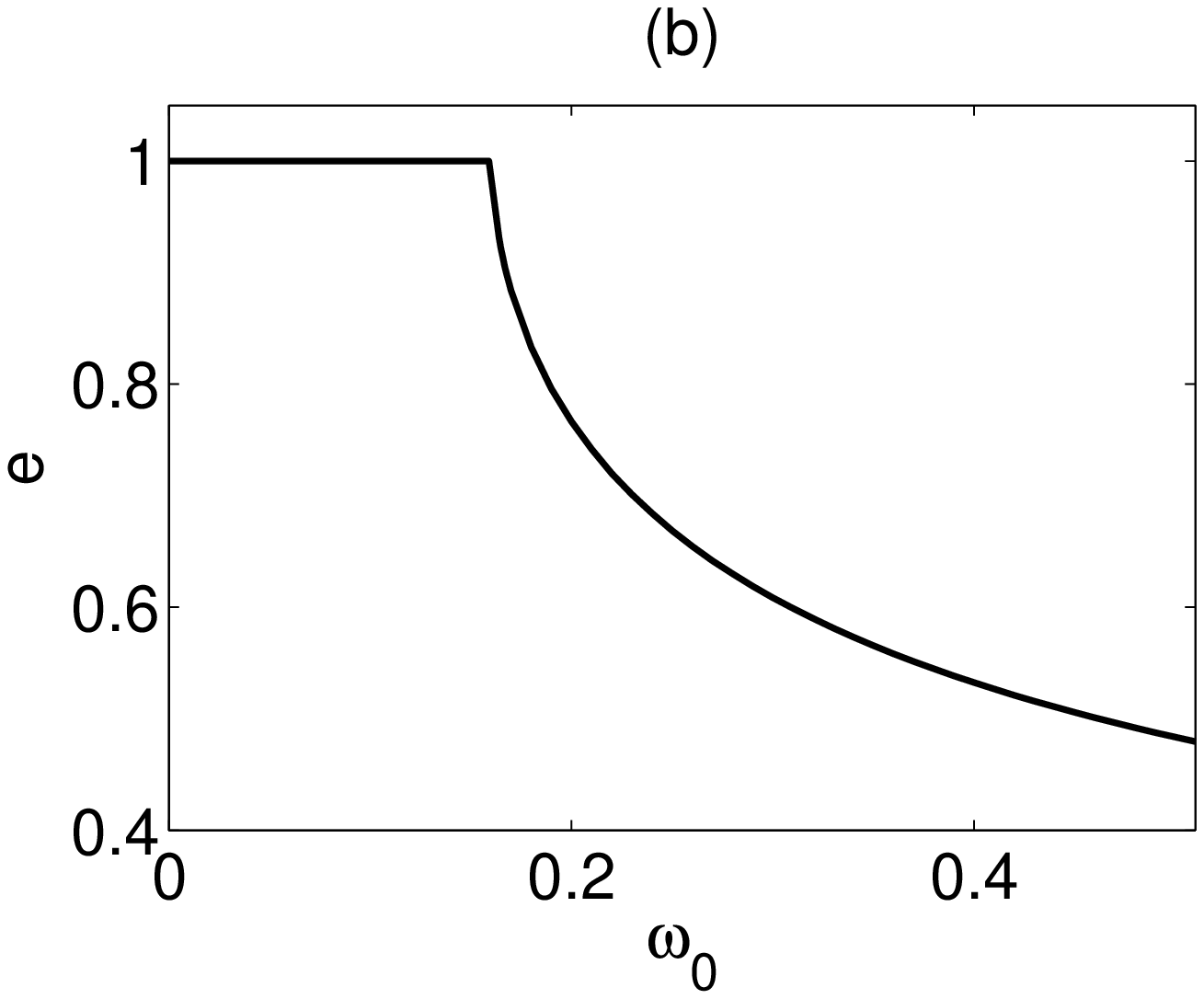}
\caption{The dependence of (a) $\Omega$ and (b) the fraction $e$ of the domain occupied by the coherent oscillators on $\omega_0$, with $\omega(x)=\omega_0\cos(2x)$ and $\beta=0.05$.}
\label{fig:splay2}
\end{figure}

\begin{figure}
\includegraphics[height=4cm]{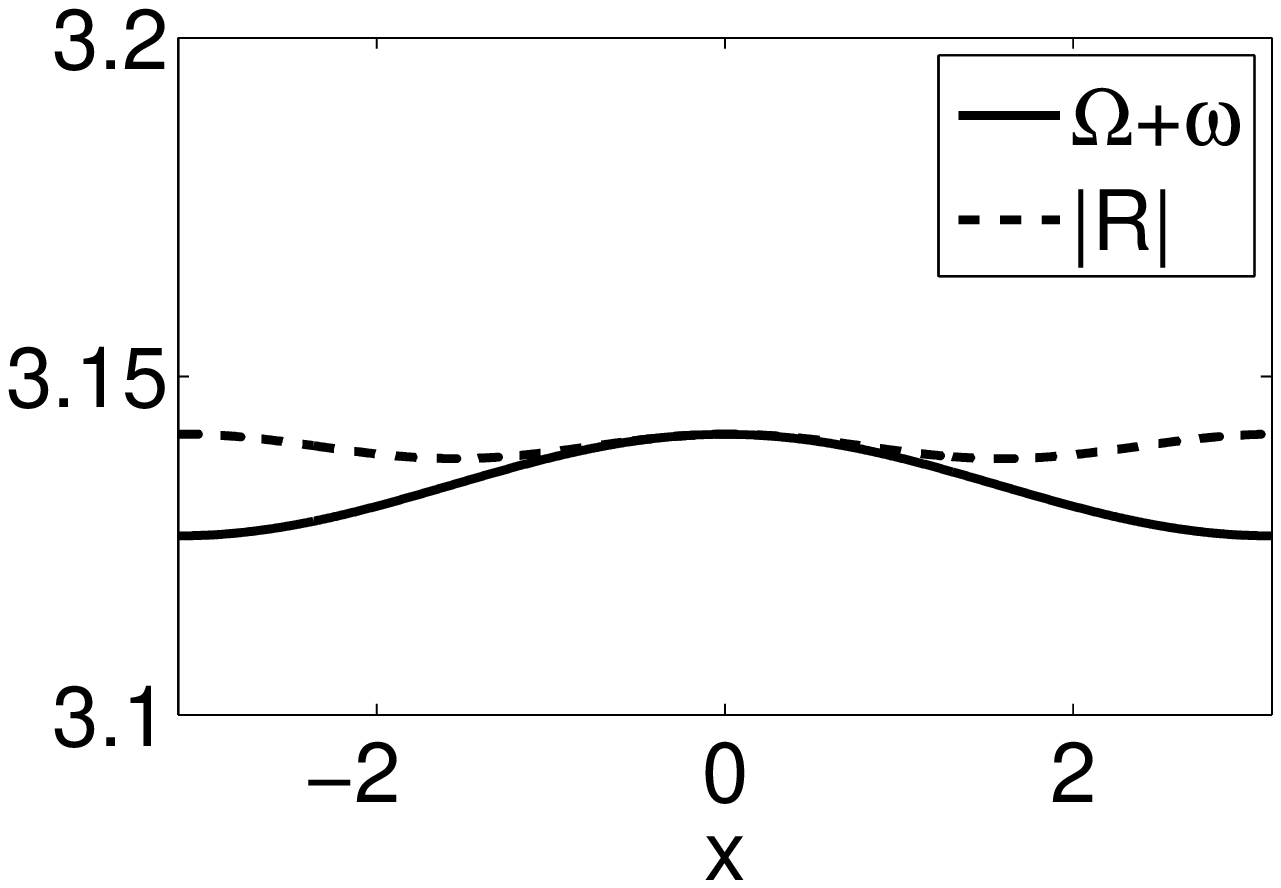}
\includegraphics[height=4cm]{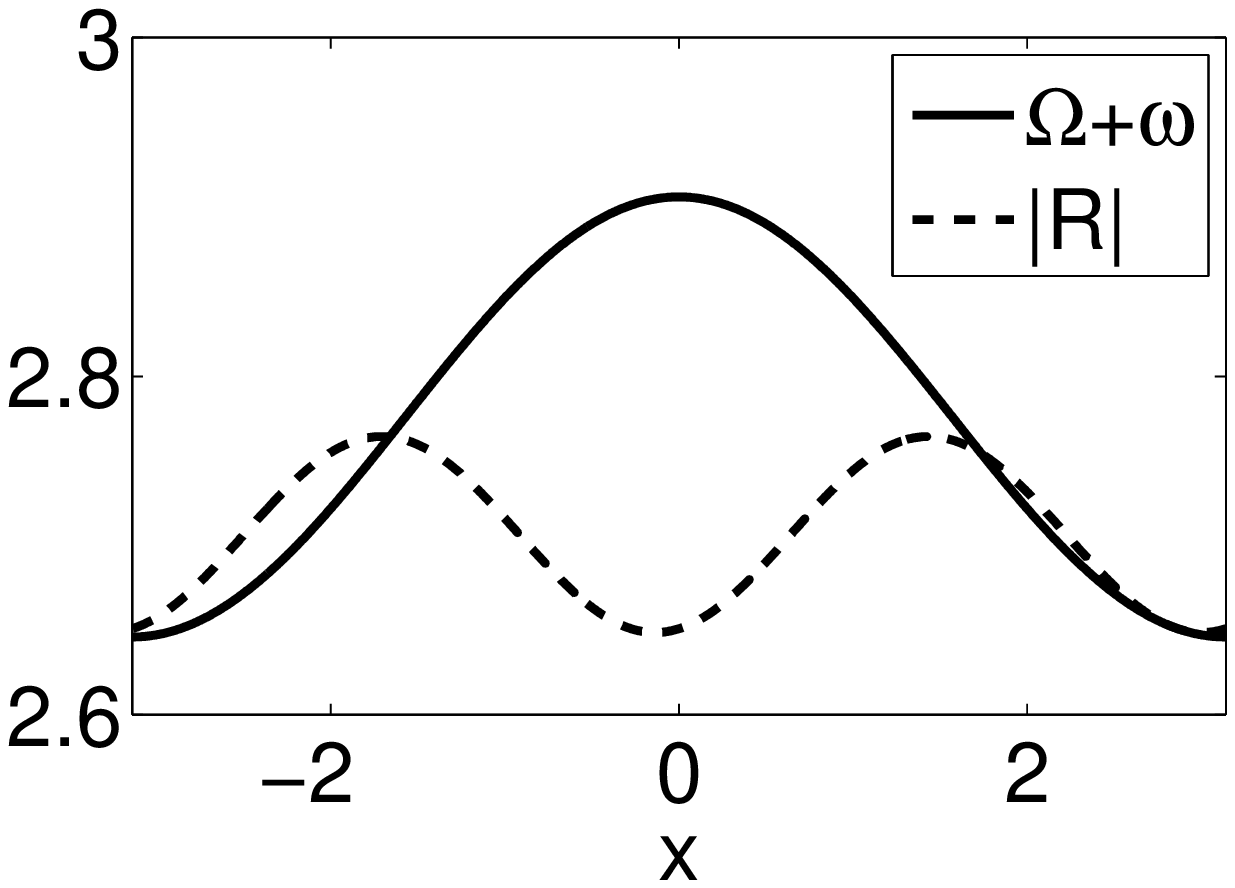}
\caption{Comparison of $\Omega + \omega(x)$ and $R(x)$ at the critical values $\omega_0$ for the appearance of a new region of incoherence around (a) $x=0$ for $\omega_0 = 0.0075$ and (b) $x=2.8$ for $\omega_0 = 0.13$. Parameters: $l=1$ and $\beta = 0.05$.}
\label{fig:cosx_cosx_tangent}
\end{figure}

\begin{figure}
\includegraphics[height=4cm]{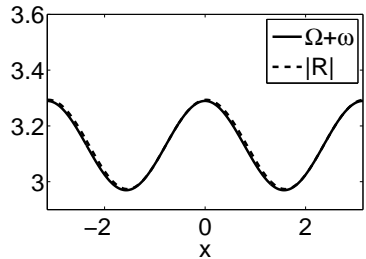}
\caption{Comparison of $\Omega + \omega(x)$ and $R(x)$ at the critical value $\omega_0 \approx 0.16$. Parameters: $l=2$ and $\beta = 0.05$.}
\label{fig:tangent_l2}
\end{figure}

To compute the solution branches and identify thresholds for additional transitions, we continue the solutions of Eqs.~(\ref{sc-full12}) and ({\ref{sc-full22}}) with respect to the parameter $\omega_0$. Figures~\ref{fig:splay1} and \ref{fig:splay2} show the dependence of $\Omega$ and of the coherent fraction $e$ on $\omega_0$ when $l=1$ and $2$, respectively. The figures indicate that the coherent fraction $e$ falls below 1 at $\omega_0\approx 0.0075$ ($l=1$), $\omega_0\approx 0.16$ ($l=2$), and $\omega_0\approx 0.0048$ ($l=3$, 4 and 5). These values coincide with the parameter values at which an incoherent region emerges. In addition, Fig.~\ref{fig:splay1}(b) reveals the presence of a second transition, at $\omega_0\approx 0.13$, corresponding to the emergence of a 2-cluster state from a 1-cluster state. Figures~\ref{fig:cosx_cosx_tangent} and \ref{fig:tangent_l2} show the profiles of $\Omega+\omega(x)$ and $R(x)$ at the critical values of $\omega_0$ at which incoherent regions emerge.

\begin{figure}
\includegraphics[height=3.5cm]{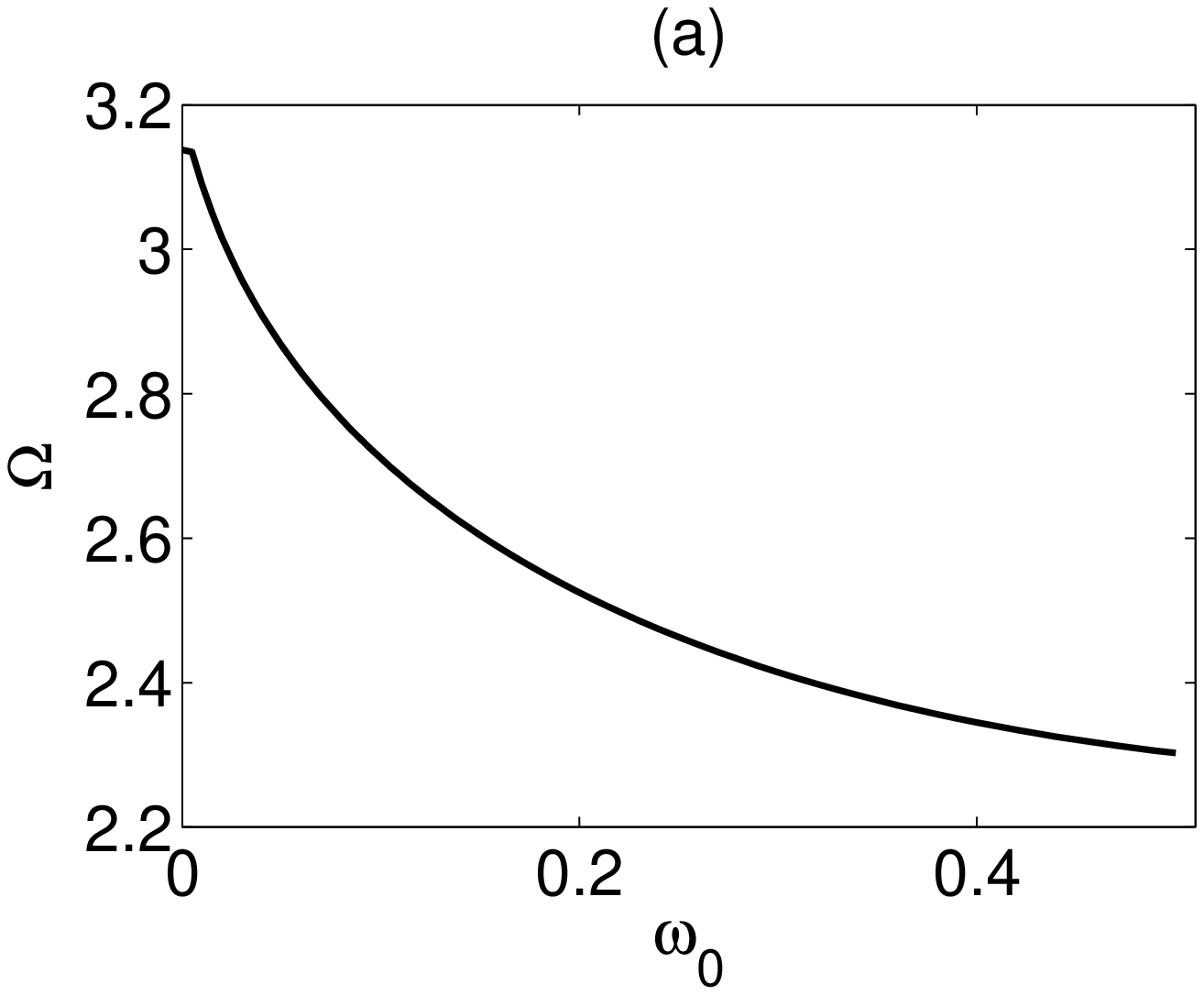}
\includegraphics[height=3.5cm]{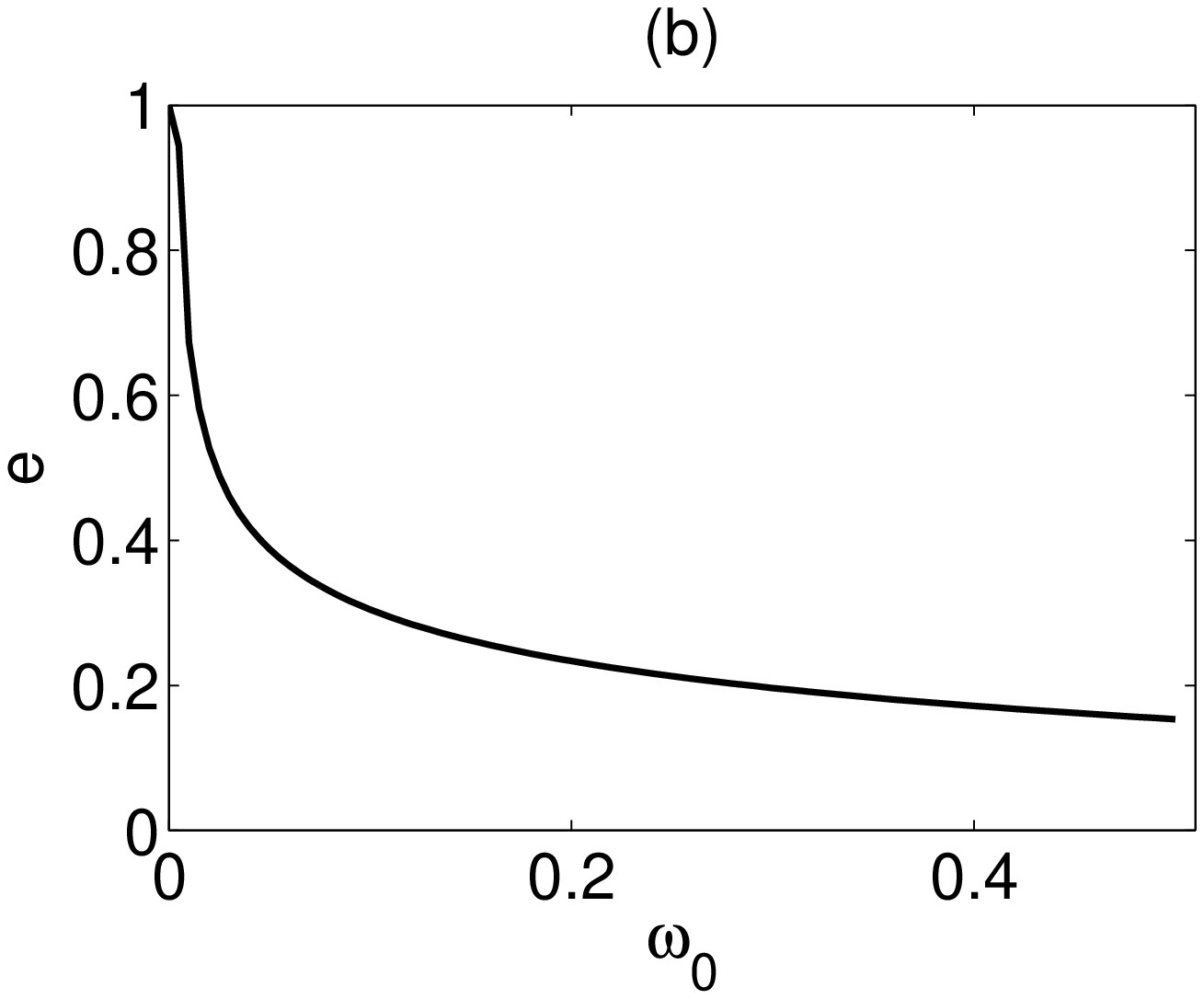}
\caption{Dependence of (a) $\Omega$ and (b) the coherent fraction $e$ on $\omega_0$, with $\omega(x)=\omega_0\cos(3x)$ and $\beta=0.05$.}
\label{fig:3cluster}
\end{figure}

To understand the $l$-cluster chimera states reported in Fig.~\ref{fig:cosx_inho_l_cluster_345}, we computed the local order parameter $\tilde{Z}(x)$ and found that $R(x)$ is approximately constant while the phase $\Theta(x)$ varies at a constant rate, suggesting the Ansatz $\tilde{Z}(x) = a e^{ix}$. With this Ansatz, Eqs.~(\ref{sc-full12}) and (\ref{sc-full22}) reduce to a single equation,
\begin{equation}
2\pi\Omega-\left<\sqrt{(\Omega+\omega_0\cos(ly))^2-a^2}\right>=2e^{-i\beta}a^2,\label{sc_Lcluster}
\end{equation} 
for all positive integers $l\geq 3$. Figure~\ref{fig:3cluster} shows the result of numerical continuation of a solution for $l=3$. The figure reveals no further transitions, indicating that the $l=3$ chimera state persists to large values of $\omega_0$. Numerical simulation shows that these states ($l=3$, 4, 5) are stable and persist up to $\omega_0=1$.

\subsubsection{1-cluster chimera states}

\begin{figure}
\includegraphics[height=3.5cm]{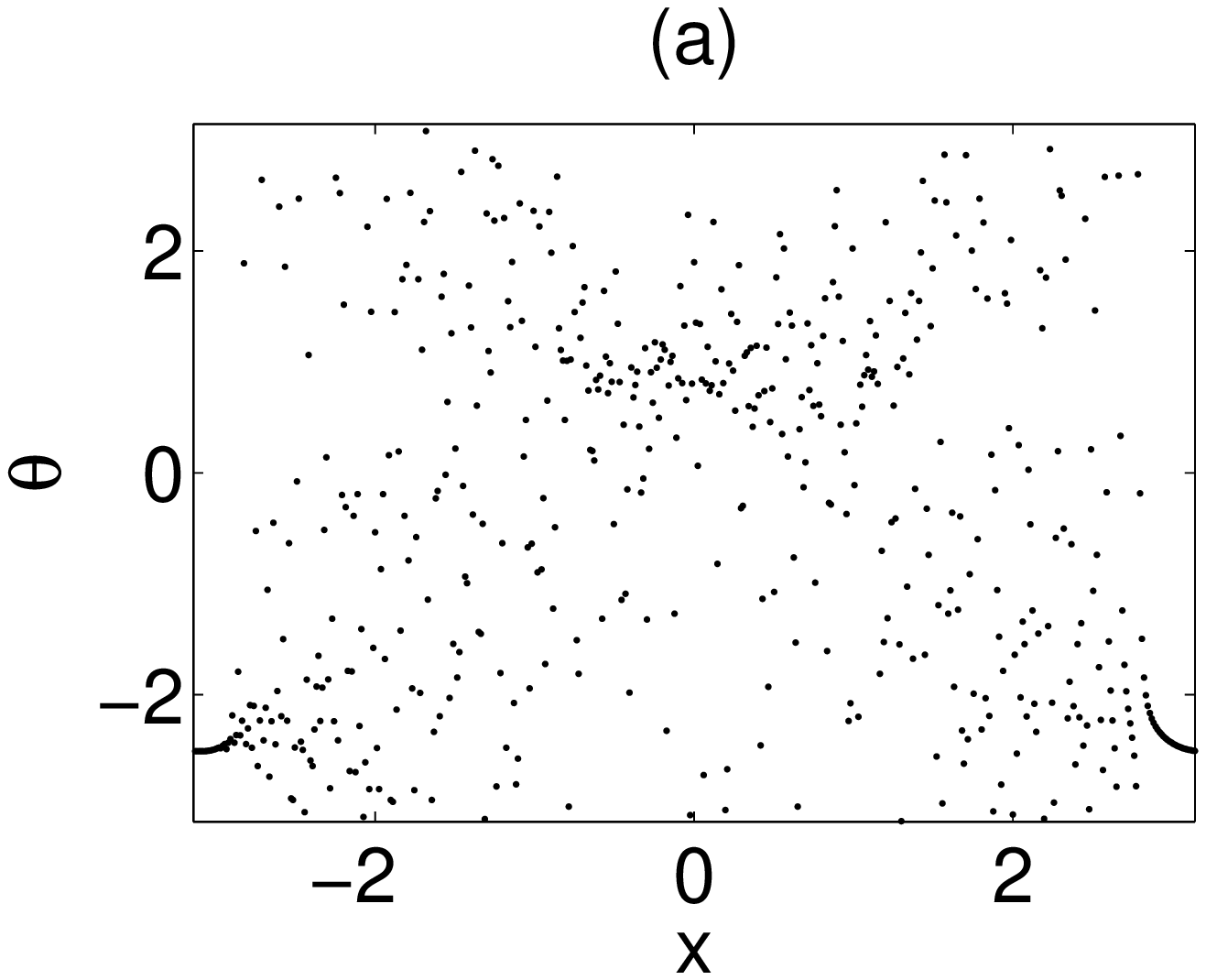}
\includegraphics[height=3.5cm]{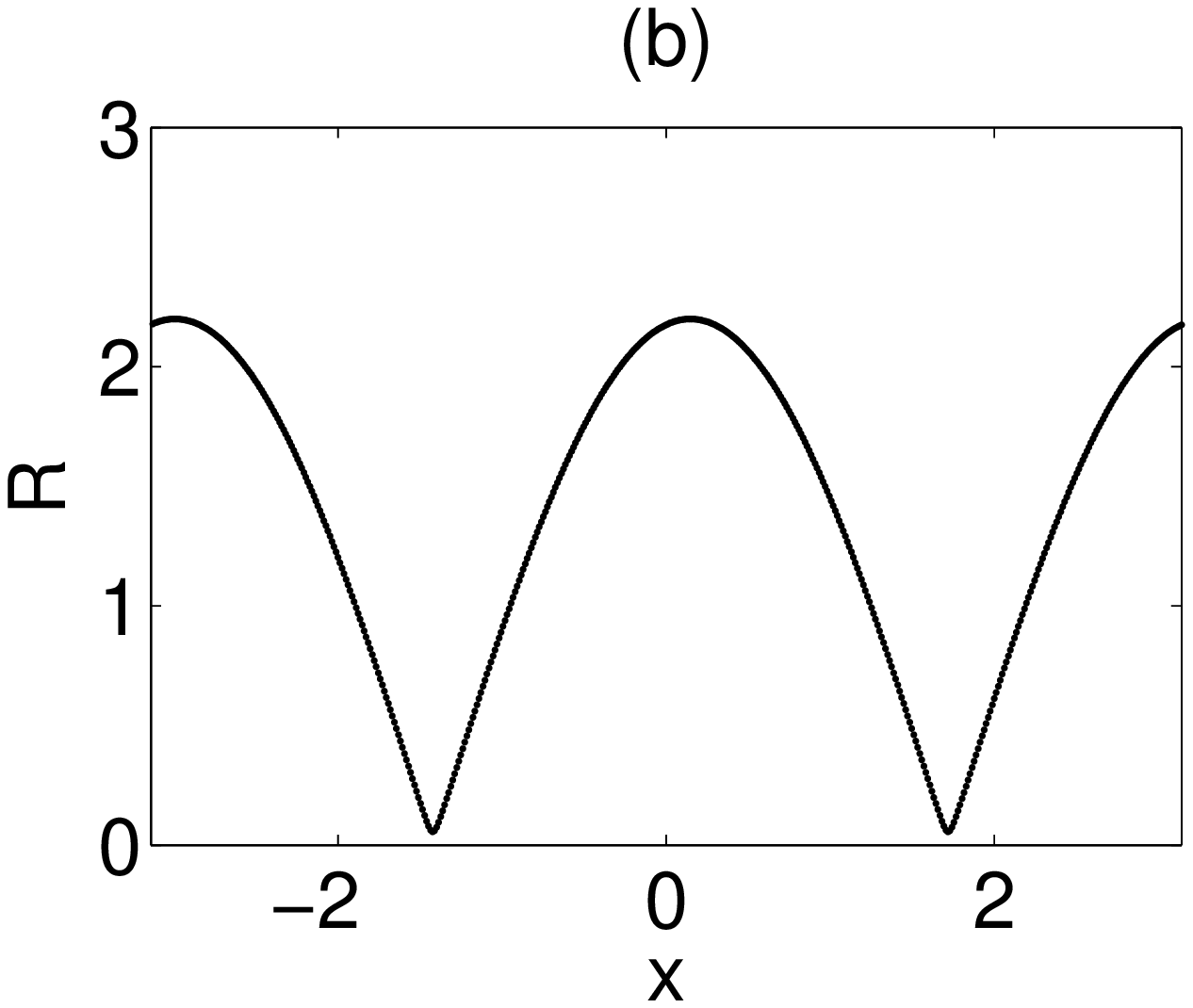}
\includegraphics[height=3.5cm]{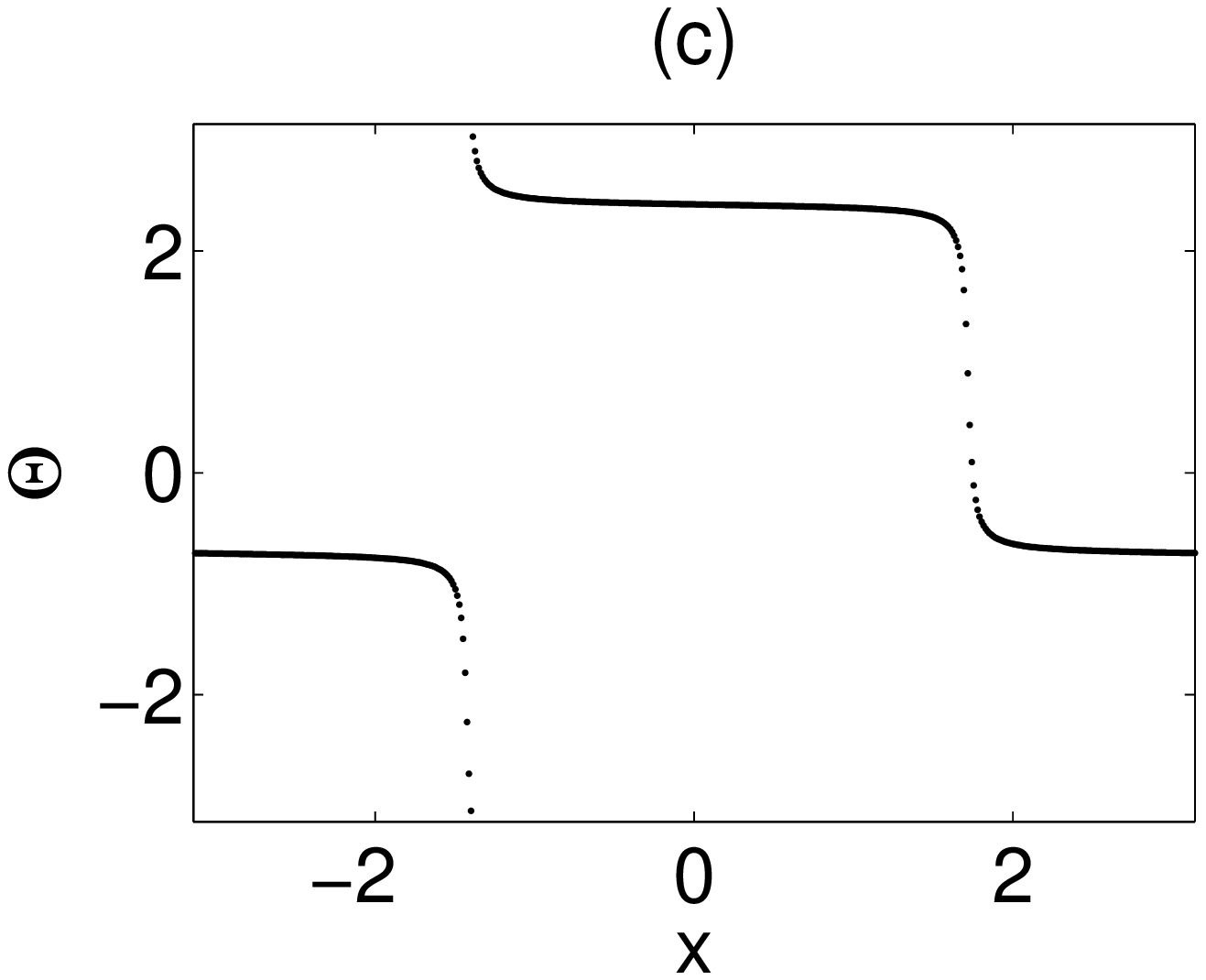}
\caption{(a) A snapshot of the phase distribution $\theta(x,t)$ in a 1-cluster chimera state for $G(x) = \cos(x)$ and $\omega(x) = 0.1 \cos(x)$. (b) The corresponding order parameter $R(x)$. (c) The corresponding order parameter $\Theta(x)$. Figure (a) shows that the presence of a nearly coherent region near $x=0$ with oscillators that oscillate $\pi$ out of phase with the coherent cluster. The calculation is done with $\beta = 0.05$ and $N=512$.}
\label{fig:cosx_inho_1cluster}
\end{figure}
\begin{figure}
\includegraphics[height=4cm]{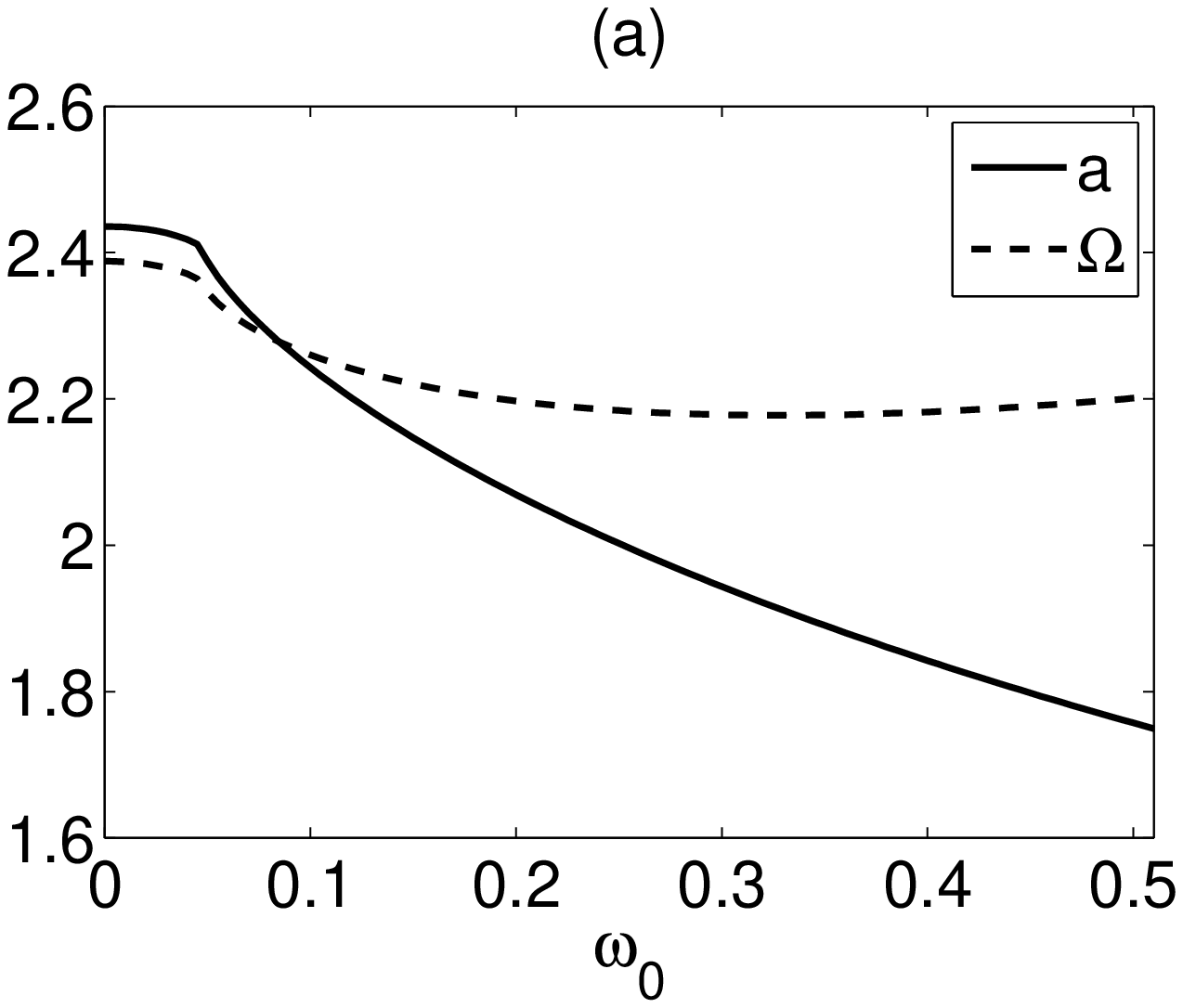}
\includegraphics[height=4cm]{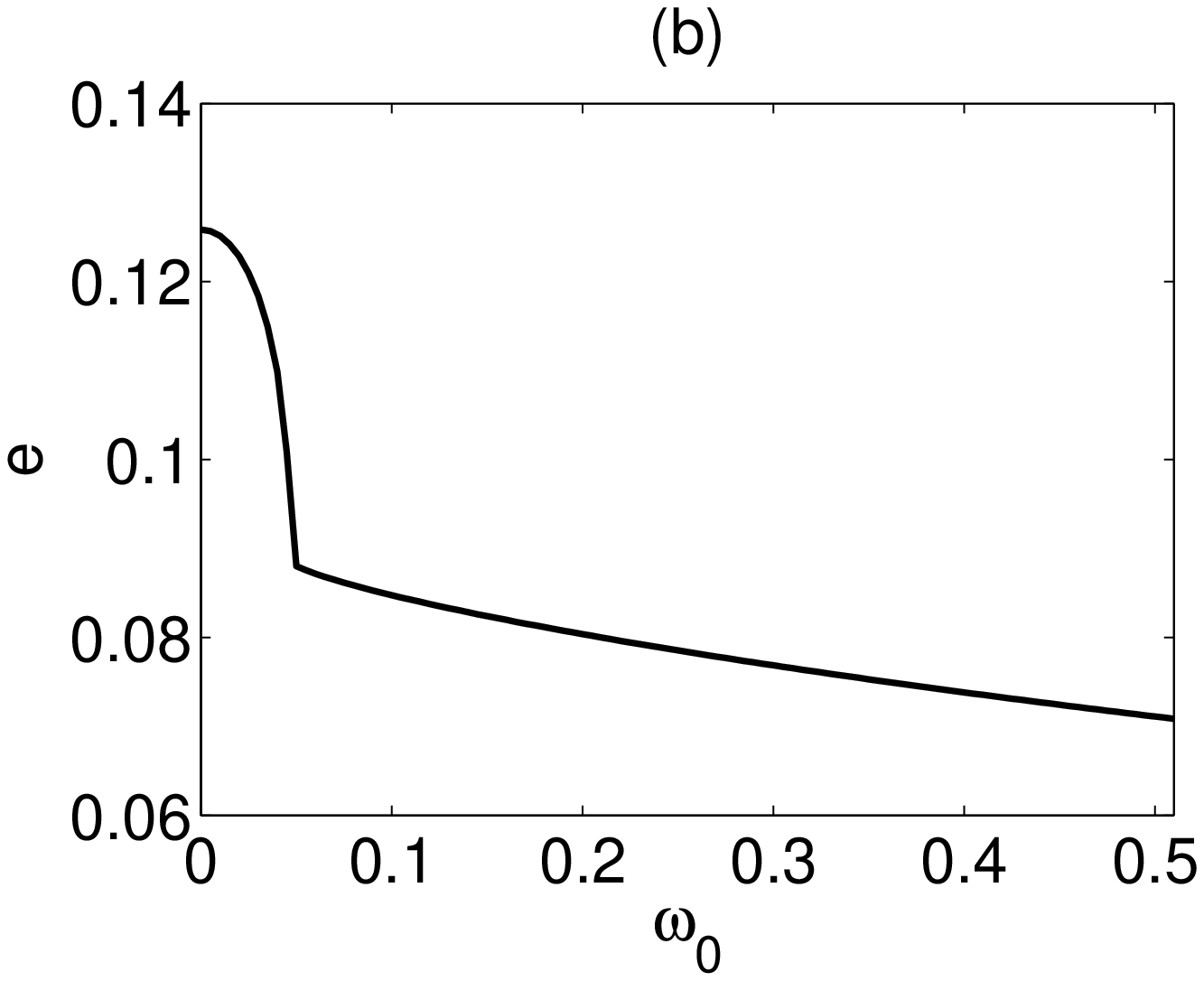}
\includegraphics[height=4cm]{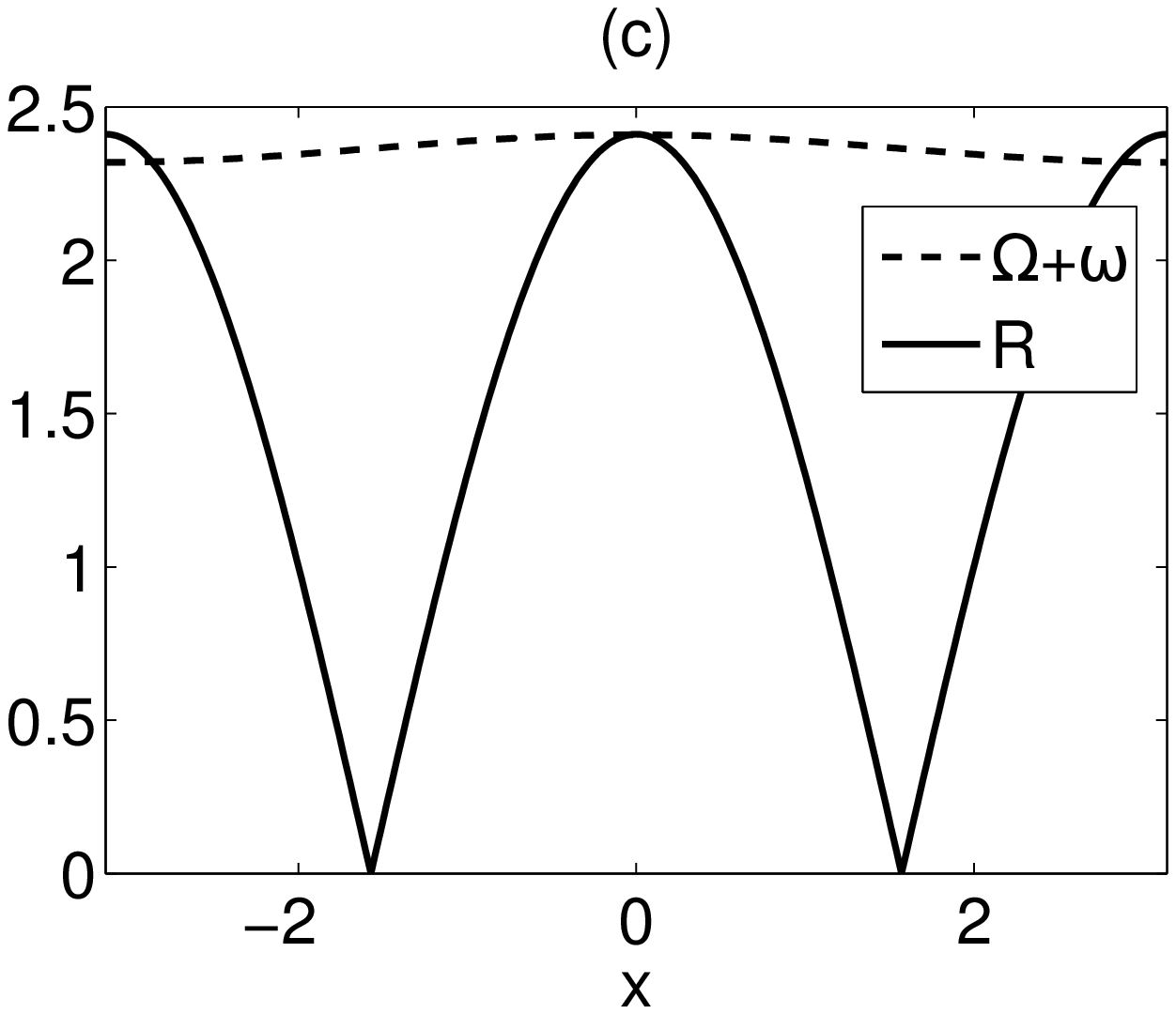}
\caption{(a) The quantities $a$ and $\Omega$, and (b) the coherent fraction $e$, all as functions of $\omega_0$ for the chimera state shown in Fig.~\ref{fig:cosx_inho_1cluster}. (c) $\Omega+\omega(x)$ and $R(x)$ as functions of $x$ when $\omega_0=0.0463$.}
\label{fig:romegae}
\end{figure}
Figure~\ref{fig:cosx_inho_1cluster} shows an example of a 1-cluster chimera state when $\omega(x)=0.1\cos(x)$; 1-cluster chimera states of this type have not thus far been reported for $G(x) = \cos(x)$, $\omega_0=0$, where computations always result in 2-cluster states \cite{XKK2014}. To understand the origin of this unexpected 1-cluster state we use Fig.~\ref{fig:cosx_inho_1cluster}(b) to conclude that $\tilde{Z}(x)$ is of the form $a\cos(x)$, and hence that Eqs.~(\ref{sc-full12}) and (\ref{sc-full22}) reduce to the single equation
\begin{equation}
2\pi\Omega-\left<\sqrt{(\Omega+\omega(y))^2-a^2\cos^2y}\right>=e^{-i\beta}a^2.\label{chimera1_integro1}
\end{equation}
Figure~\ref{fig:romegae} shows the result of numerical continuation of the solution of this equation as a function of $\omega_0$. The figure reveals a transition at $\omega_0=0.0463$ (Figs.~\ref{fig:romegae}(a,b)). When $\omega_0<0.0463$, the solution is a 2-cluster chimera; as $\omega_0$ increases through $\omega_0\approx 0.0463$ the coherent region around $x=0$ disappears, leaving a single cluster chimera state (Fig.~\ref{fig:romegae}(c)).

\subsubsection{2-cluster chimera states}

\begin{figure}
\includegraphics[height=3.5cm]{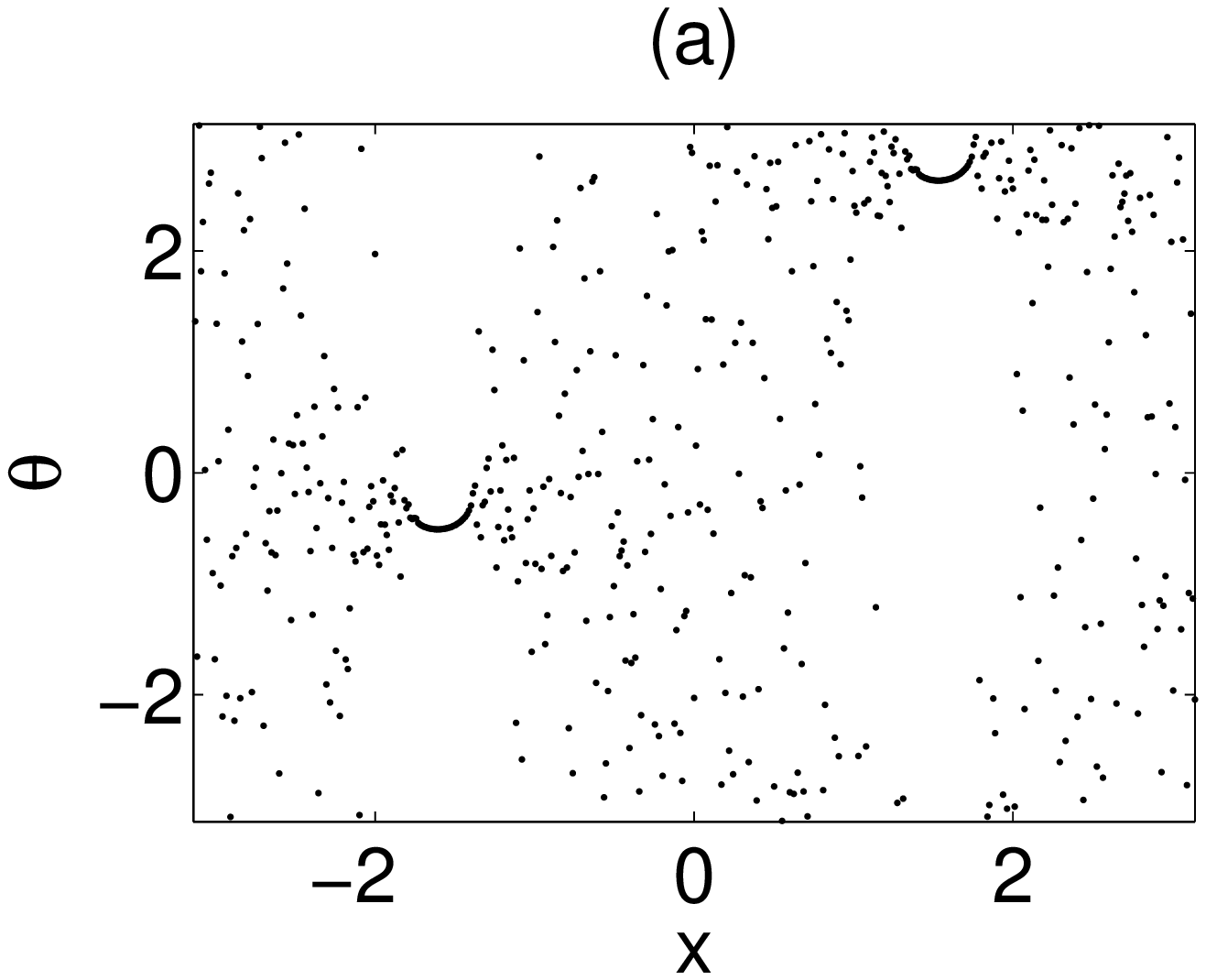}
\includegraphics[height=3.5cm]{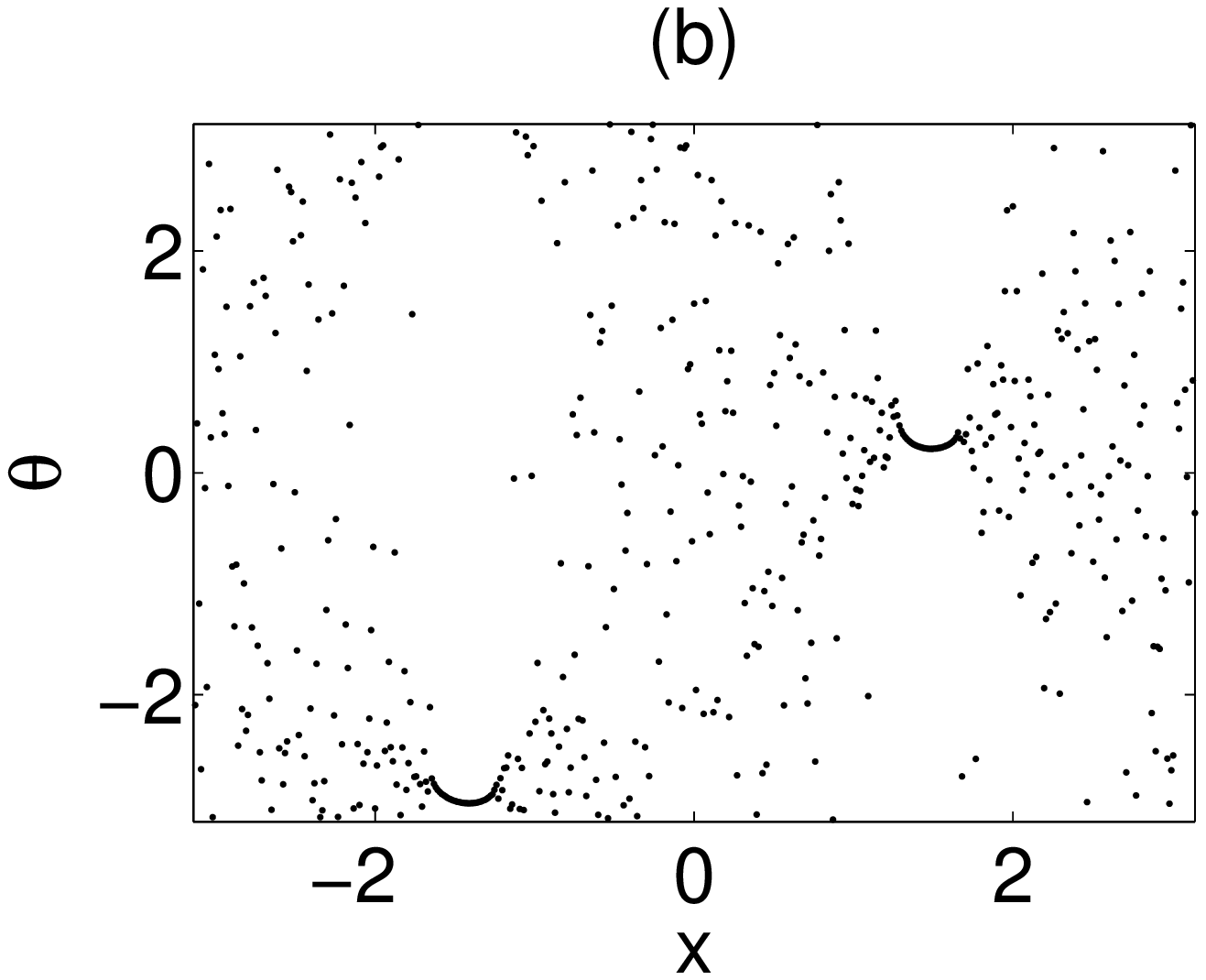}
\caption{2-cluster chimera states for $G(x) = \cos(x)$ and (a) $\omega(x)=0.1 \cos(2x)$ and (b) $\omega(x) = 0.1 \cos(3x)$. In both cases $\beta = 0.05$ and $N=512$.}
\label{fig:cosx_inho_2cluster}
\end{figure}

For $G(x) = \cos(x)$ and constant $\omega$ simulations always evolve into either a 2-cluster chimera state or a splay state \cite{XKK2014}. It is expected that when $\omega_0$ is small, the 2-cluster chimera is not destroyed. Figures \ref{fig:cosx_inho_2cluster}(a,b) gives examples of this type of state with $l=2$ and $l=3$ inhomogeneities, respectively. In both cases the clusters are located at specific locations selected by the inhomogeneity. 

\begin{figure}
\includegraphics[height=3.5cm]{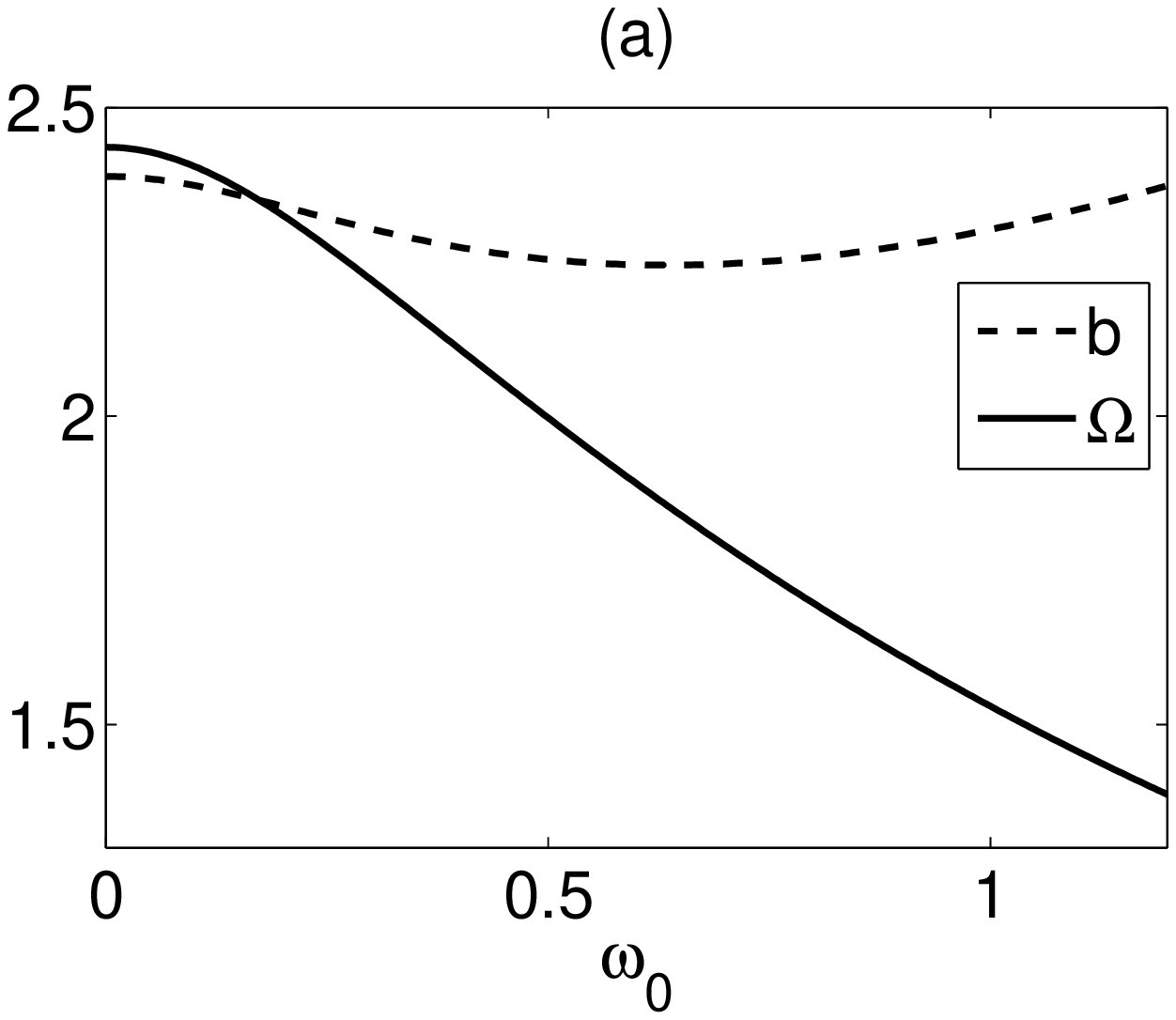}
\includegraphics[height=3.5cm]{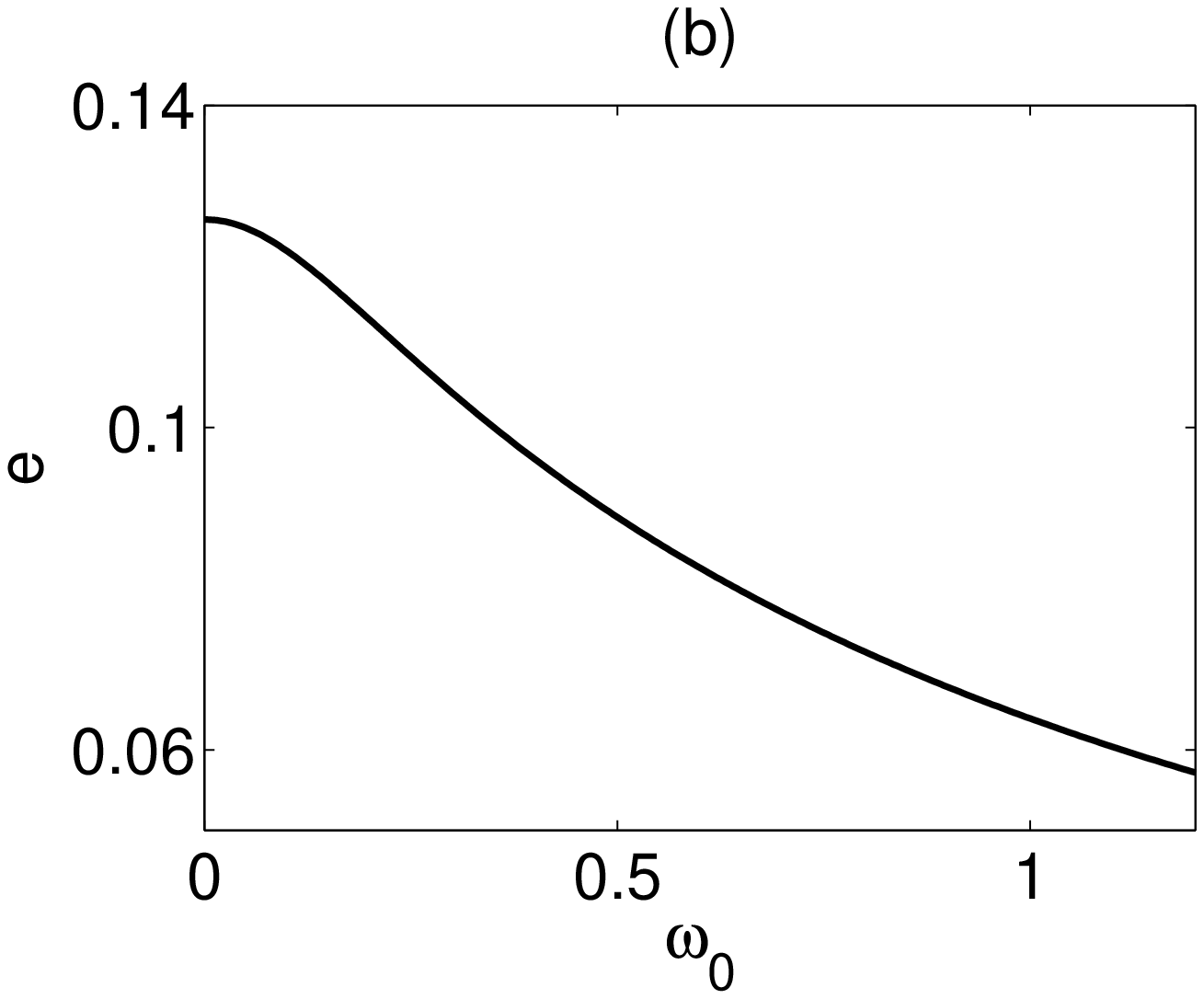}
\caption{The dependence of (a) $\Omega$ (solid line) and $b$ (dashed line), and (b) the coherent fraction $e$ on $\omega_0$ when $l=3$ and $\beta = 0.05$. }
\label{fig:chimera2-1}
\end{figure}

\begin{figure}
\includegraphics[height=3.5cm]{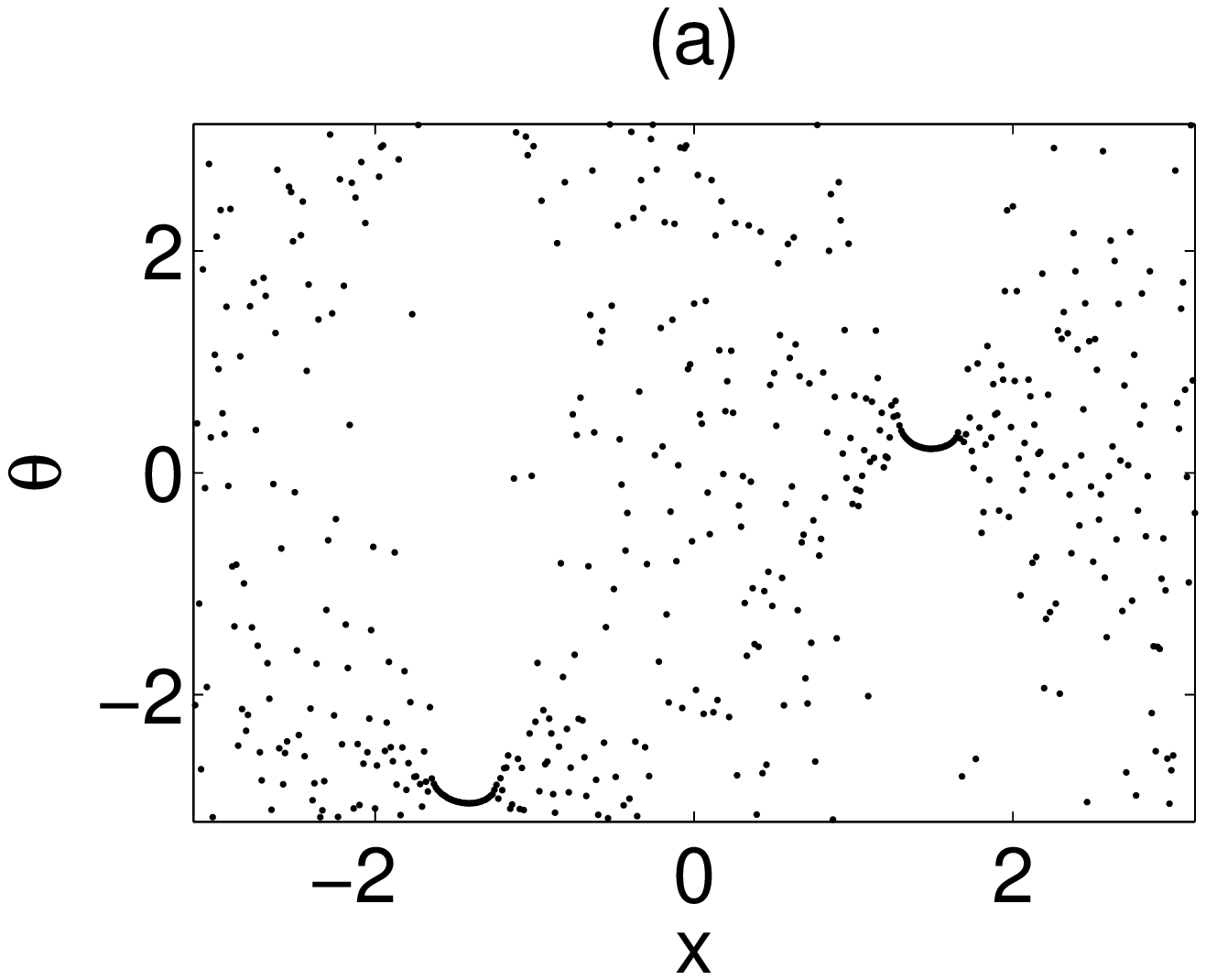}
\includegraphics[height=3.5cm]{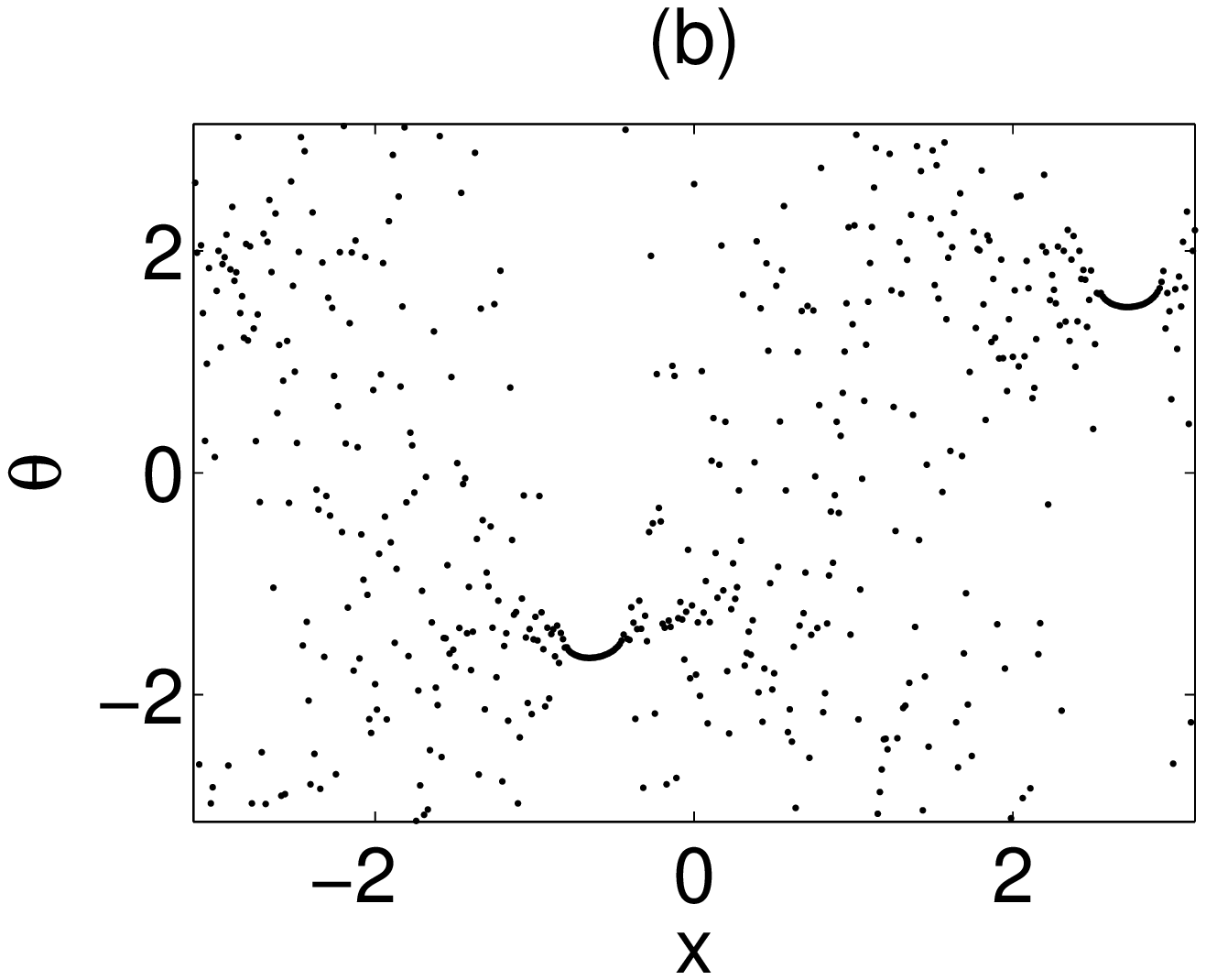}
\includegraphics[height=3.5cm]{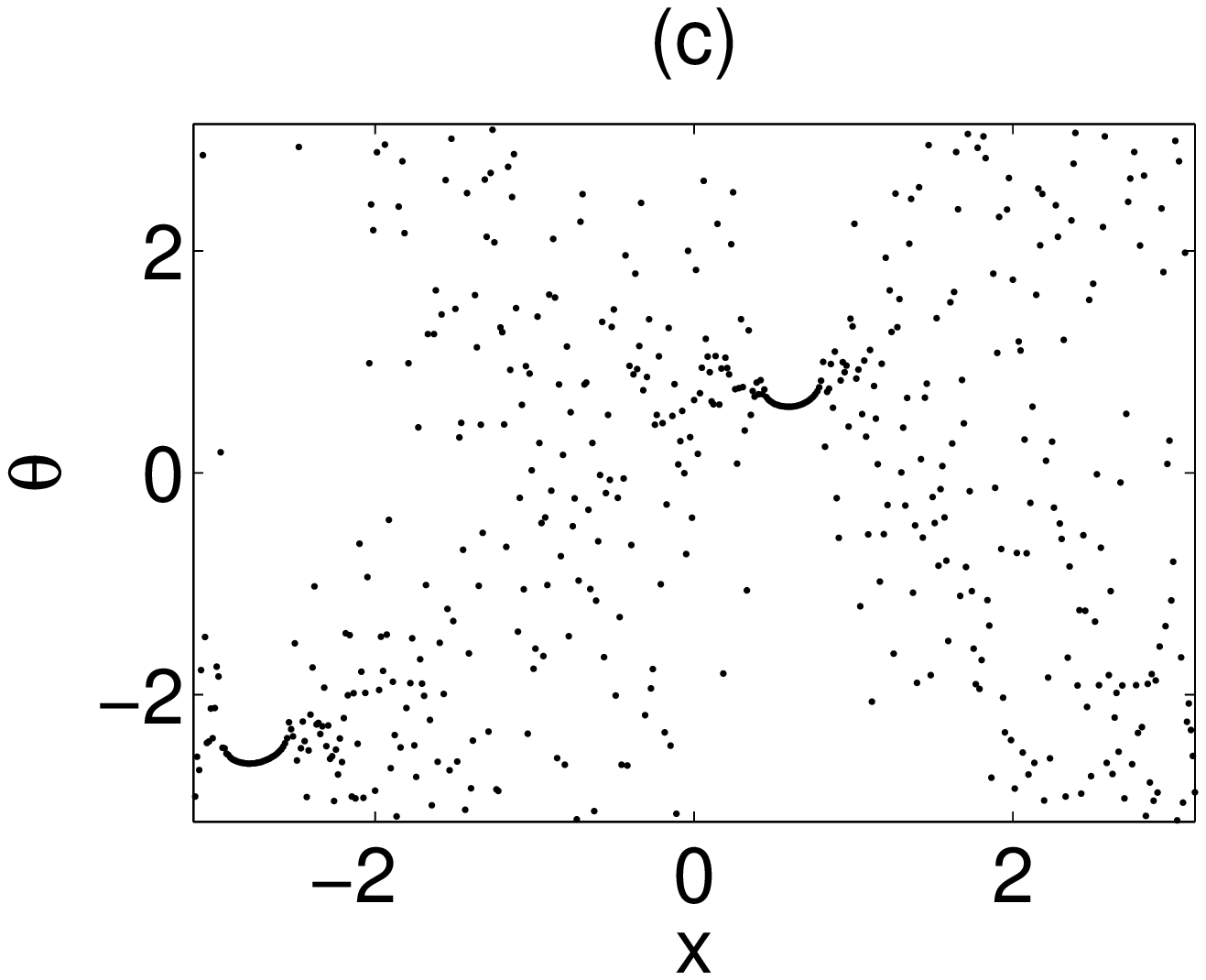}
\caption{The three possible locations of the 2-cluster chimera state when $\omega(x)= 0.1 \cos(3x)$. The simulation is done with $\beta = 0.05$ and $N = 512$.}
\label{fig:cosx_inho_wa01_wp3_2cluster}
\end{figure}

The order parameter profile with $l=3$ corresponding to the numerical solution in Fig.~\ref{fig:cosx_inho_2cluster}(b) has the same symmetry properties as that in Figs.~\ref{fig:cosx_bump_2cluster}(b,c). It follows that we may set $a=0$, $b=b_r>0$ in the order parameter representation (\ref{ansatz0}), resulting once again in the self-consistency relation (\ref{bump_scequation_2cluster}). We have continued the solutions of this relation for $l=3$ using the simulation in Fig.~\ref{fig:cosx_inho_2cluster}(b) with $\omega_0=0.1$ to initialize continuation in $\omega_0$. Figure~\ref{fig:chimera2-1} shows the result of numerical continuation of the order parameter for this state. No further transitions are revealed. There are three preferred locations for the coherent clusters, related by the translation symmetry $x\to x +\frac{2m\pi}{3} $, $m = 0,1,2$, as shown in Fig.~\ref{fig:cosx_inho_wa01_wp3_2cluster}.

\section{Traveling coherent solutions}\label{trav_coh}

\begin{figure}
\includegraphics[height=3cm]{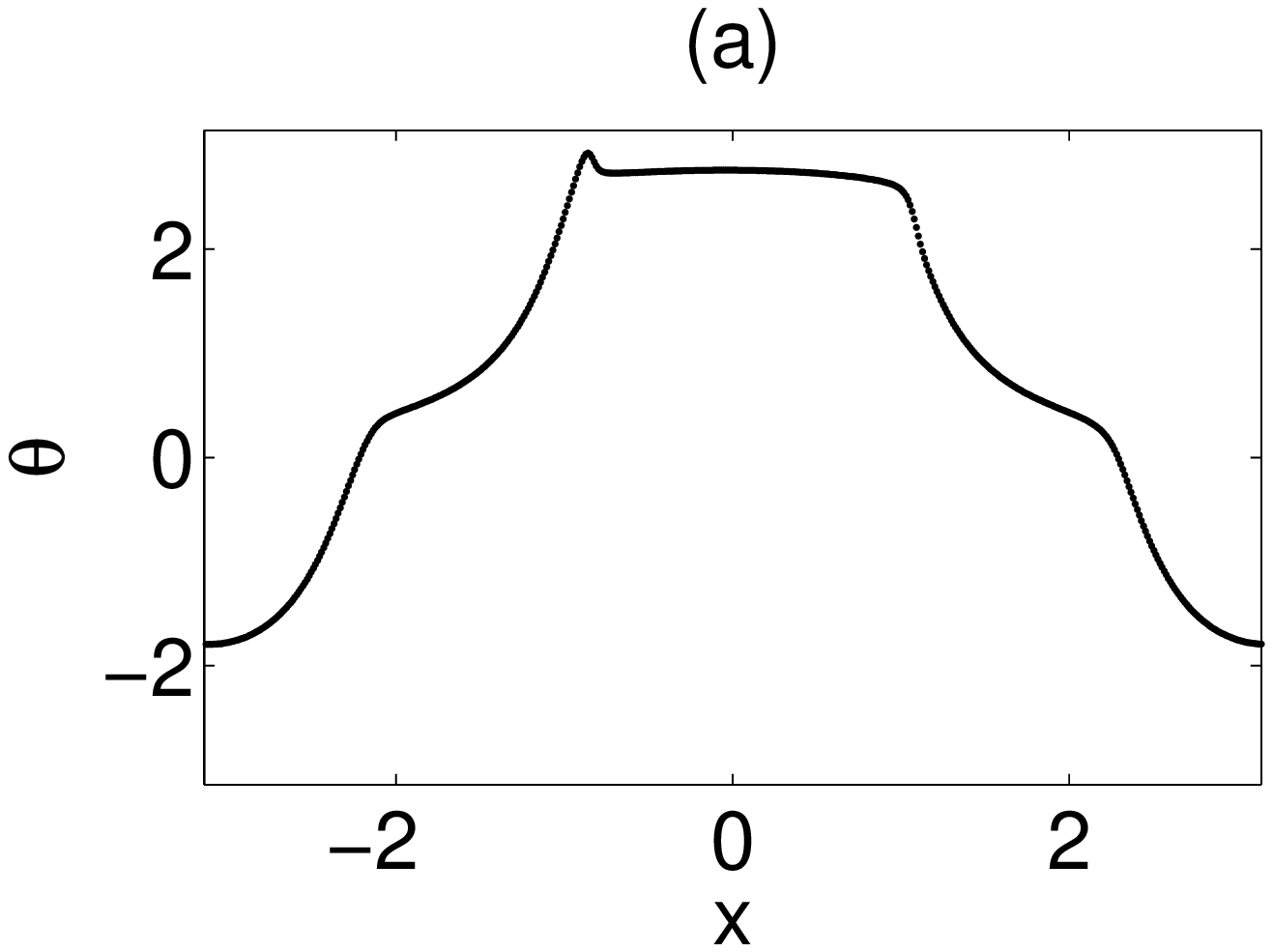}
\includegraphics[height=3cm]{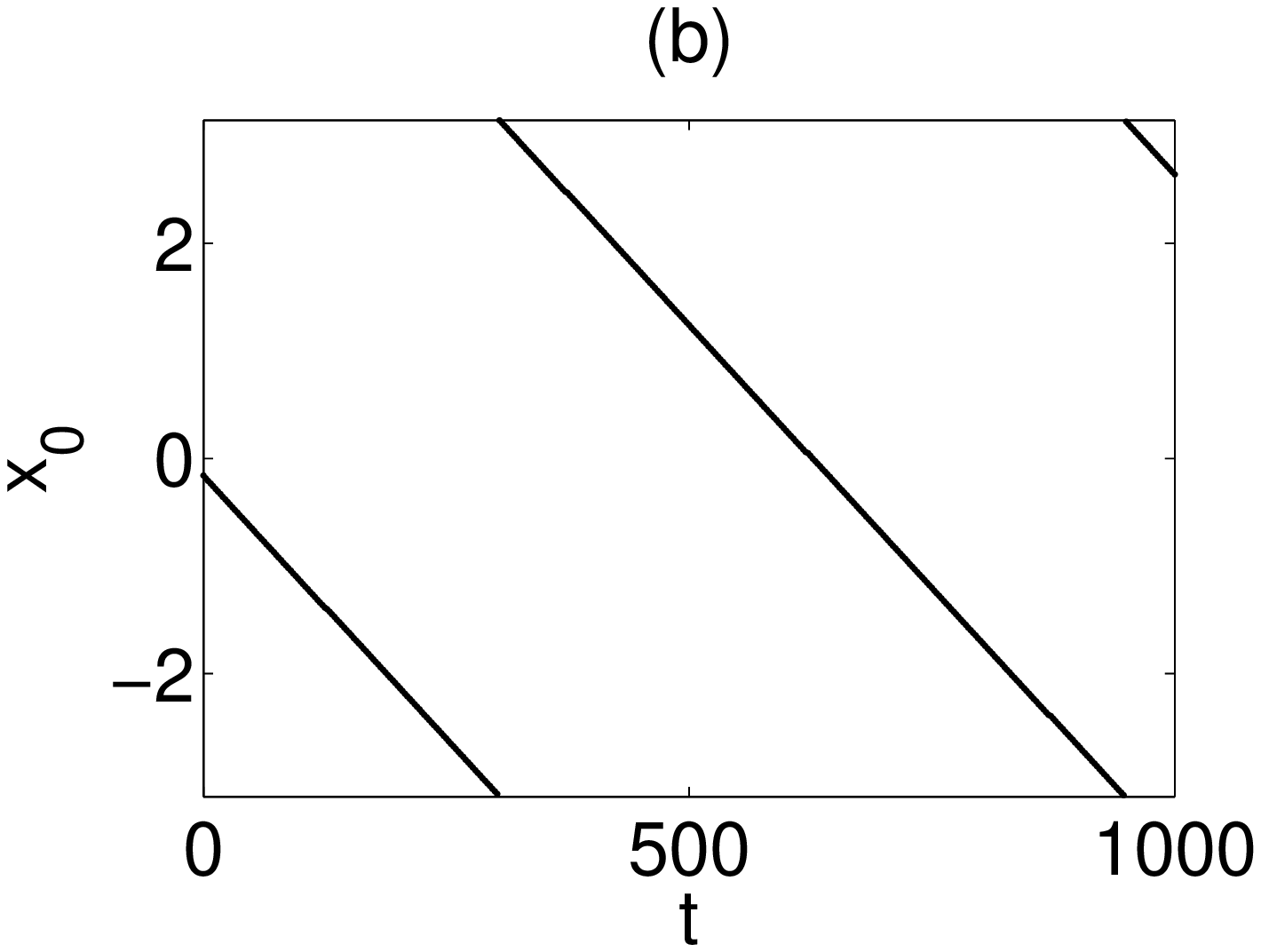}
\caption{(a) A snapshot of the phase pattern in a traveling coherent state when $\omega$ is constant. (b) The position $x_0$ of this state as a function of time. The simulation is done for $G(x) = \cos(x) + \cos(2x)$ with $\beta = 0.75$ and $N = 512$.}
\label{fig:phase_cos12x}
\end{figure}

We now turn to states with a spatially structured order parameter undergoing translation. With $G(x) = \cos(x) + \cos(2x)$ and constant $\omega$ the system (\ref{phase_eq}) exhibits a fully coherent but non-splay state that travels with a constant speed $c(\beta)$ when $0.646\lesssim\beta\lesssim 0.7644$ \cite{XKK2014}. Figure~\ref{fig:phase_cos12x}(a) shows a snapshot of this traveling coherent state while Fig.~\ref{fig:phase_cos12x}(b) shows its position $x_0$ as a function of time. The speed of travel is constant and can be obtained by solving a nonlinear eigenvalue problem \cite{XKK2014}.  In this section, we investigate how the inhomogeneities $\omega(x) = \omega_0\exp{(-\kappa|x|)}$ and $\omega(x)=\omega_0\cos(lx)$ affect the dynamical behavior of this state.

\subsection{Bump inhomogeneity: $\omega(x)\equiv \omega_0\exp{(-\kappa|x|)}$}

In this section we perform two types of numerical experiments. In the first, we fix $\beta$ and start with the traveling coherent solution for $\omega_0 = 0$. We then gradually increase $\omega_0$ gradually at fixed values of $\kappa$ and $\beta$. In the second we fix $\omega_0 > 0$, $\kappa>0$, and vary $\beta$.

\begin{figure}
\includegraphics[height = 7cm]{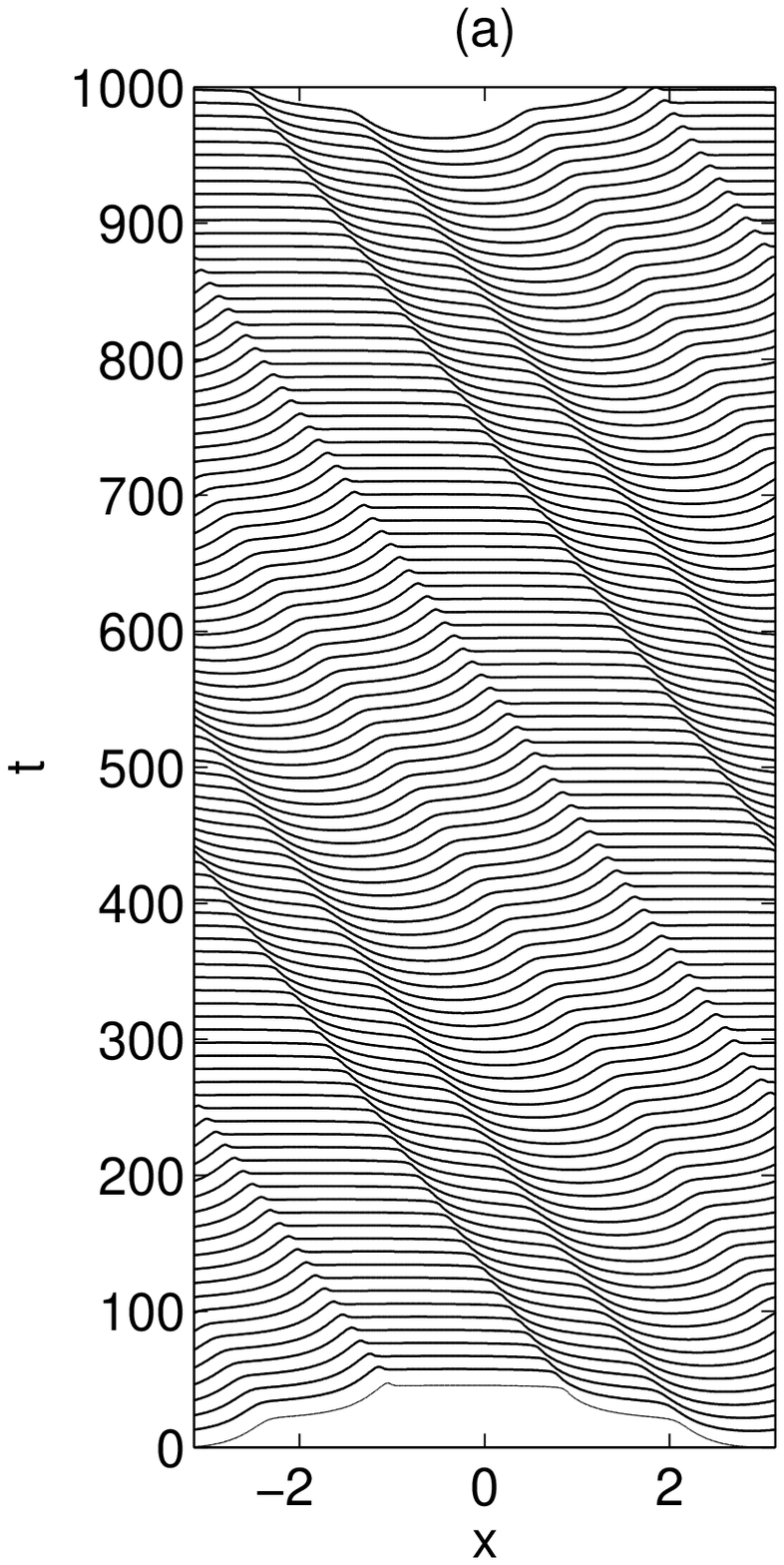}
\includegraphics[height = 7cm]{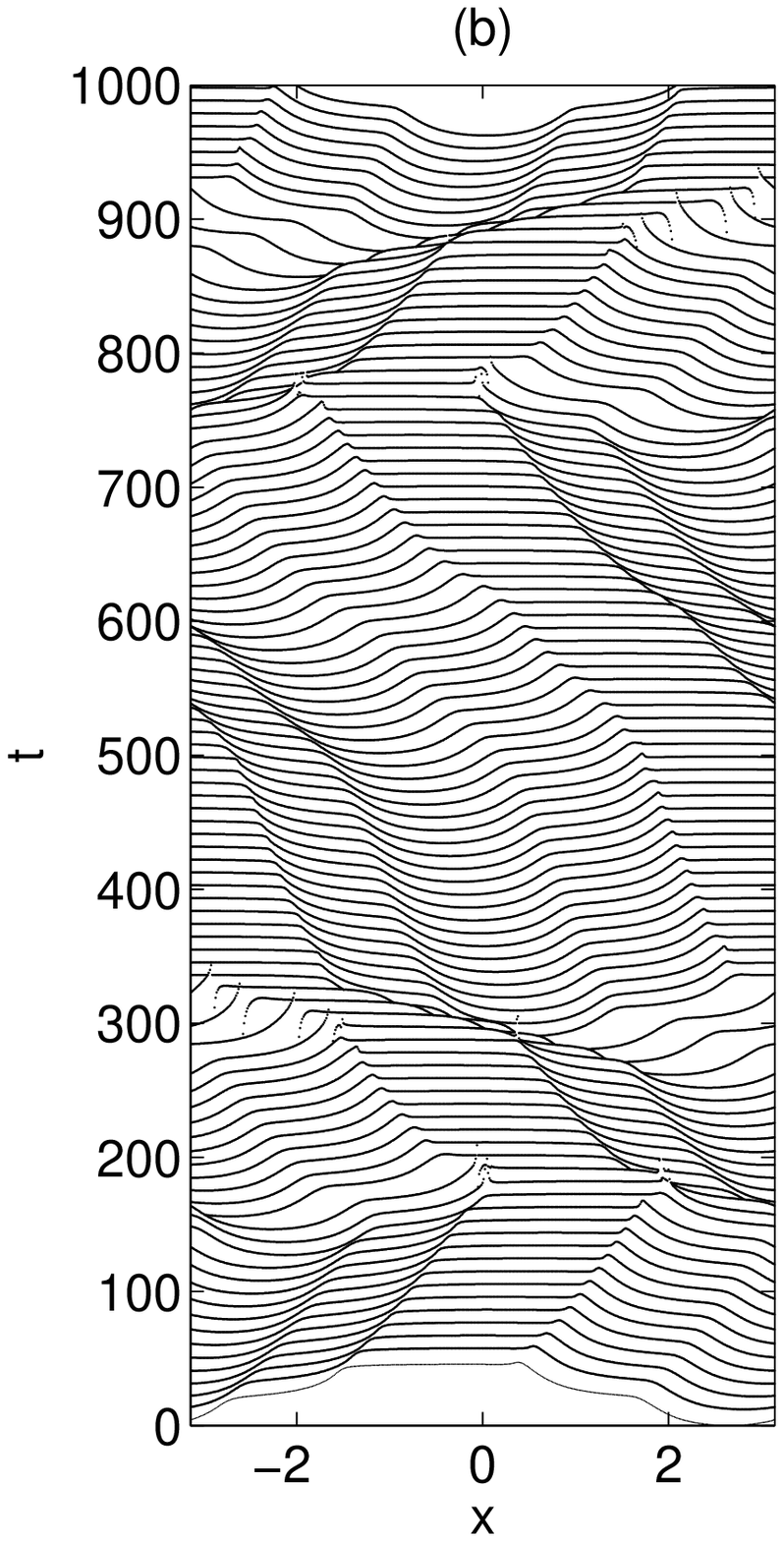}
\caption{Hidden line plot of the phase distribution $\theta(x,t)$ when (a) $\omega_0 =0$. (b) $\omega_0 =0.04$. In both cases, $\kappa = 2$, $\beta = 0.75$ and $N = 512$.}
\label{fig:hidden1}
\end{figure}

\begin{figure}
\includegraphics[height = 7cm]{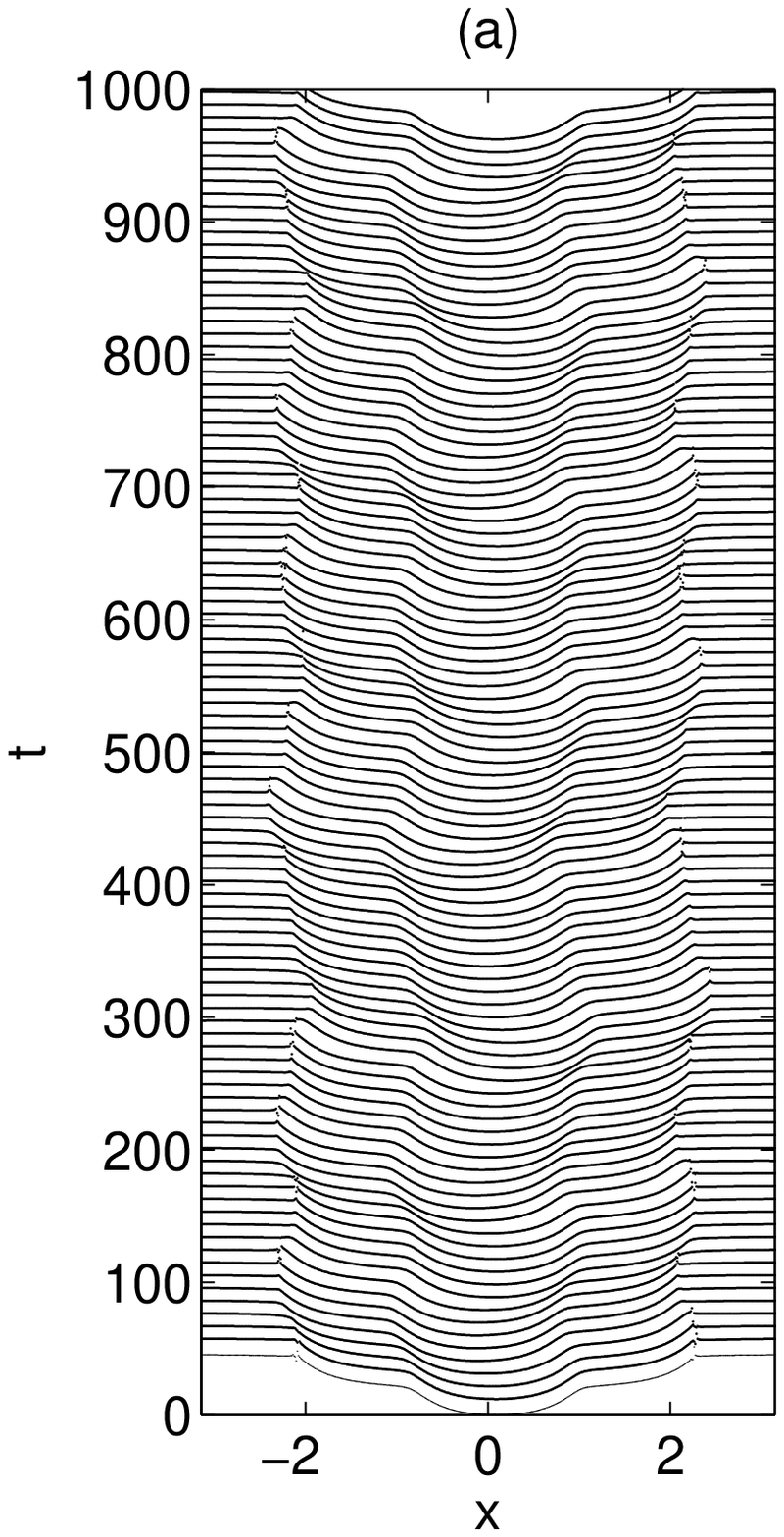}
\includegraphics[height = 7cm]{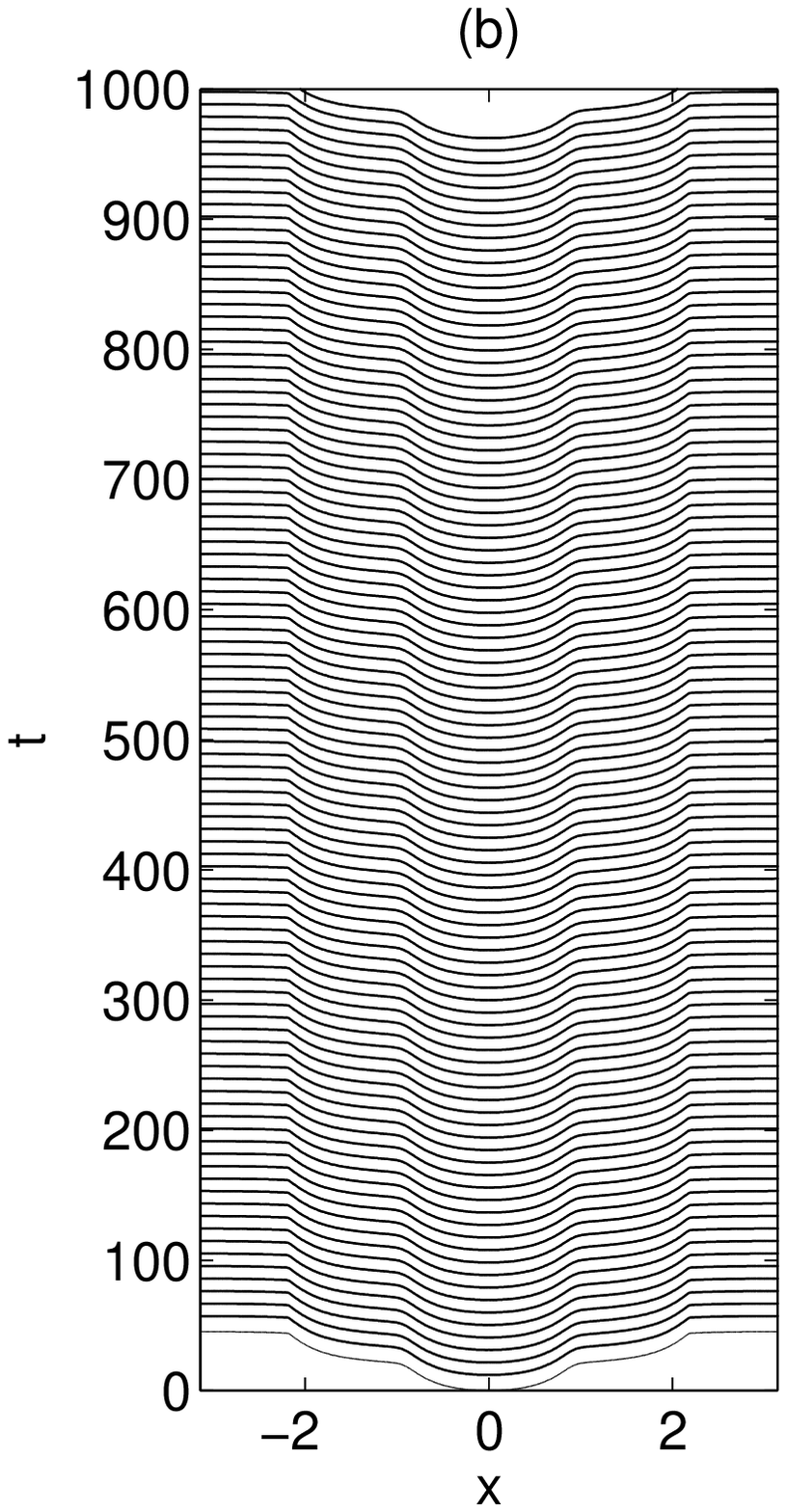}
\caption{Hidden line plot of the phase distribution $\theta(x,t)$ when (a) $\omega_0 =0.08$. (b) $\omega_0 =0.12$. In both cases, $\kappa = 2$, $\beta = 0.75$ and $N = 512$. }
\label{fig:hidden2}
\end{figure}

\begin{figure}
\includegraphics[height=3cm]{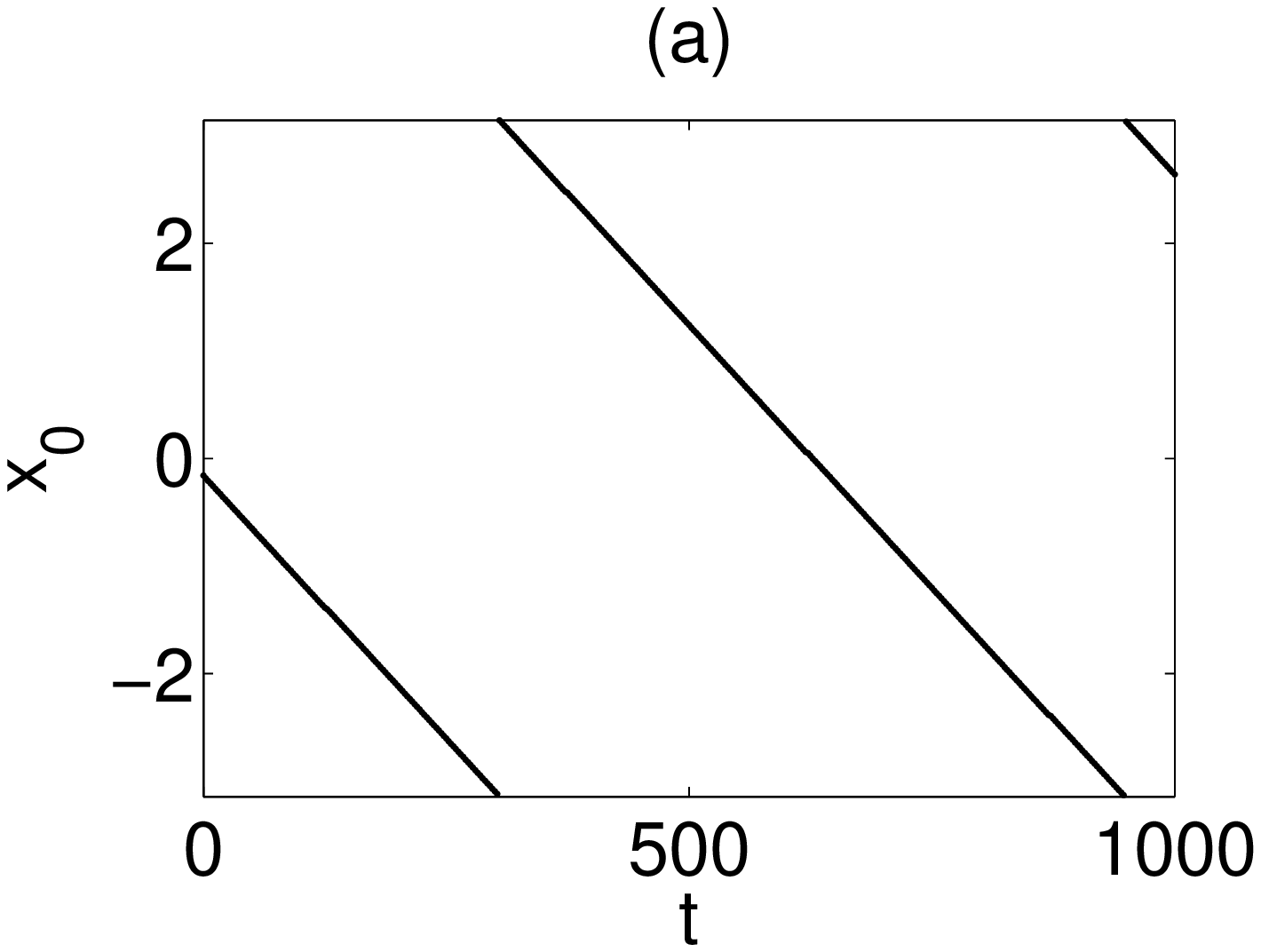}
\includegraphics[height=3cm]{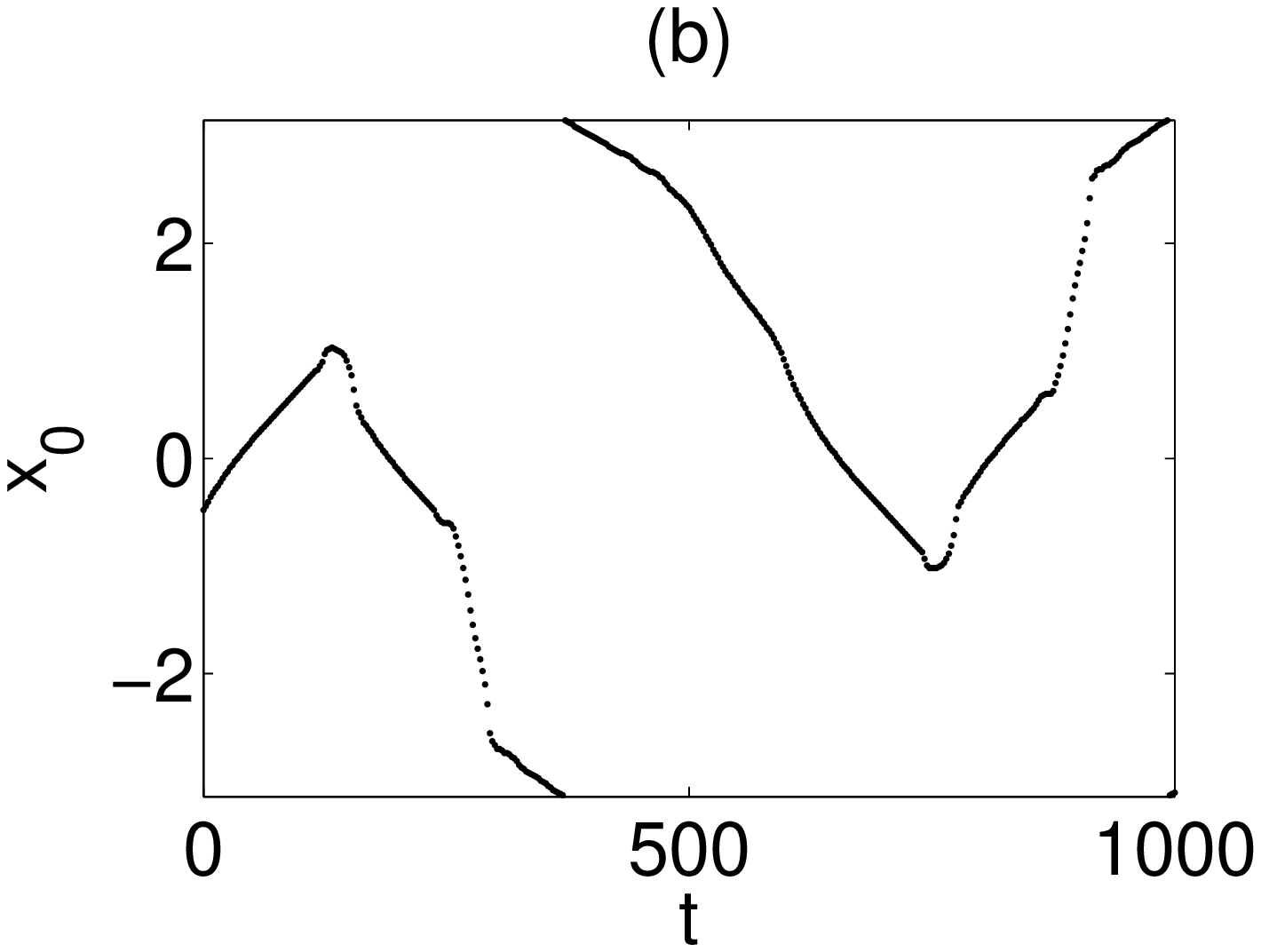}
\includegraphics[height=3cm]{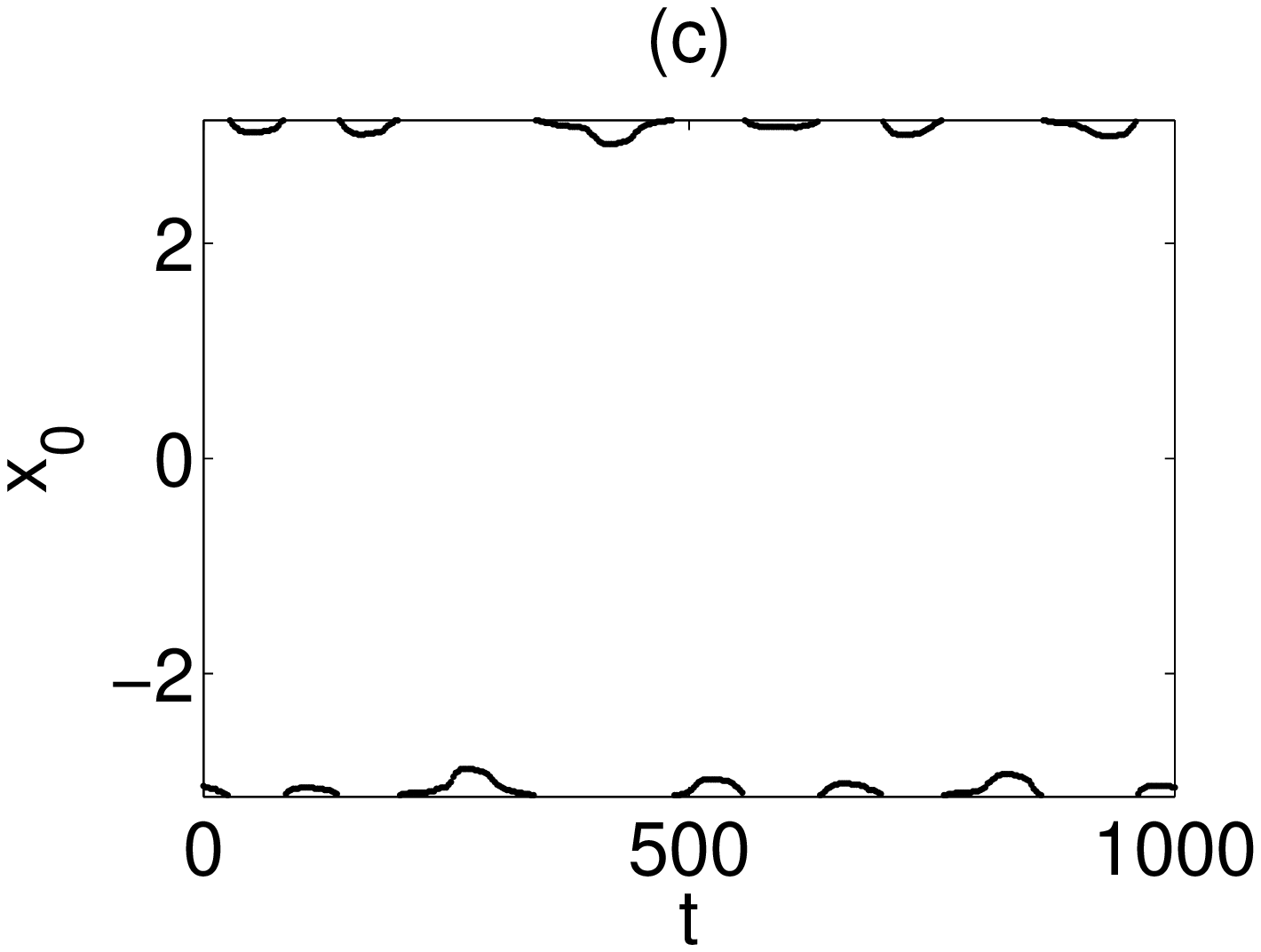}
\includegraphics[height=3cm]{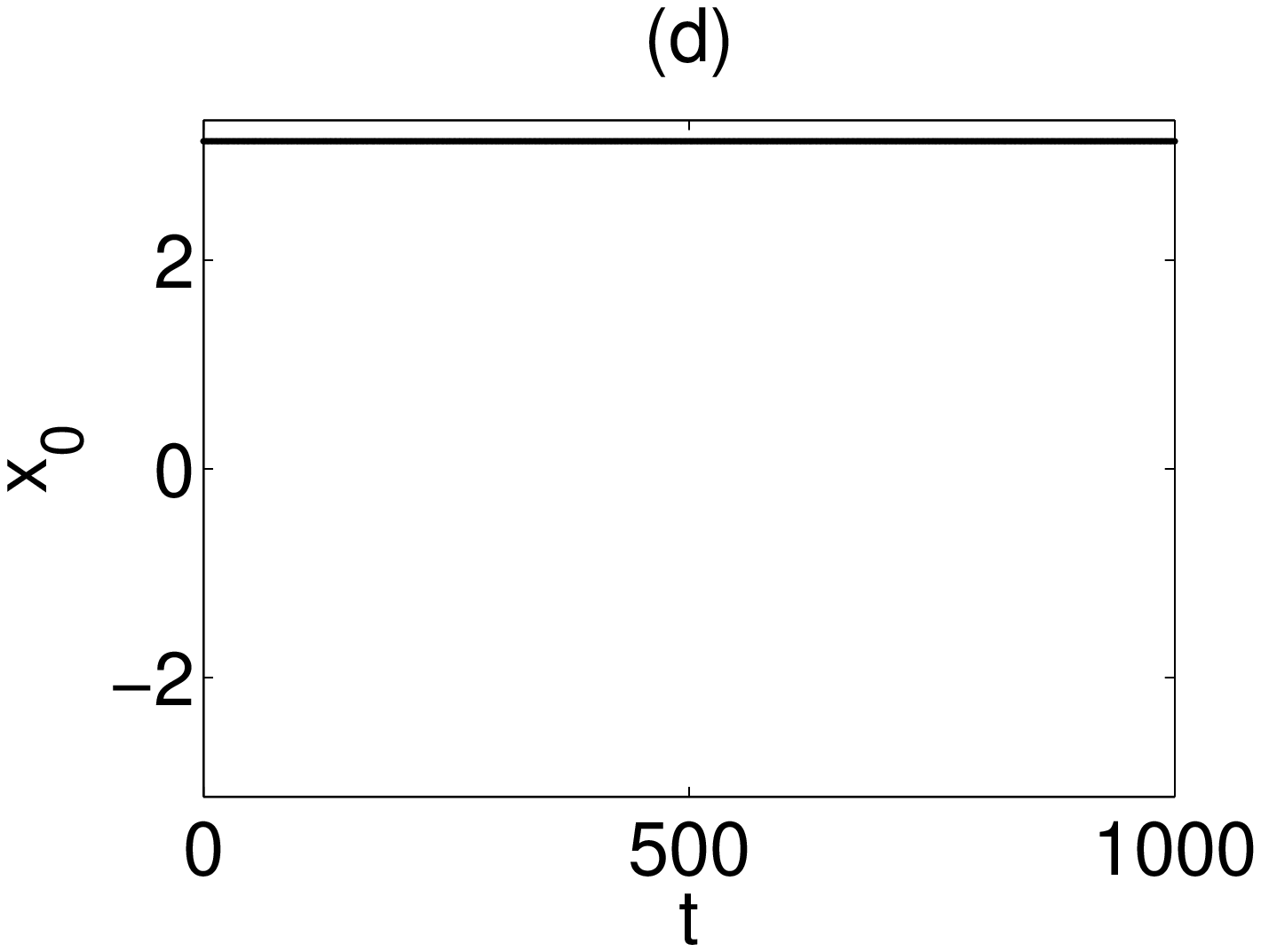}
\caption{The position $x_0$ of the coherent state in Figs.~\ref{fig:hidden1} and \ref{fig:hidden2} as a function of time for (a) $\omega_0 =0$, (b) $\omega_0 =0.04$, (c) $\omega_0 =0.08$, and (d) $\omega_0 =0.12$. In all cases $\kappa = 2$, $\beta = 0.75$ and $N = 512$.}
\label{fig:xvst}
\end{figure}

For each $0.646\lesssim\beta\lesssim 0.7644$ we find that the coherent state continues to travel, albeit nonuniformly, until $\omega_0$ reaches a threshold value that depends of the values of $\beta$ and $\kappa$. The case $\beta = 0.75$ provides an example. Figure~\ref{fig:hidden1}(a) shows the traveling coherent state in the homogeneous case ($\omega_0 = 0$) while Fig.~\ref{fig:hidden1}(b) shows the corresponding state in the presence of a frequency bump $\omega(x) \equiv \omega_0\exp{(-\kappa|x|)}$ with $\omega_0=0.04$ and $\kappa=2$. In this case the presence of inhomogeneity leads first to a periodic fluctuation in the magnitude of the drift speed followed by, as $\omega_0$ continues to increase, a transition to a new state in which the direction of the drift oscillates periodically (Fig.~\ref{fig:hidden1}(b)). We refer to states of this type as direction-reversing waves, by analogy with similar behavior found in other systems supporting the presence of such waves \cite{Landsberg,Hettel}. With increasing $\omega_0$ the reversals become localized in space (and possibly aperiodic, Fig.~\ref{fig:hidden2}(a)) and then cease, leading to a stationary pinned structure at $\omega_0=0.12$ (Fig.~\ref{fig:hidden2}(b)). Figure~\ref{fig:xvst} shows the position $x_0$ of the maximum of the local order parameter $R(x,t)$ of the coherent state as a function of time for the cases in Figs.~\ref{fig:hidden1} and \ref{fig:hidden2}, showing the transition from translation to pinning as $\omega_0$ increases, via states that are reflection-symmetric {\it on average}. The final state is a steady reflection-symmetric pinned state aligned with the imposed inhomogeneity.


In fact, the dynamics of the present system may be more complicated than indicated above since a small group of oscillators located in regions where the order parameter undergoes rapid variation in space may lose coherence in a periodic fashion even when $\omega_0=0$ thereby providing a competing source of periodic oscillations in the magnitude of the drift speed. As documented in \cite{XKK2014} this is the case when $0.7570<\beta<0.7644$. For $\beta=0.75$, however, the coherent state drifts uniformly when $\omega_0=0$ and this is therefore the case studied in greatest detail. 




\begin{figure}[!htpb]
\includegraphics[height = 4cm]{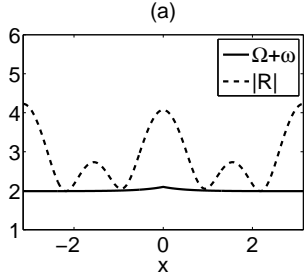}
\caption{The profiles of $\Omega +\omega(x)$ and $R = |a\cos(x) + c\cos(2x)|$ for $\omega_0 = 0.11$, $\kappa = 2$ and $\beta=0.75$.}
\label{fig:ROmega_omega0_011}
\end{figure}

The profile of the pinned coherent state can also be determined from a self-consistency analysis. For $G(x) = \cos(x)+ \cos(2x)$, the local order parameter $\tilde{Z}(x)$ can be written in the form
\begin{equation}
\tilde{Z}(x) = a \cos(x) + b\sin(x) +c\cos(2x) + d\sin(2x).
\end{equation}
Owing to the reflection symmetry of the solution $ b = d = 0$; on applying a rotation in $\theta$ we may take $a$ to be real. The self-consistency equation then becomes 
\begin{eqnarray}
ae^{-i\beta}&=&\left<\frac{\cos(y)(\Omega+\omega(y)-\mu(y))}{a\cos y+c^*\cos (2y)}\right>,\label{sc-cos12x1}\\
ce^{-i\beta}&=&\left<\frac{\cos(2y)(\Omega+\omega(y)-\mu(y))}{a\cos y+c^*\cos (2y)}\right>.
\label{sc-cos12x2}
\end{eqnarray}
When $\omega_0 = 0.12$, $\kappa=2$ and $\beta=0.75$ the solution of these equations is $\Omega = 1.9879, a = 3.1259, c_r =-0.1007, c_i = -2.7300 $, results that are consistent with the values obtained from numerical simulation. For these parameter values, $\Omega +\omega(x) < |a\cos(x) + c\cos(2x)|$ for $x\in (-\pi, \pi]$, indicating that all phases rotate with the same frequency $\Omega$.  As we decrease $\omega_0$ to $\omega_0 \approx 0.11$ the profiles $\Omega +\omega(x)$ and $ |a\cos(x) + c\cos(2x)|$ start to touch (Fig.~\ref{fig:ROmega_omega0_011}) and for yet lower $\omega_0$ the stationary coherent state loses stability and begins to oscillate as described in the previous paragraph. 

\begin{figure}
\includegraphics[height = 7cm]{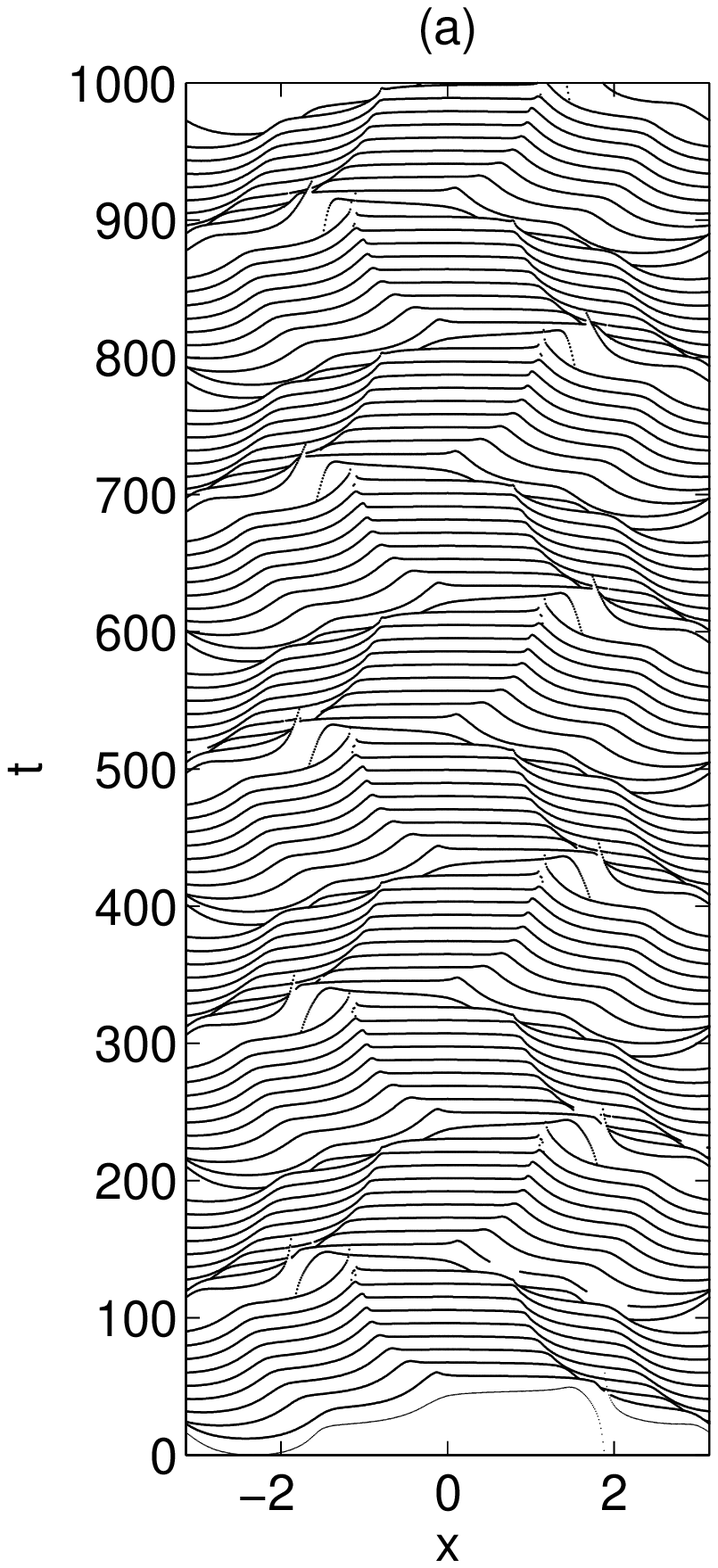}
\includegraphics[height = 7cm]{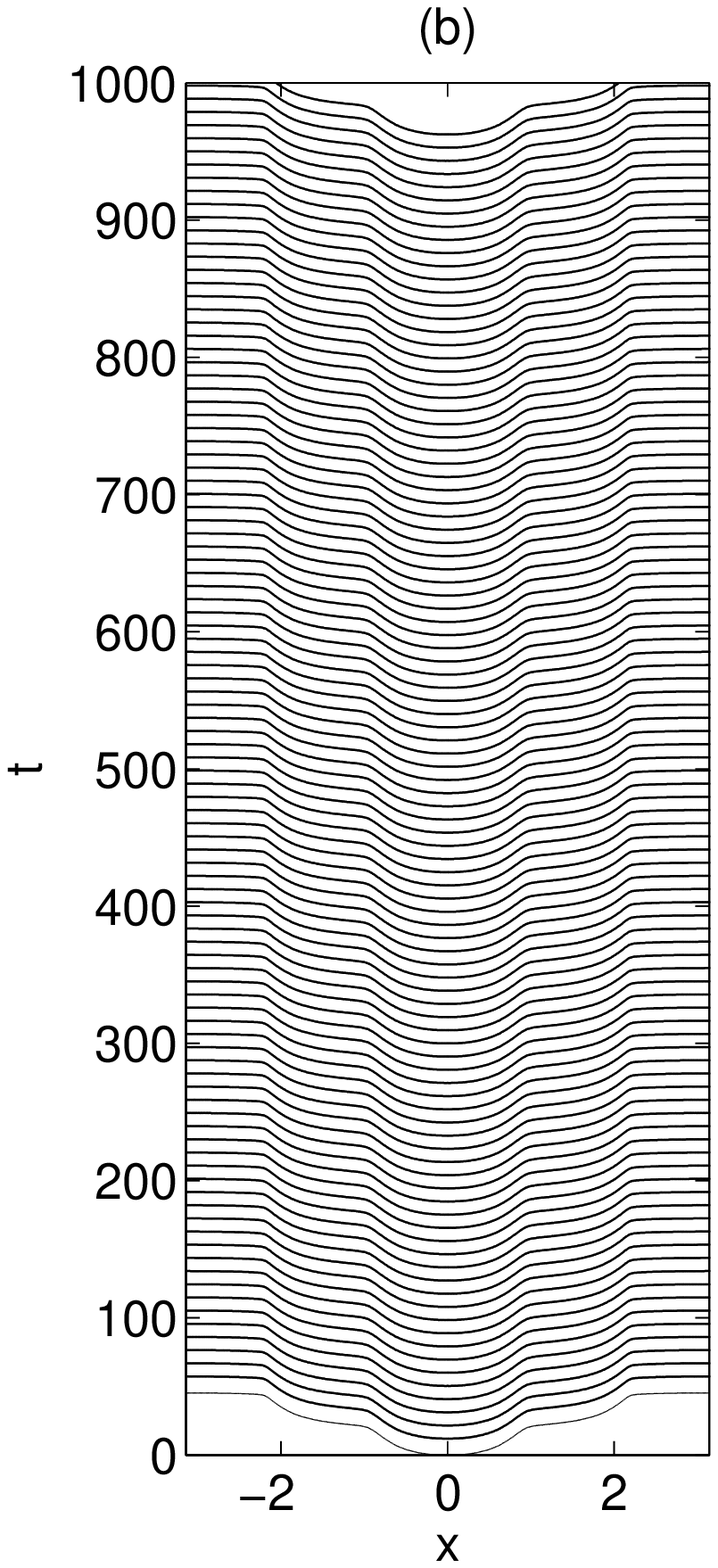}
\caption{Hidden line plot of the phase distribution $\theta(x,t)$ when $\omega_0 =0.12$ and $\beta = 0.76$. (a) Unpinned state obtained from a traveling coherent state with $\omega_0 = 0$ and $\beta = 0.76$ on gradually increasing $\omega_0$ to 0.12. (b) Pinned state obtained from a traveling coherent with $\omega_0 = 0$ and $\beta = 0.75$ on gradually increasing $\omega_0$ to 0.12, and then increasing $\beta $ to 0.76. In both cases $\kappa = 2$ and $N = 512$. }
\label{fig:hidden3}
\end{figure}

We have also conducted experiments at a fixed value of $\omega_0>0$ and $\kappa>0$ while changing $\beta$. For example, at fixed $\omega_0 = 0.12$, $\kappa=2$ and $\beta=0.75$ the system is in the pinned state shown in Fig.~\ref{fig:hidden2}(b). Since the speed of the coherent state with $\omega_0=0$ gradually increases as $\beta$ decreases, we anticipate that a given inhomogeneity will find it harder and harder to pin the state as $\beta$ decreases. This is indeed the case, and we find that there is a critical value of $\beta$ at which the given inhomogeneity is no longer able to pin the structure, with depinning via back and forth oscillations of the structure \cite{Hettel}. For yet smaller values of $\beta$, this type of oscillation also loses stability and evolves into near-splay states. When we increase $\beta$ again we uncover hysteresis in each of these transitions. Figure \ref{fig:hidden3} shows two distinct states at identical parameter values: $\omega_0 = 0.12$, $\kappa=2$ and $\beta = 0.76$ generated using different protocols: Fig.~\ref{fig:hidden3}(a) shows a direction-reversing state evolved from a traveling coherent state when we increase $\omega_0$ from 0 to 0.12 at $\beta = 0.76$, while Fig.~\ref{fig:hidden3}(b) shows a pinned state generated from the pinned state at $\beta = 0.75$ when we change $\beta$ from 0.75 to 0.76 at fixed $\omega_0=0.12$.

\subsection{Periodic inhomogeneity: $\omega(x)\equiv \omega_0\cos(lx)$}


\begin{figure}
\includegraphics[height=3cm]{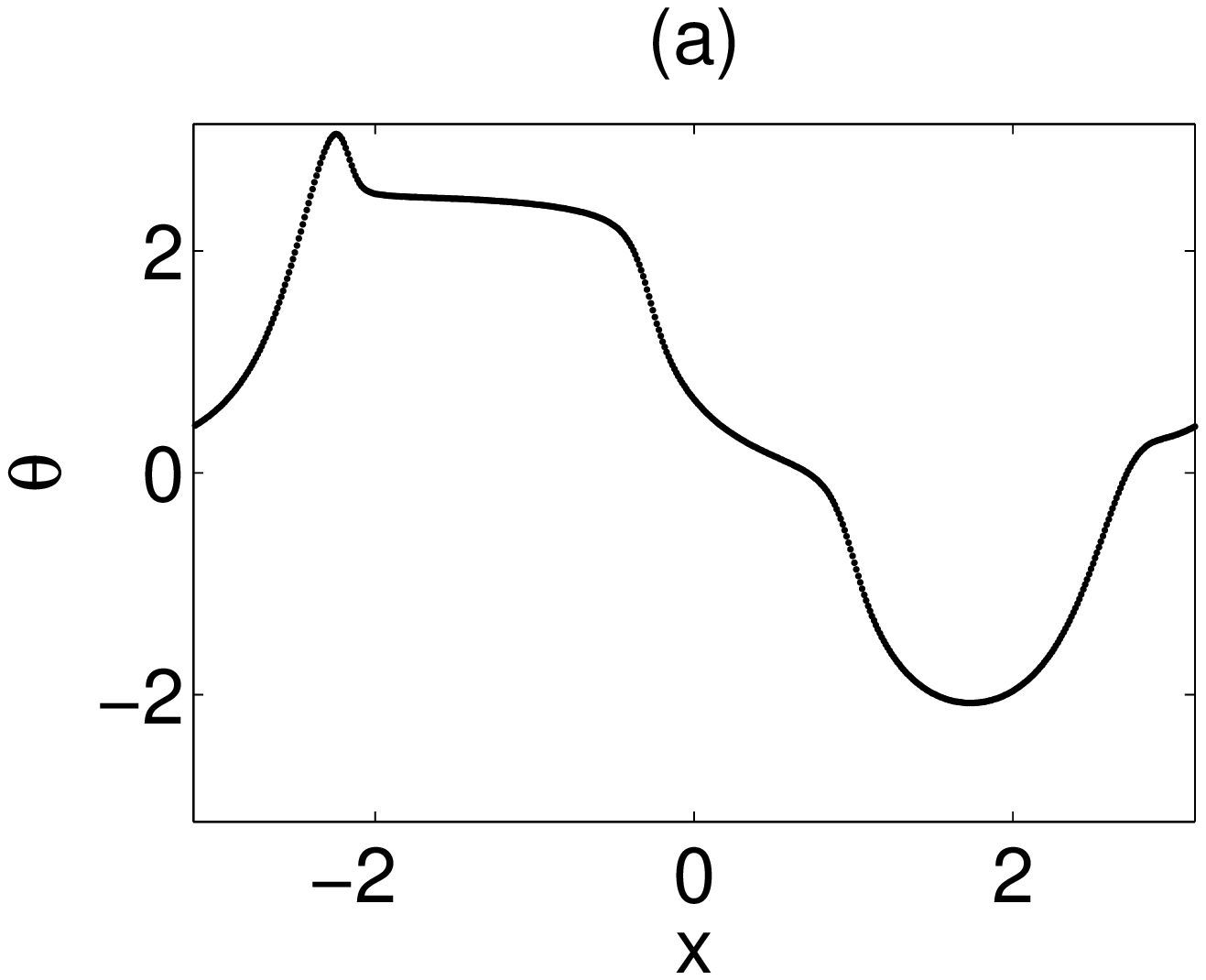}
\includegraphics[height=3cm]{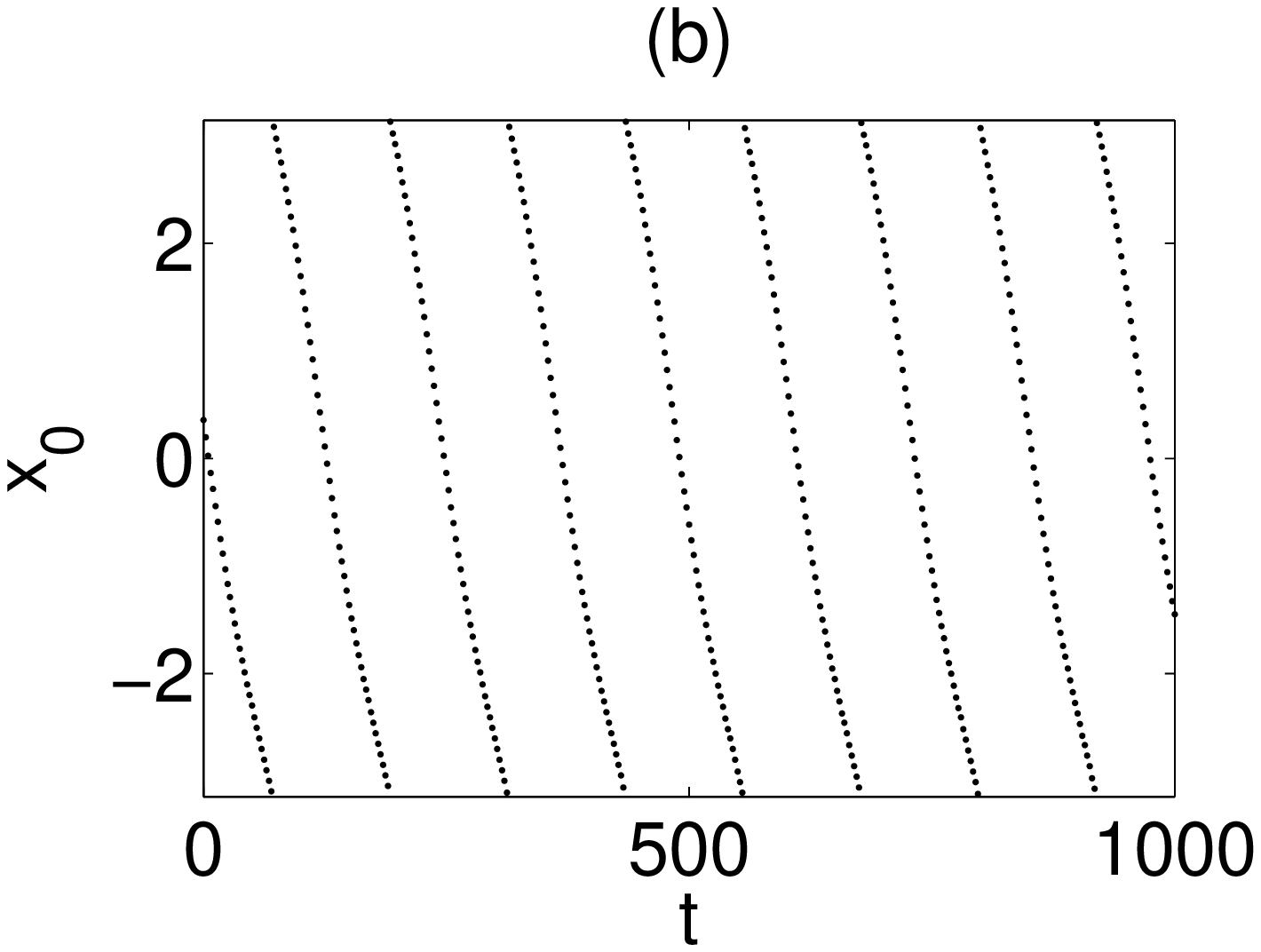}
\caption{(a) A snapshot of the phase distribution $\theta(x,t)$ of a coherent solution when $\omega_0 = 0.028$. (b) The position $x_0$ of the coherent solution as a function of time. In both cases $l=1$, $\beta = 0.05$ and $N = 512$.}
\label{fig:cos12x_l1_omega0_0028}
\end{figure}

\begin{figure}
\includegraphics[height=3.5cm]{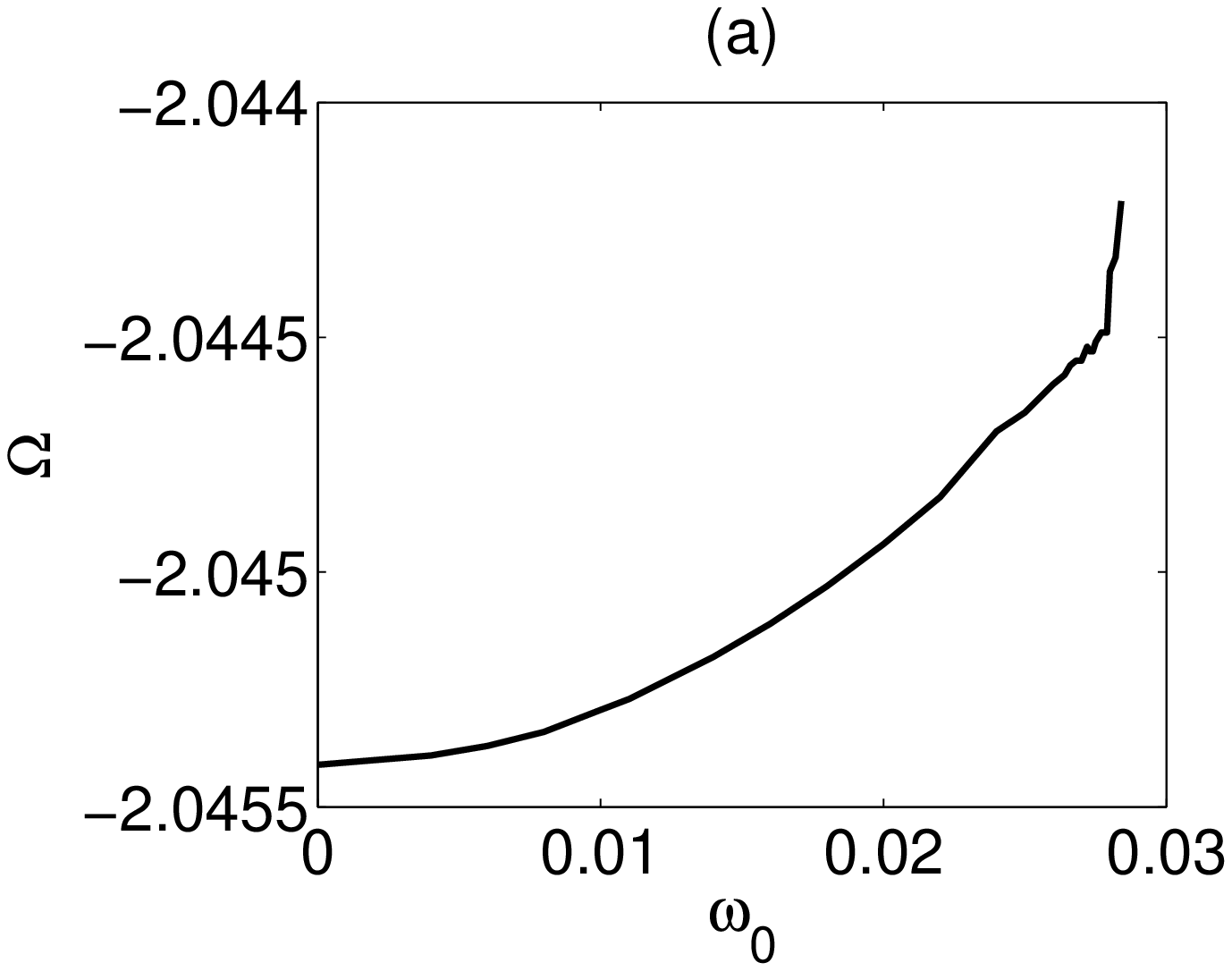}
\includegraphics[height=3.5cm]{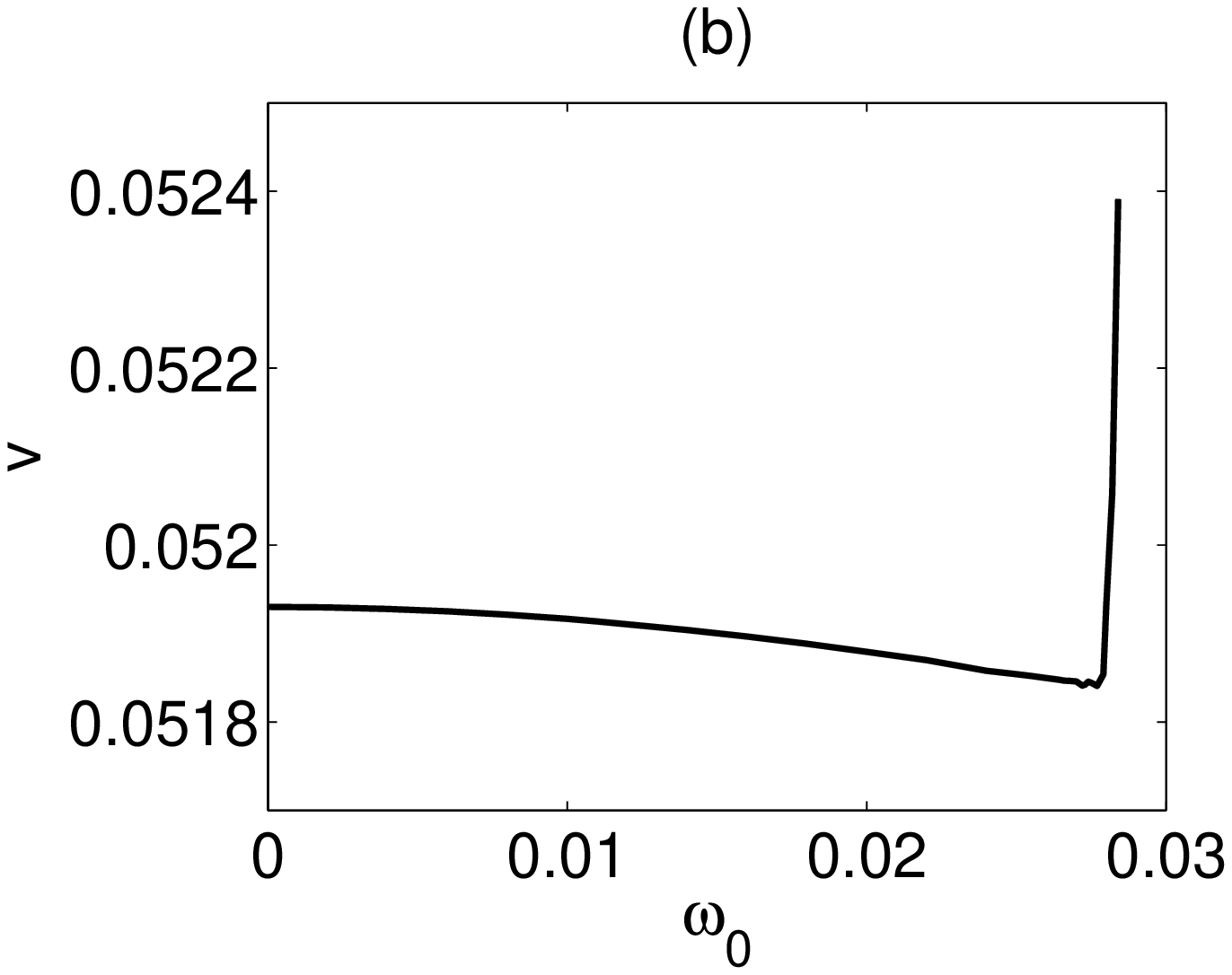}
\caption{(a) The mean angular velocity $\Omega$ and (b) the mean drift speed $v$, both as functions of $\omega_0$ when $l=1$, $\beta=0.7$.}
\label{fig:cos12x_Omegav}
\end{figure}

We now turn to the case $\omega(x) \equiv \omega_0\cos(lx)$. Similar to the bump inhomogeneity case, when $\beta$ is fixed, the traveling coherent state will continue to travel until $\omega_0$ reaches certain threshold. However, we did not find the pinned state as in the previous subsection around $\beta = 0.75$. Here we focus on the case $\beta = 0.7$ for which the traveling coherent state has a reasonable speed when $\omega_0=0$ and take $l=1$. When $\omega_0$ is small, the state remains coherent (Fig.~\ref{fig:cos12x_l1_omega0_0028}(a)) and continues to drift, albeit no longer with a uniform speed of propagation. Figure \ref{fig:cos12x_l1_omega0_0028}(b) shows that the speed executes slow, small amplitude oscillations about a well-defined mean value $v(\omega_0)$ shown in Fig.~\ref{fig:cos12x_Omegav}(b); the corresponding time-averaged oscillation frequency $\Omega(\omega_0)$ is shown in Fig. \ref{fig:cos12x_Omegav}(a). When $\omega_0$ is increased in sufficiently small increments the oscillations grow in amplitude but the solution continues to travel to the left until $\omega_0 \approx 0.285$ where a hysteretic transition to a near-splay state takes place. Figure~\ref{fig:cos12x_Omegav}(b) shows that prior to this transition the average speed first decreases as a consequence of the inhomogeneity, but then increases abruptly just before the transition owing to the loss of coherence on the part of a group of oscillators and the resulting abrupt increase in asymmetry of the order parameter. 

\section{Traveling chimera states}\label{trav}

\begin{figure}
\includegraphics[height=3cm]{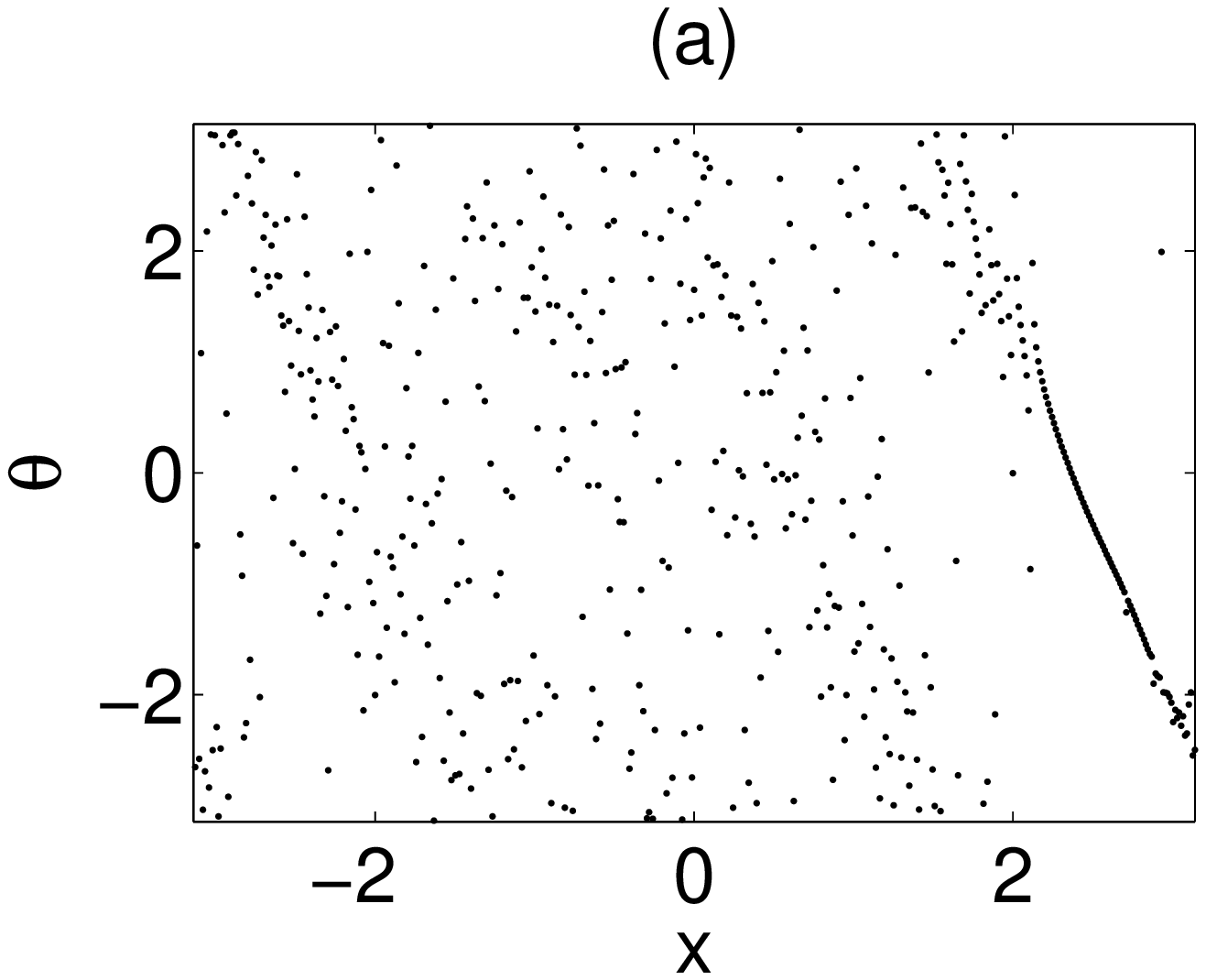}
\includegraphics[height=3cm]{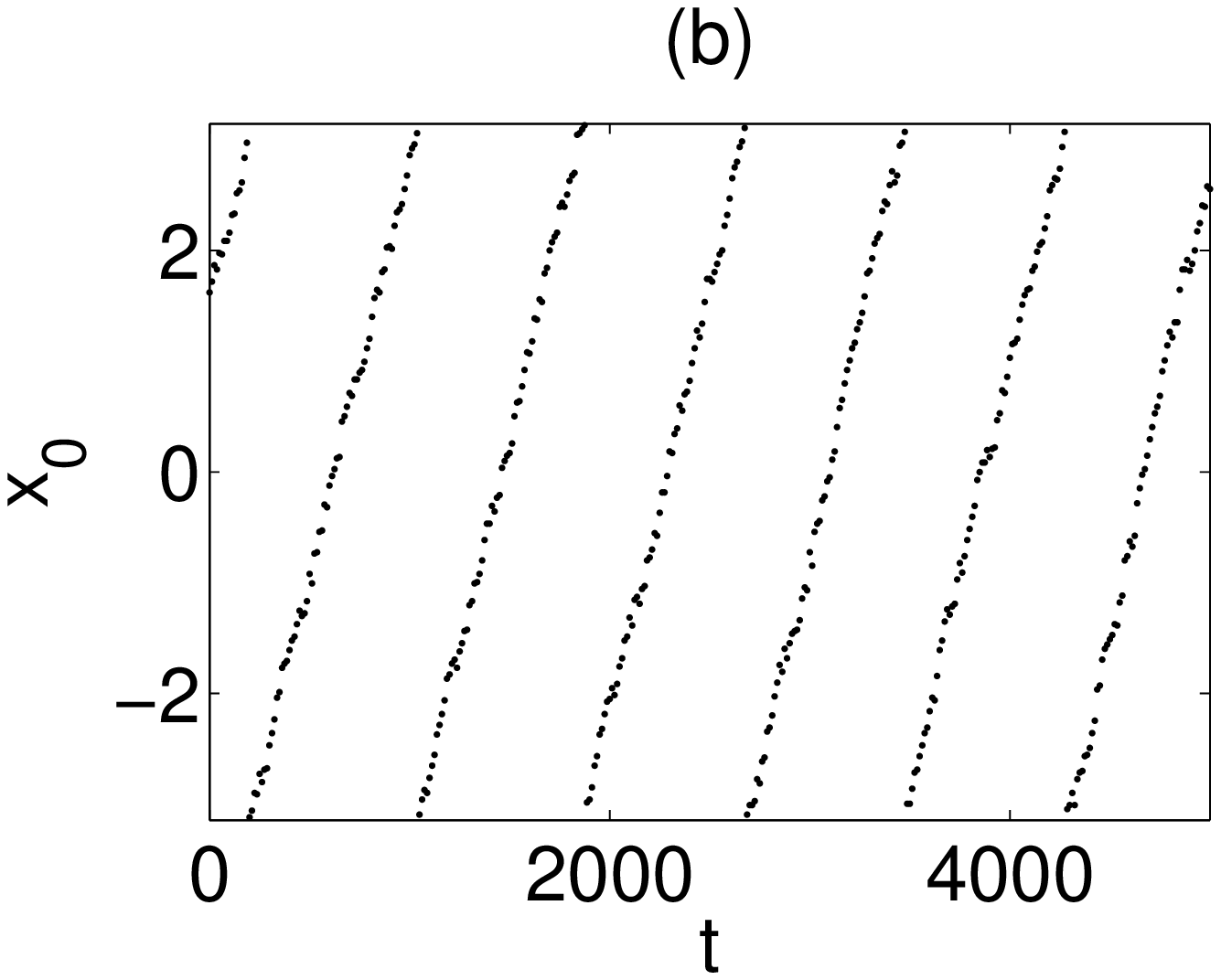}
\caption{(a) A snapshot of the phase distribution $\theta(x,t)$ for a traveling chimera state in a spatially homogeneous system. (b) The position $x_0$ of the coherent cluster as a function of time. The simulation is done for $G(x) \equiv \cos(3x) + \cos(4x)$ with $\beta = 0.03$ and $N = 512$.}
\label{fig:phase_cos34x}
\end{figure}

In addition to the states discussed in the previous sections, Eq.~(\ref{phase_eq}) with constant $\omega$ admits traveling one-cluster chimera states when $G(x)=G^{(2)}_n(x) \equiv \cos(nx) + \cos[(n+1)x]$ and appropriate values of the the phase lag $\beta$. This state consists of a single coherent cluster that drifts through an incoherent background as time evolves at more or less a constant speed. Figure \ref{fig:phase_cos34x}(a) shows a snapshot of such a state when $n = 3$. The direction of motion is determined by the gradient of the phase in the coherent region: the cluster travels to the left when the gradient is positive and to the right when the gradient is negative. Figure \ref{fig:phase_cos34x}(b) shows the position $x_0$ of the coherent cluster as a function of time and confirms that the cluster moves to the right at an almost constant speed. In \cite{XKK2014}, we use numerical simulations to conclude that for $n = 3$ the traveling chimera is stable in the interval $0.015 \lesssim \beta \lesssim 0.065$. We therefore focus on the effects of spatial inhomogeneity on the traveling chimera state when $\beta = 0.03$. In fact the traveling chimera state is more complex than suggested in Fig. \ref{fig:phase_cos34x}(a,b): unlike the states discussed in the previous sections, the profile of the local order parameter fluctuates in time, suggesting that the state does not drift strictly as a rigid object. 

\subsection{Bump inhomogeneity: $\omega(x) \equiv \omega_0 \exp{(-\kappa |x|)}$}

\begin{figure}[!htpb]
\includegraphics[height=4cm]{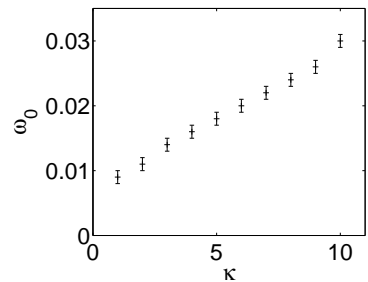}
\caption{Dependence of pinning threshold $\omega_0$ on $\kappa$ when $\beta = 0.03$.}
\label{fig:kappa_omega0}
\end{figure}

In this section we investigate the effect of a bump-like inhomogeneity $\omega(x) \equiv \omega_0 \exp{(-\kappa |x|)}$ on the motion of the traveling chimera state. Starting with the traveling chimera state for $\omega_0 = 0$, we increase $\omega_0$ in steps of $\Delta \omega_0 = 0.01$. To describe the motion of the coherent cluster, we follow the method in \cite{Omelchnko2010} and determine the instantaneous position $x_0$ of the cluster by minimizing the function $F(x^*) = \frac{1}{N} \sum\limits_{k}^{N}[\theta_t(x_k,t) - f(x_k,x^*)]^2$, where $f(x,x^*) = -\cos(x-x^*)$ is a reference profile, and using the minimizer $x^*(t)$ as a proxy for $x_0(t)$. We find that even small $\omega_0$ suffices to stop a traveling chimera from moving: Fig.~\ref{fig:kappa_omega0} shows that the threshold $\omega_0 \approx 0.01$ for $\kappa=1$ and that it increases monotonically to $\omega_0 \approx 0.03$ for $\kappa=10$. The resulting pinned state persists to values of $\omega_0$ as large as $\omega_0 = 1$.

\begin{figure}
\includegraphics[height=3cm]{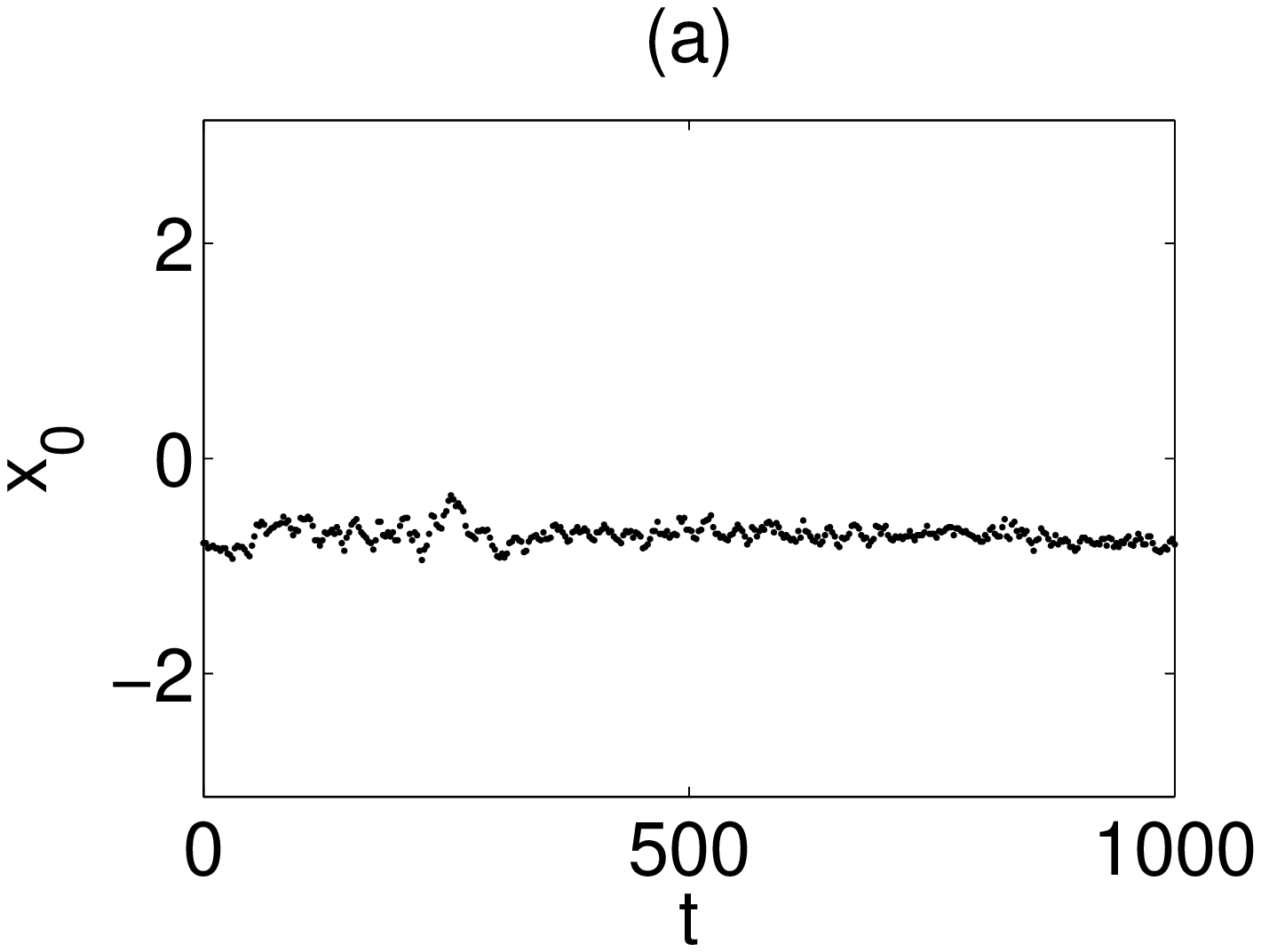}
\includegraphics[height=3cm]{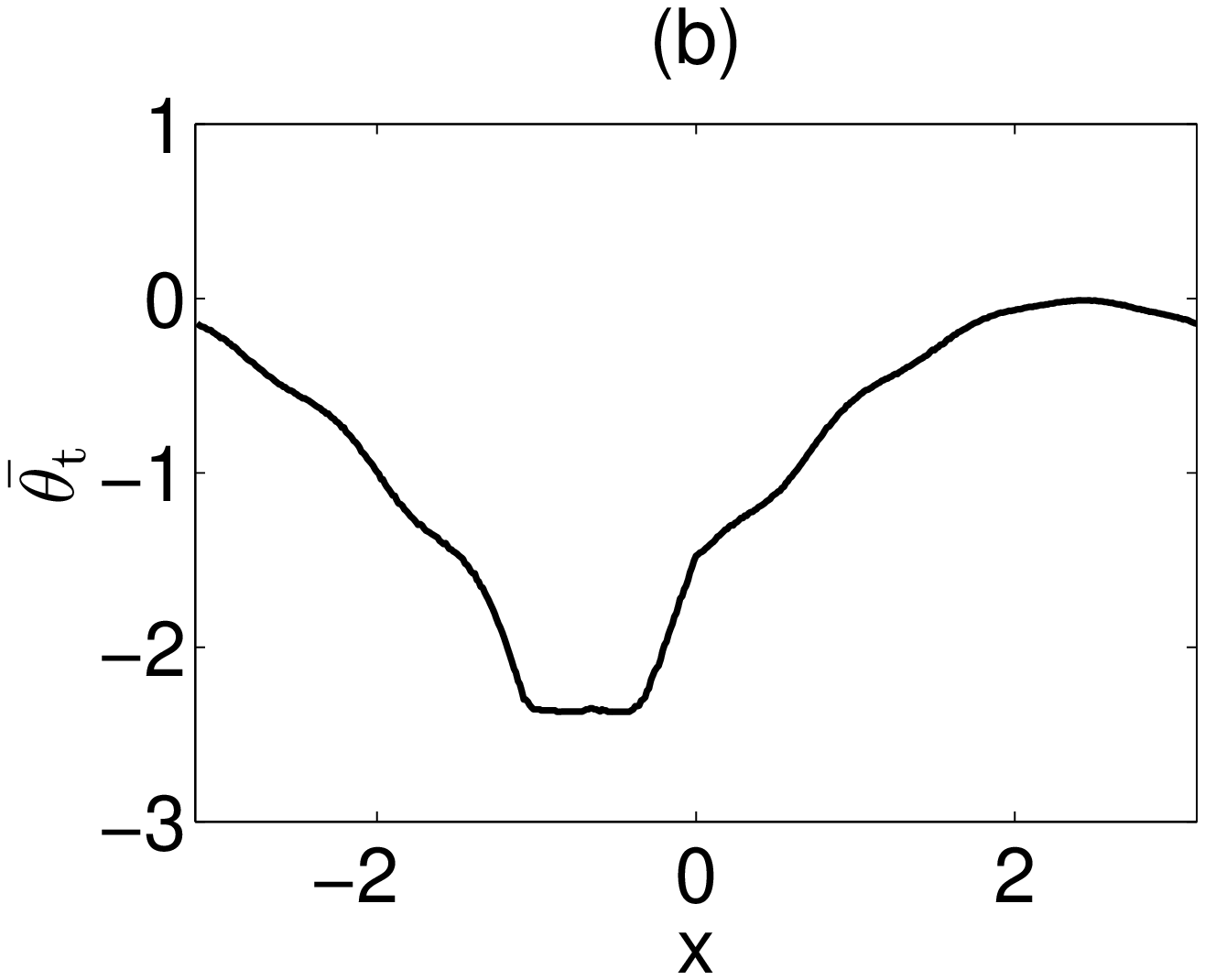}
\includegraphics[height=3cm]{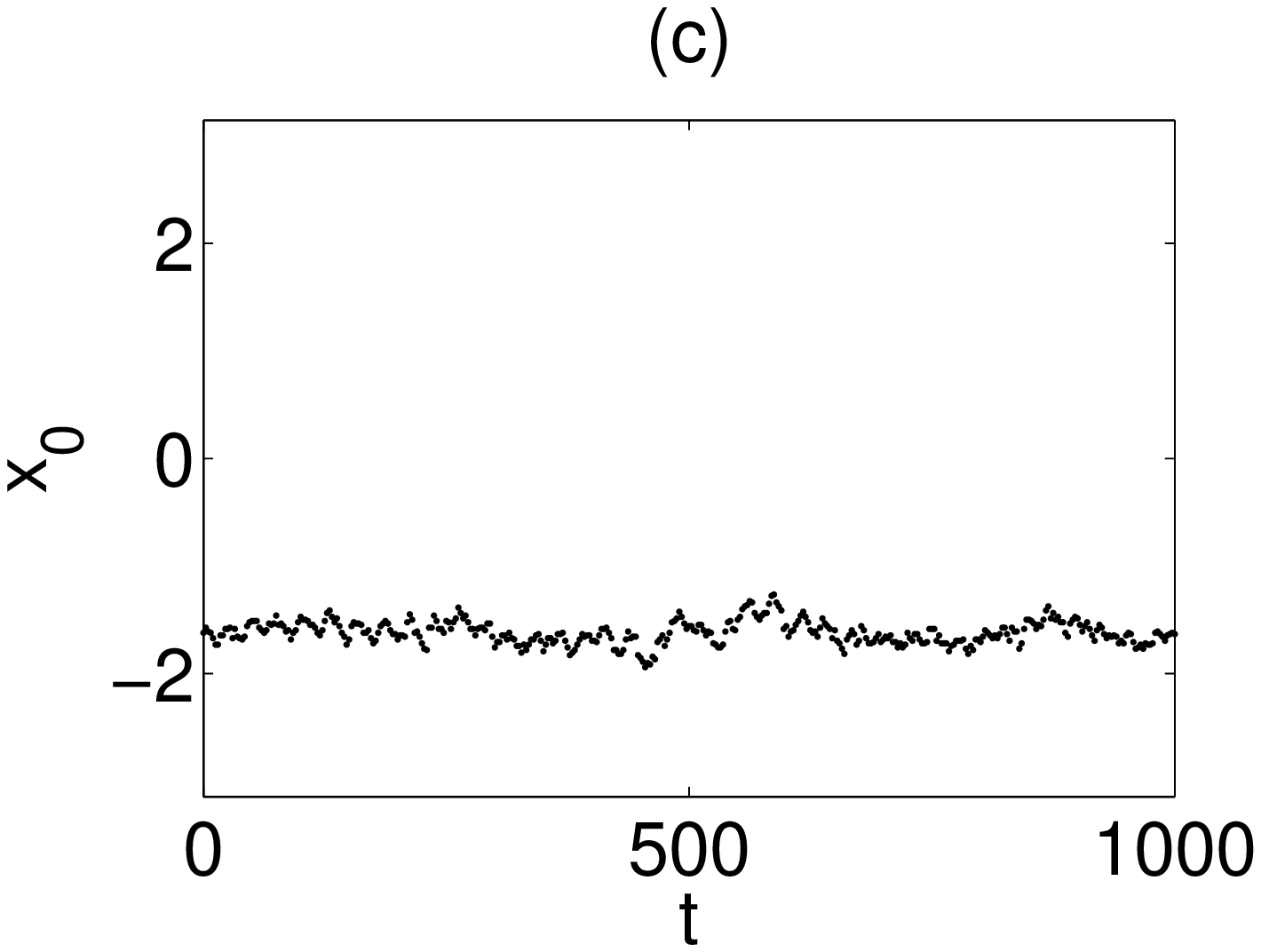}
\includegraphics[height=3cm]{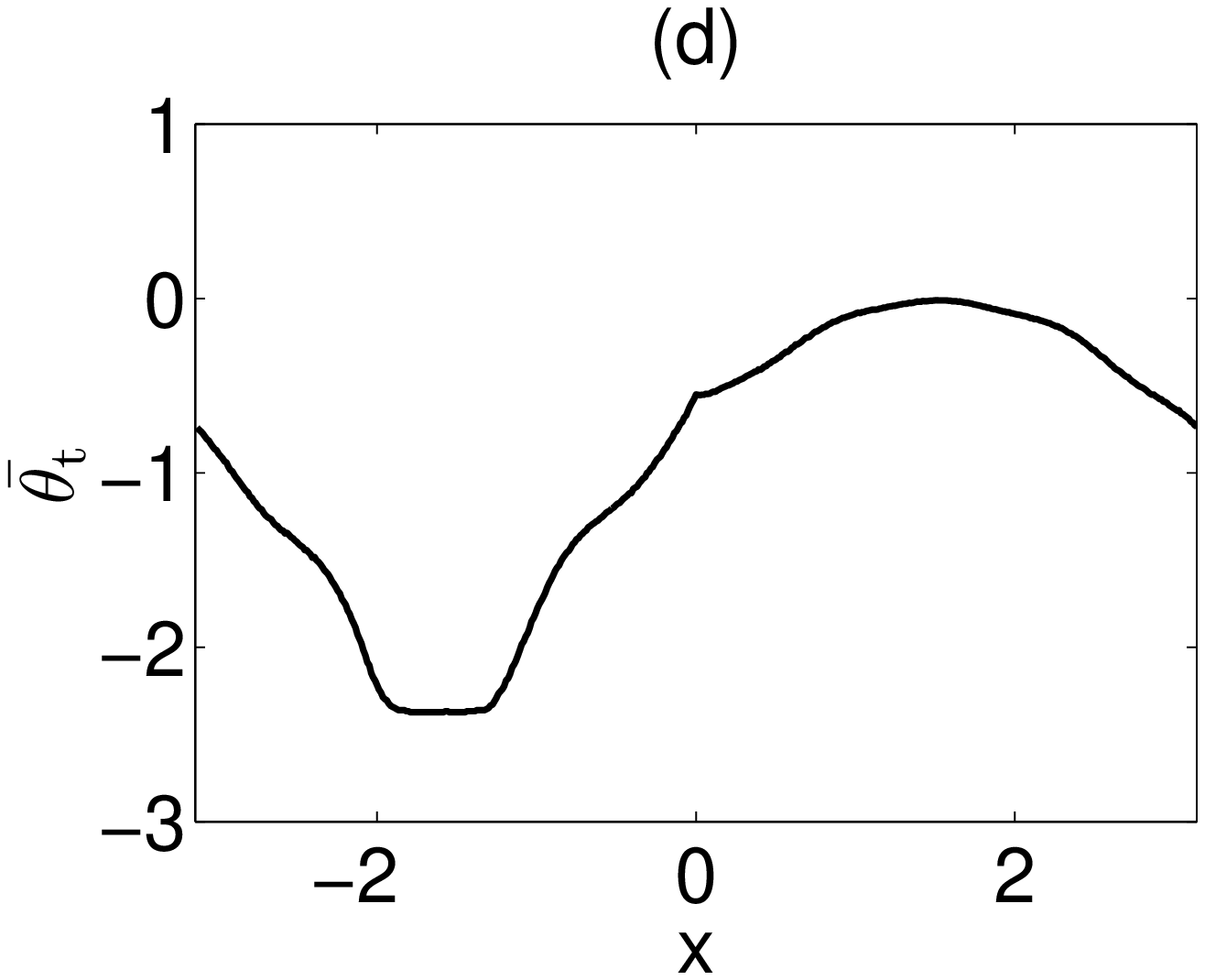}
\includegraphics[height=3cm]{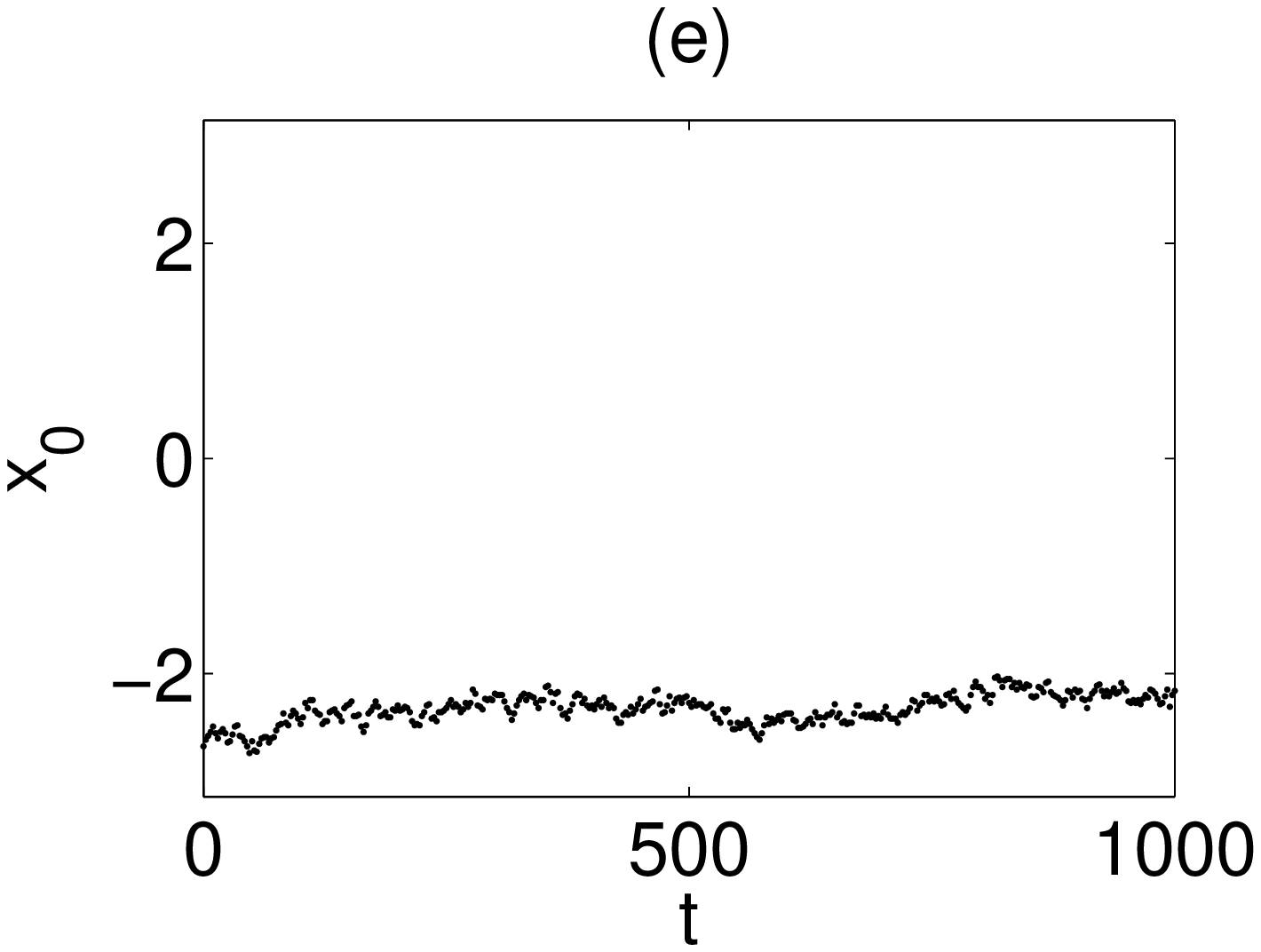}
\includegraphics[height=3cm]{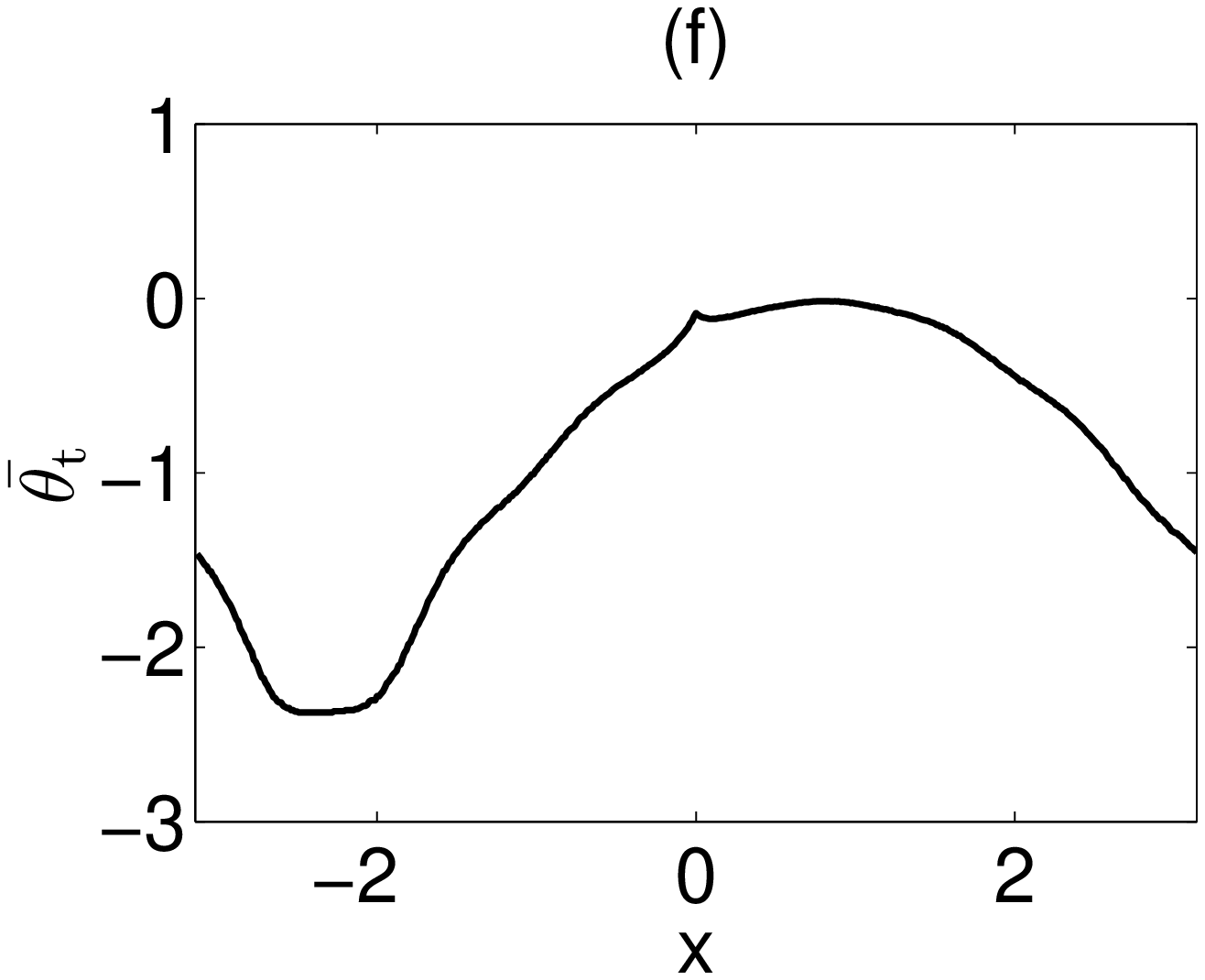}
\caption{The position $x_0$ of the pinned coherent cluster in a traveling chimera state as a function of time when (a) $\omega_0 =0.04$, (c) $\omega_0 =0.08$, (e) $\omega_0 =0.12$. The average rotation frequency ${\bar \theta}_t$ for (b) $\omega_0 =0.04$, (d) $\omega_0 =0.08$, (f) $\omega_0 =0.12$.  In all cases $\beta = 0.03$ and $\kappa = 10$, $N = 512$.}
\label{fig:xvst_travel_bump}
\end{figure} 

Figures \ref{fig:xvst_travel_bump}(a,c,e) show the position $x_0$ of the coherent cluster as a function of time obtained using the above procedure for $\omega_0 = 0.04,0.08,0.12$, respectively, i.e., in the pinned regime. The figures show that the equilibrium position of the coherent region is located farther from the position $x=0$ of the inhomogeneity peak as $\omega_0$ increases. The bump in $\omega(x)$ thus exerts a "repelling force'' on the coherent cluster, whose strength increases with the height of the bump. We interpret this observation as follows. The coherent cluster can only survive when the frequency gradient is sufficiently small, and is therefore repelled by regions where $\omega(x)$ varies rapidly. In the present case this implies that the coherent cluster finds it easiest to survive in the wings of the bump inhomogeneity, and this position moves further from $x=0$ as $\omega_0$ increases. This interpretation is confirmed in Figs. \ref{fig:xvst_travel_bump}(b,d,f) showing the average rotation frequency, ${\bar \theta}_t(x)$, of the oscillators. The plateau in the profile of $\bar{\theta_t}(x)$ indicates frequency locking and hence the location of the coherent cluster; the fluctuations in the position of the coherent cluster are smoothed out by the time-averaging.

\subsection{Periodic inhomogeneity: $\omega(x)\equiv \omega_0 \cos(lx)$}

\begin{figure}
\includegraphics[height=3cm]{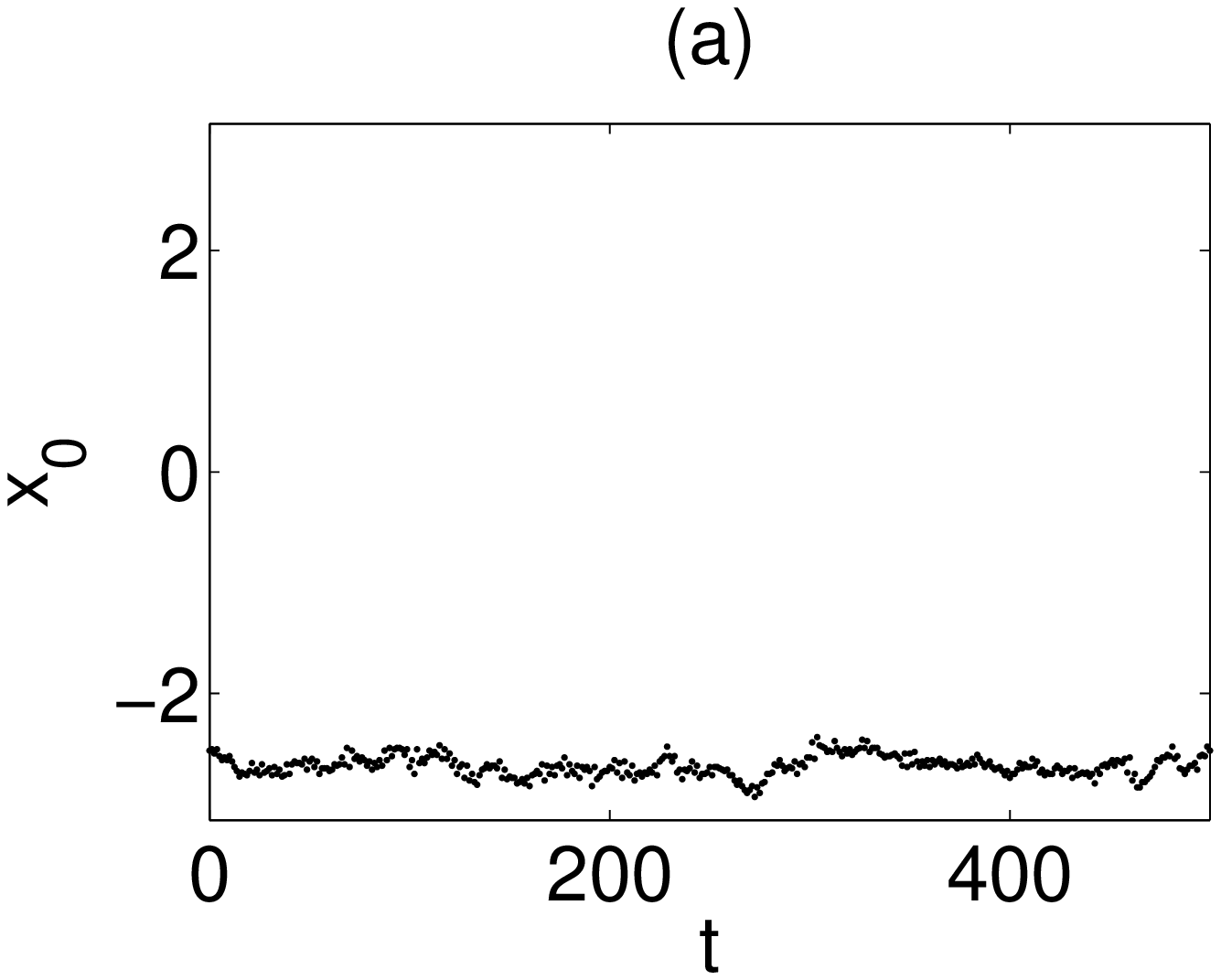}
\includegraphics[height=3cm]{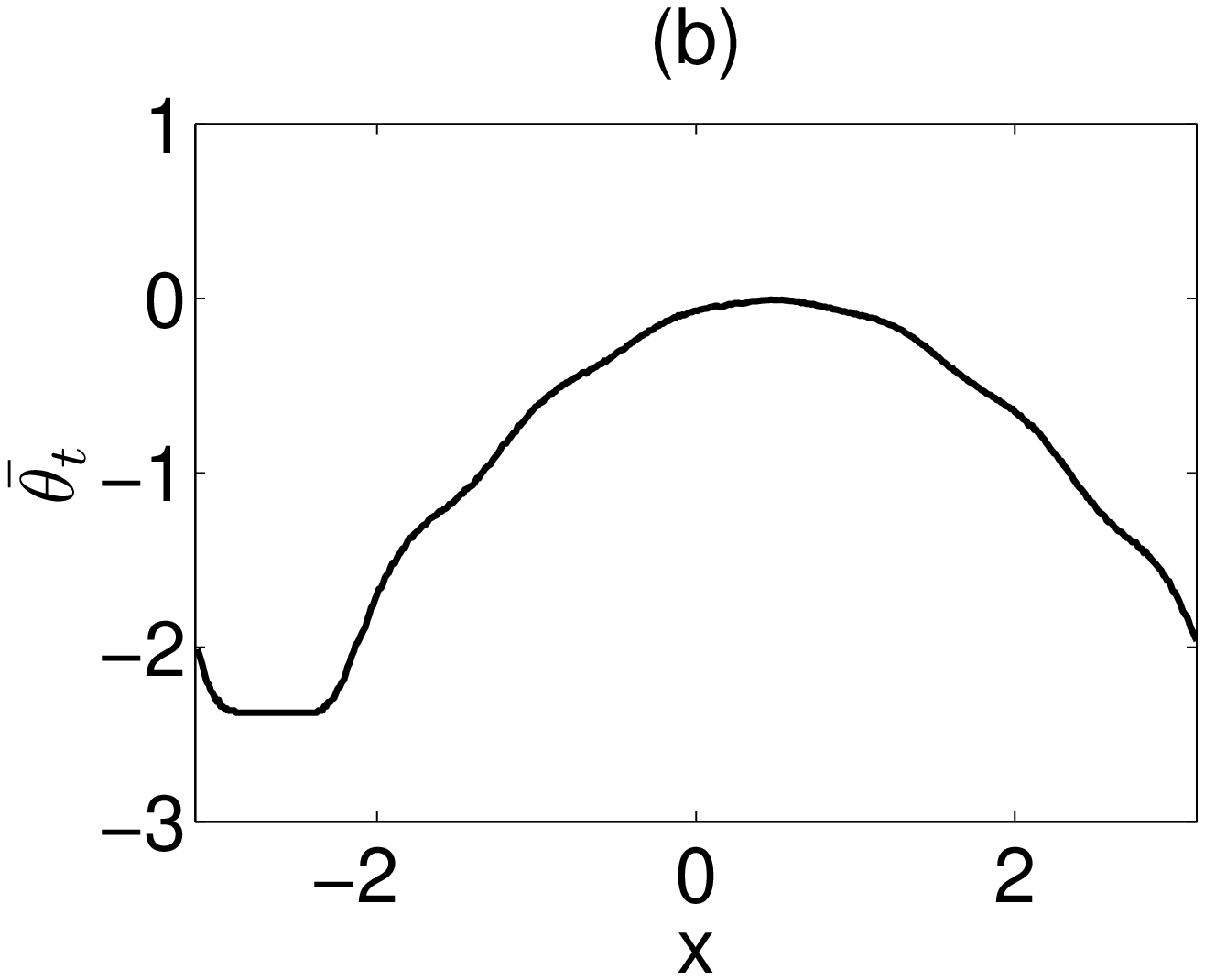}
\includegraphics[height=3cm]{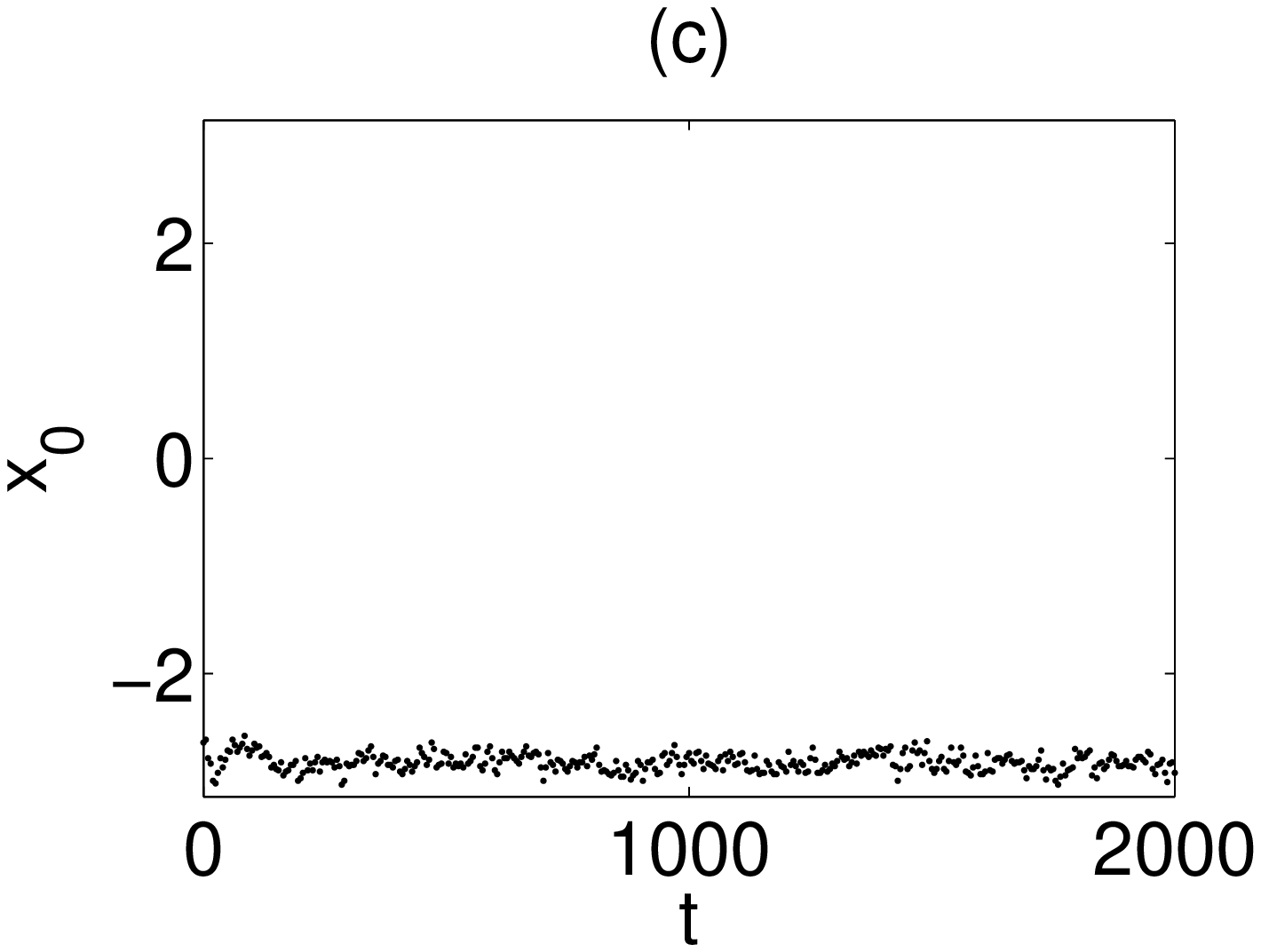}
\includegraphics[height=3cm]{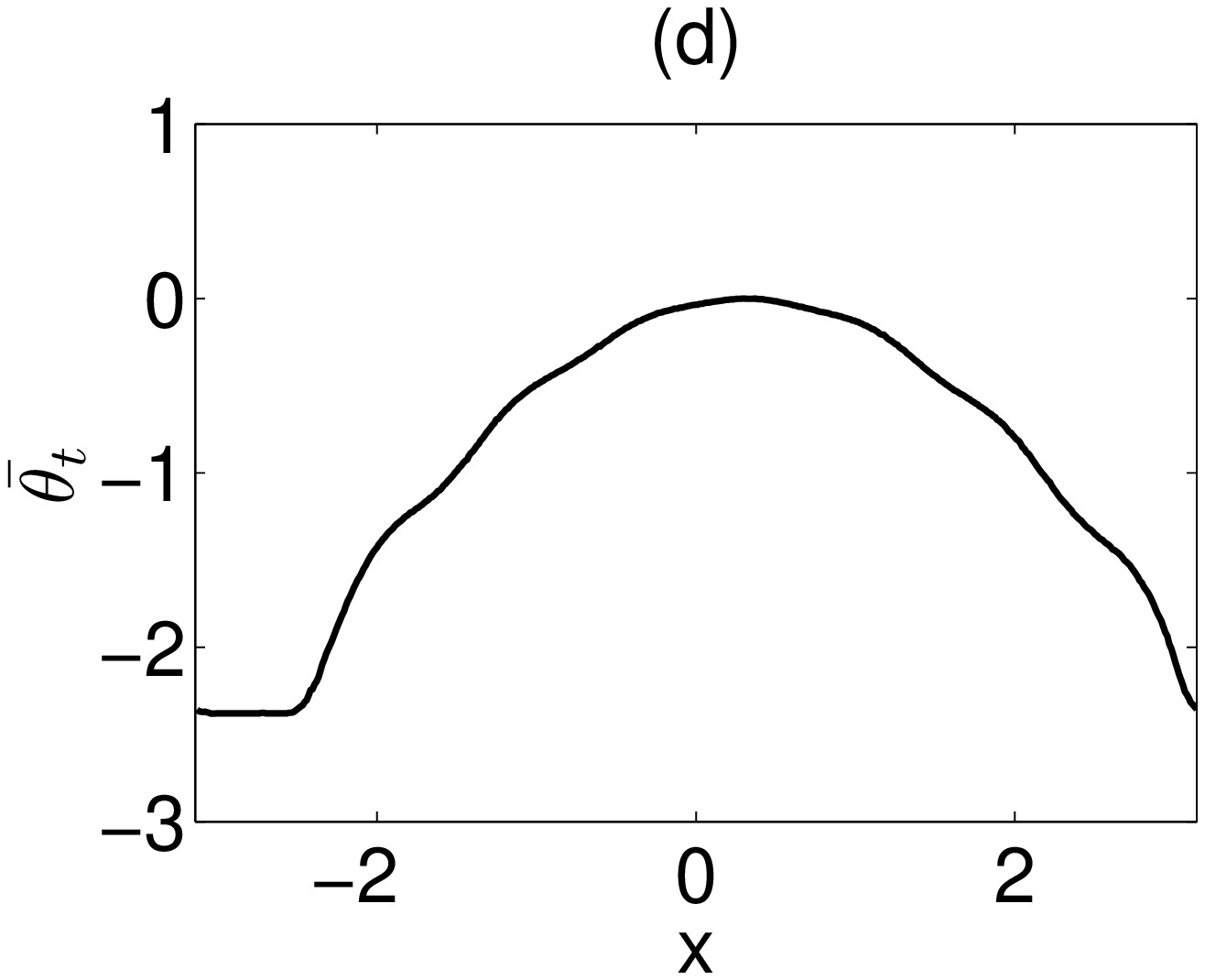}
\caption{The position $x_0$ of the pinned coherent cluster in a traveling chimera state as a function of time when (a) $\omega_0 =0.005$, (c) $\omega_0 =0.01$. The average rotation frequency ${\bar \theta}_t$ for (b) $\omega_0 =0.005$, (d) $\omega_0 =0.01$. In all cases $\beta = 0.03$, $l = 1$ and $N = 512$.}
\label{fig:xvst_travel_periodic_l1}
\end{figure}

\begin{figure}
\includegraphics[height=3cm]{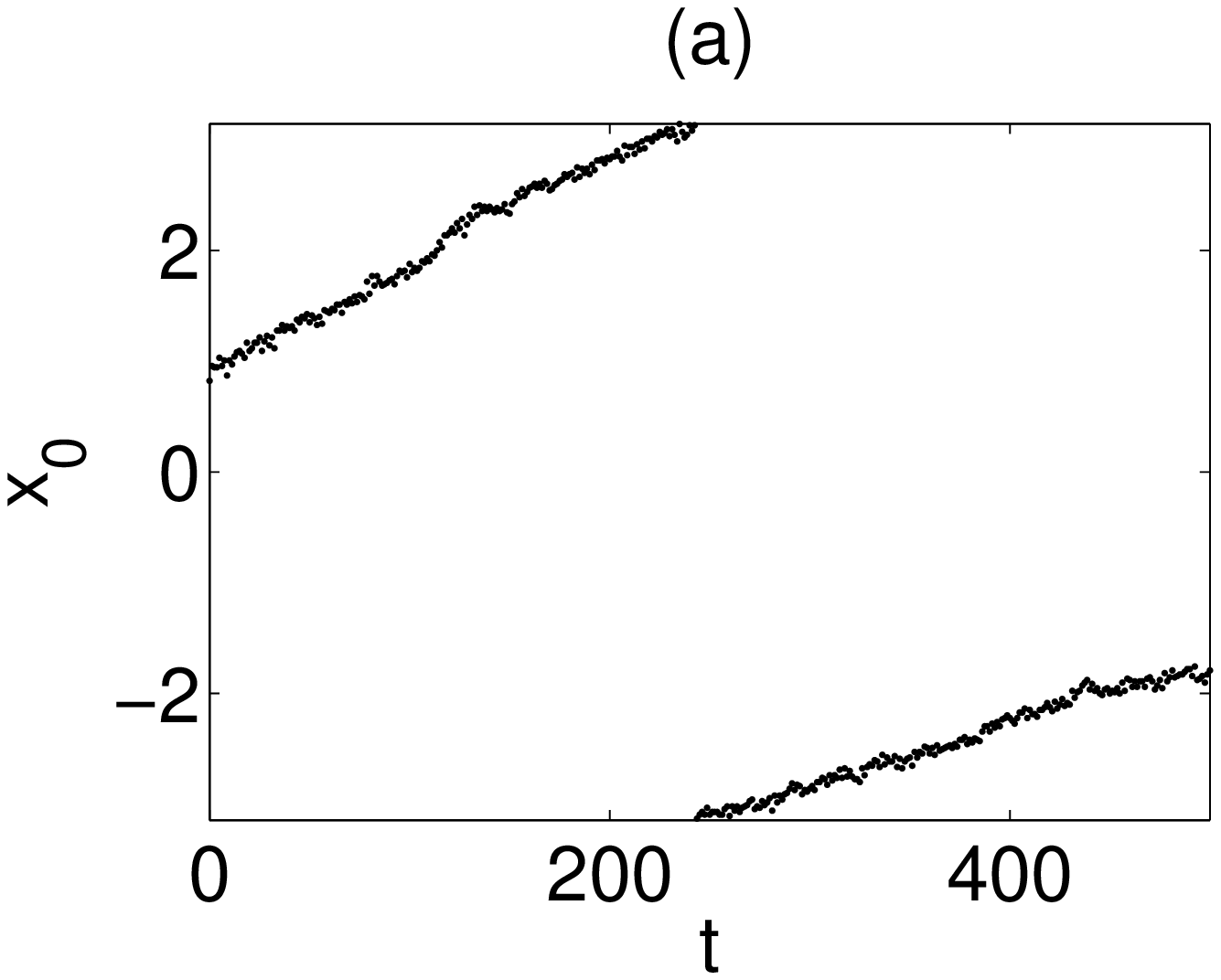}
\includegraphics[height=3cm]{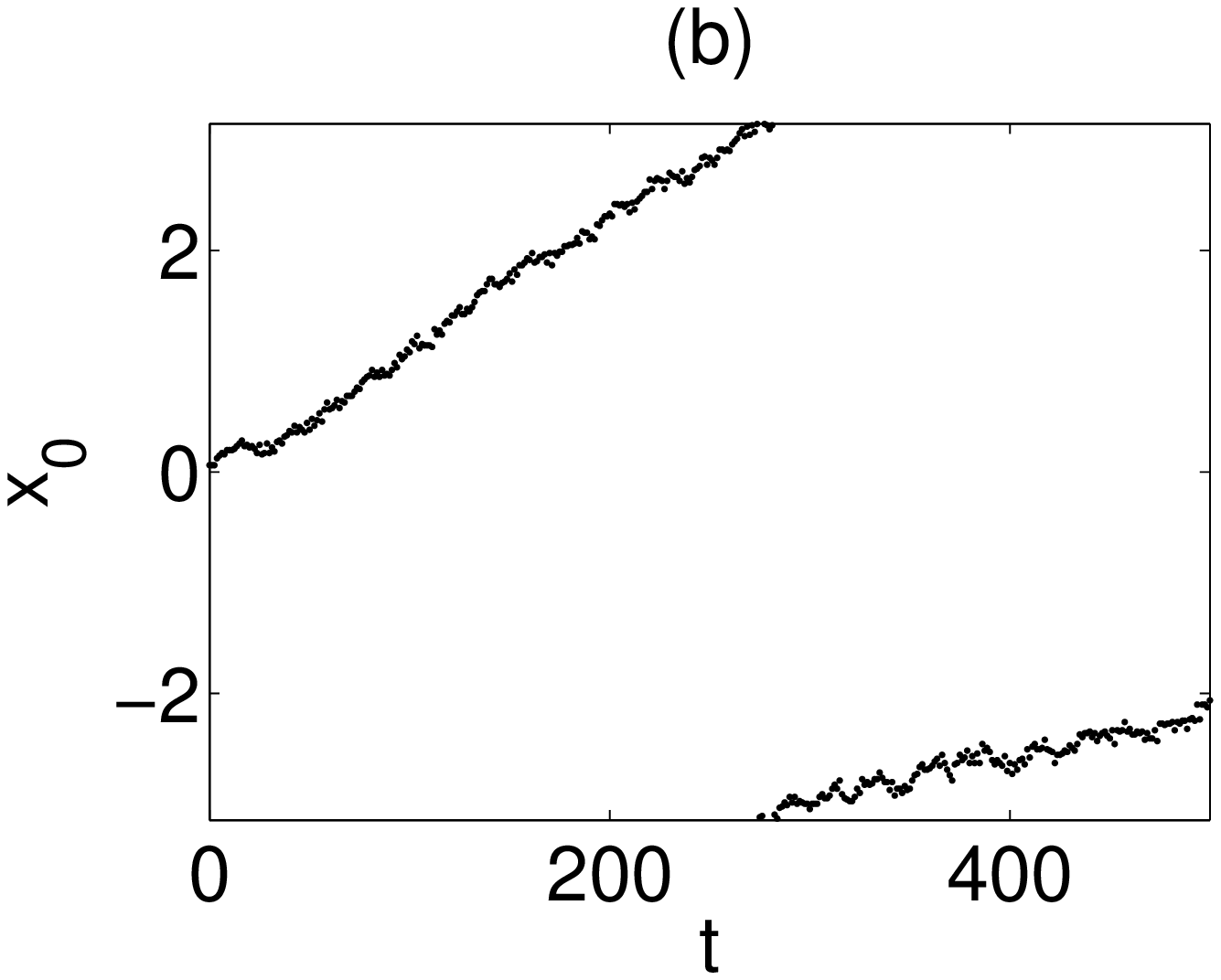}
\caption{The position $x_0$ of the coherent cluster in a traveling chimera state as a function of time when (a) $\omega_0 =0.001$, (b) $\omega_0 =0.002$. In all cases $\beta = 0.03$, $l = 1$ and $N = 512$.}
\label{fig:xvst_travel_periodic_l1_2}
\end{figure}

We now turn to the effects of a periodic inhomogeneity $\omega(x) \equiv \omega_0 \cos(lx)$. When $l=1$ and $\omega_0$ increases the coherent cluster initially travels with a non-constant speed but then becomes pinned in place; as in the case of the bump inhomogeneity quite small values of $\omega_0$ suffice to pin the coherent cluster in place (for $l = 1$ the value $\omega_0 \approx 0.003$ suffices). As shown in Figs.~\ref{fig:xvst_travel_periodic_l1}(a,c) the position $x_0$ of the pinned cluster relative to the local maximum of the inhomogeneity (i.e., $x=0$) depends on the value of $\omega_0>0.003$. As shown in Figs. \ref{fig:xvst_travel_periodic_l1_2}(a,b) the coherent cluster travels to the right and does so with a speed that is larger when $x_0(t)>0$ than when $x_0(t)<0$. This effect becomes more pronounced as $\omega_0$ increases. This is because the coherent structure is asymmetric, with a preferred direction of motion, and this asymmetry increases with $\omega_0$. Evidently, the speed of the synchronization front at the leading edge is enhanced when $\omega'(x)<0$ but suppressed when $\omega'(x)>0$ and likewise for the desynchronization front at the rear.

\begin{figure}
\includegraphics[height=3cm]{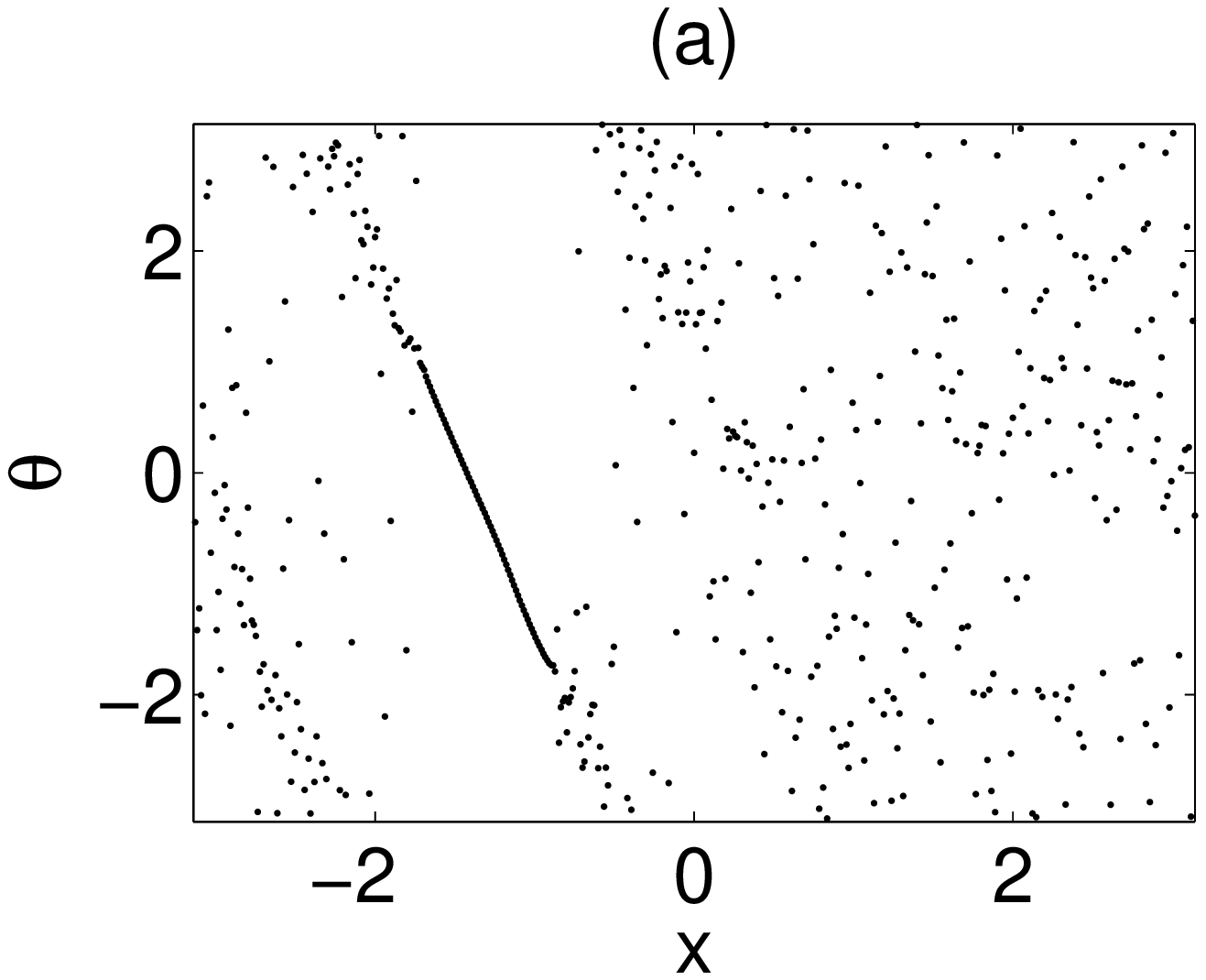}
\includegraphics[height=3cm]{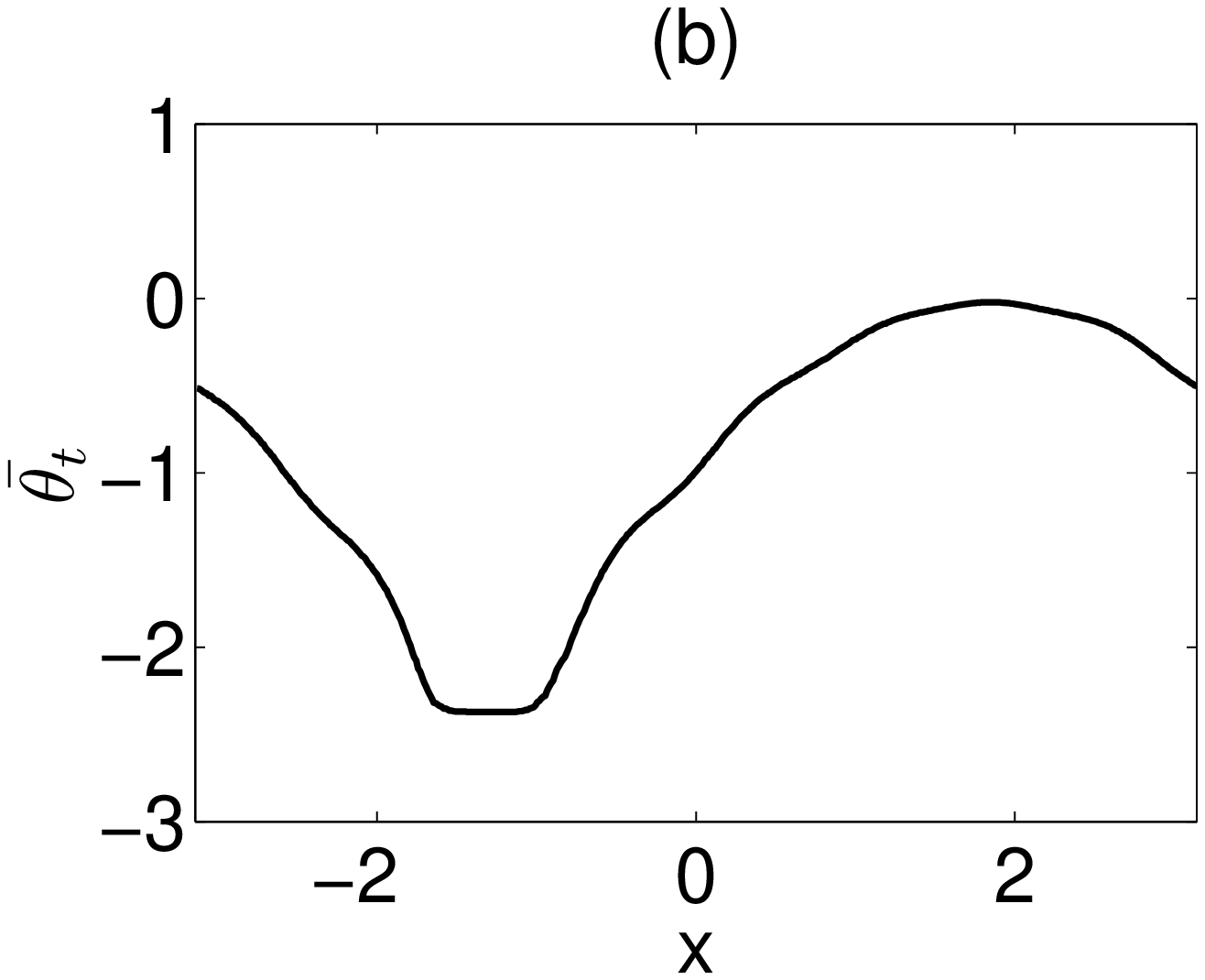}
\includegraphics[height=3cm]{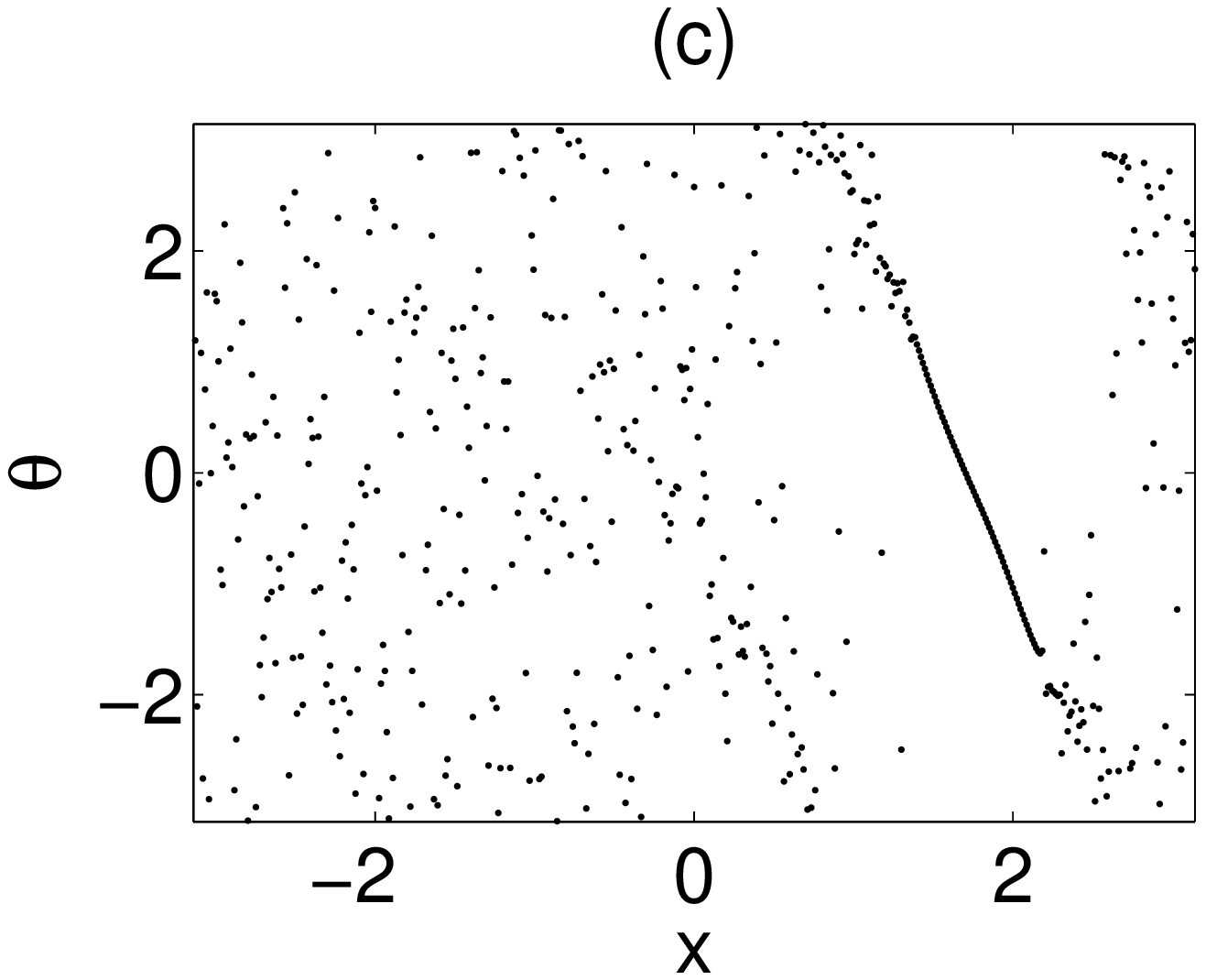}
\includegraphics[height=3cm]{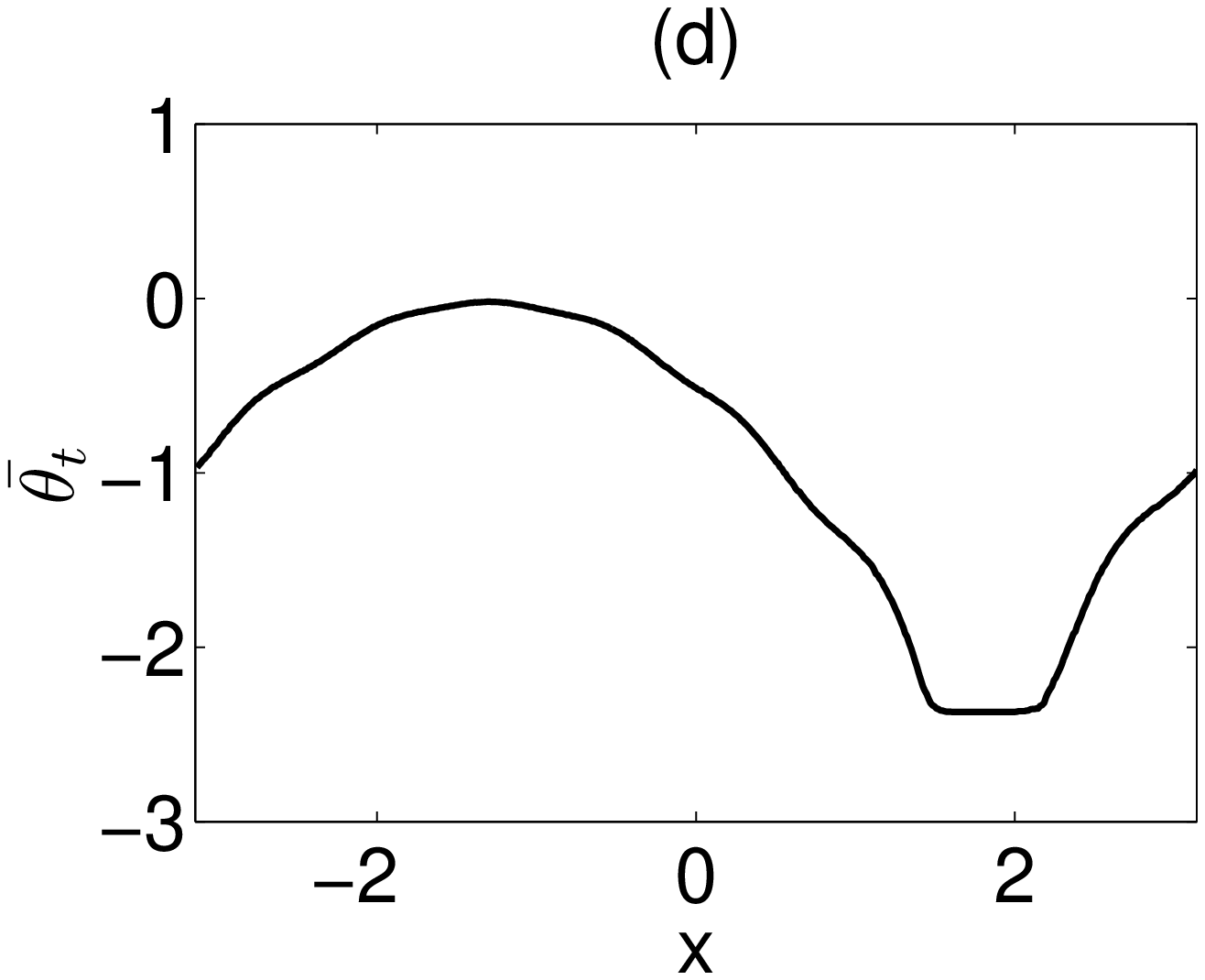}
\caption{(a,c) The two possible phase distributions $\theta(x,t)$ of pinned traveling chimera states when $\omega_0 =0.01$, $\beta = 0.03$ and $l = 2$. (b,d) The corresponding average rotation frequencies ${\bar \theta}_t$. In both cases $N = 512$.}
\label{fig:xvst_travel_periodic_l2}
\end{figure}

For $l>1$ we observe similar results. The coherent cluster is pinned in space already at small values of $\omega_0$. Since the inhomogeneous system has the discrete translation symmetry $x\to x + \frac{2\pi}{l}$ the coherent cluster has $l$ possible preferred positions. Figure \ref{fig:xvst_travel_periodic_l2} shows an example for $l = 2$. Panels (a,c) show snapshots of the phase distribution $\theta(x,t)$ for $\omega_0 = 0.01$ in the two preferred locations (separated by $\Delta x=\pi$), while panels (b,d) show the corresponding average rotation frequency ${\bar \theta}_t$. 

\section{Conclusion}\label{conclusion}

In this paper we have investigated a system of non-identical phase oscillators with nonlocal coupling, focusing on the effects of weak spatial inhomogeneity in an attempt to extend earlier results on identical oscillators to more realistic situations. Two types of inhomogeneity were considered, a bump inhomogeneity in the frequency distribution specified by $\omega(x) = \omega_0 \exp(-\kappa |x|)$ and a periodic inhomogeneity specified by $\omega(x) = \omega_0 \cos(lx)$. In each case we examined the effect of the amplitude $\omega_0$ of the inhomogeneity and its spatial scale $\kappa^{-1}$ ($l^{-1}$) on the properties of states known to be present in the homogeneous case $\omega_0=0$, viz., splay states and stationary chimera states, traveling coherent states and traveling chimera states \cite{XKK2014}.

We have provided a fairly complete description of the effects of inhomogeneity on these states for the coupling functions $G(x) = \cos(x)$, $\cos(x)+ \cos(2x)$ and $\cos(3x)+ \cos(4x)$ employed in \cite{XKK2014}. Specifically, we found that as the amplitude of the inhomogeneity increased a splay state turned into a near-splay state, characterized by a nonuniform spatial phase gradient, followed by the appearance of a stationary incoherent region centered on the location of maximum inhomogeneity amplitude. With further increase in $\omega_0$ additional intervals of incoherence opened up, leading to states resembling the stationary multi-cluster chimera states also present in the homogeneous system. These transitions, like many of the transitions identified in this paper, could be understood with the help of a self-consistency analysis based on the Ott-Antonsen Ansatz \cite{OA2008}, as described in the Appendix. The effect of inhomogeneity on multi-cluster chimera states was found to be similar: the inhomogeneity trapped the coherent clusters in particular locations, and eroded their width as its amplitude $\omega_0$ increased, resulting in coalescence of incoherent regions with increasing $\omega_0$. 

More significant are the effects of inhomogeneity on traveling coherent and traveling chimera states. Here the inhomogeneity predictably pins the traveling structures but the details can be complex. Figures \ref{fig:hidden1}--\ref{fig:hidden2} show one such complex pinning transition that proceeds via an intermediate direction-reversing traveling wave. These waves are generated directly as a consequence of the inhomogeneity and would not be present otherwise, in contrast to homogeneous systems undergoing a symmetry-breaking Hopf bifurcation as described in \cite{Landsberg}. Many of the pinning transitions described here are hysteretic as demonstrated in Fig.~\ref{fig:hidden3}. The traveling chimera states are particularly fragile in this respect, with small amplitude inhomogeneities sufficient to arrest the motion of these states. In all these cases the coherent regions are found in regions of least inhomogeneity, an effect that translates into an effective repulsive interaction between the coherent cluster and the inhomogeneity.

In future work we propose to explore similar dynamics in systems of more realistic nonlocally coupled oscillators and compare the results with those for similar systems with a random frequency distribution.

\begin{acknowledgments}
This work was supported in part by a National Science Foundation Collaborative Research Grant CMMI-1232902.
\end{acknowledgments}

\appendix
\section{Derivation of the self-consistency equation}\label{derive}
We suppose that an oscillator at position $x$ has intrinsic frequency $\omega(x)$, and assume that the frequency distribution $\omega(x)$ is continuous. The model equation is
\begin{equation}
\frac{\partial \theta}{\partial t}=\omega(x)-\int^{\pi}_{-\pi} G(x-y)\sin(\theta(x,t)-\theta(y,t)+\alpha)\,dy.
\end{equation}
We next introduce the probability density function $f(x,\omega,\theta,t)$ characterizing the state of the system. This function must satisfy the continuity equation
\begin{equation}
\frac{\partial f}{\partial t}+\frac{\partial}{\partial \theta}(f v)=0,
\end{equation}
where $v(x,t)$ satisfies the relation 
\begin{equation}
v(x,t)=\omega(x)-\int^{\pi}_{-\pi}G(x-y)f'(y,t)\, dy
\end{equation}
and $f'(y,t) = \int^{\infty}_{-\infty}\int_{-\pi}^{\pi}\sin(\theta-\theta'+\alpha)f(y,\omega,\theta',t)\,d\theta' d\omega$. We also define the local order parameter
\begin{equation}
Z(x,t)=\int^{\pi}_{-\pi}G(x-y)\int^{\infty}_{-\infty}\int_{-\pi}^{\pi}e^{i\theta'}f(y,\omega,\theta',t)\,d\theta'd\omega dy.
\end{equation}
Then
\begin{equation}
v=\omega-\frac{1}{2i}(Z^*e^{i(\theta +\alpha)}-Ze^{-i(\theta+\alpha)}].
\end{equation}

The above equations can be recast in a more convenient form using the Ott-Antonsen Ansatz \cite{OA2008}
\begin{equation}
f(x,\omega, \theta,t)=\frac{g(x,\omega)}{2\pi}[1+{\sum_{n=1}^{\infty}a^n(x,\omega,t)e^{-in\theta}+c.c.}].
\end{equation}
Here $g(x,\omega)$ represents the distribution of natural frequencies at each $x$. Matching terms proportional to different powers of $\exp i\theta$, we obtain
\begin{equation}
\frac{\partial a}{\partial t}=i\omega a-\frac{1}{2}[Ze^{-i\alpha}-Z^*e^{i\alpha}a^2],
\end{equation}
where the complex order parameter $Z(x,t)$ is given by
\begin{equation}
Z(x,t)=\int^{\pi}_{-\pi}G(x-y)\int^{\infty}_{-\infty}g(x,\omega)a(y,\omega,t)d\omega \,dy.
\end{equation}
If we take
\begin{equation}
g(x,\omega)=\frac{D}{\pi((\omega-\omega(x))^2+D^2)}
\end{equation}
set $z(x,t)=a(x,\omega(x)-iD,t)$ and perform the implied contour integration, we obtain
\begin{equation}
Z(x,t)=\int^{\pi}_{-\pi}G(x-y)z(y,t) \,dy.\label{Zdef}
\end{equation}
In the limit $D \to 0$ the distribution function $g$ reduces to a delta function. The corresponding quantity $z(x,t)$ satisfies 
\begin{equation}
\frac{\partial z}{\partial t}=i\omega z+\frac{1}{2}[Z\exp(-i\alpha)-Z^*\exp(i\alpha)z^2],\label{zdef}
\end{equation}
where $Z(x,t)$ is given by (\ref{Zdef}). Equations (\ref{Zdef})--(\ref{zdef}) constitute the required self-consistency description of the nonlocally coupled phase oscillator system with an inhomogeneous frequency distribution $\omega(x)$.

\end{document}